\newif\ifextended  
\extendedtrue

\ifextended
\documentclass[acmsmall,screen,authorversion,nonacm,review=false,timestamp=false]{acmart}
\else
\documentclass[acmsmall,screen]{acmart}
\fi

\setcopyright{rightsretained} 
\acmPrice{}
\acmDOI{10.1145/3586037}
\acmYear{2023}
\copyrightyear{2023}
\acmSubmissionID{oopslaa23main-p45-p}
\acmJournal{PACMPL}
\acmVolume{7}
\acmNumber{OOPSLA1}
\acmArticle{85}
\acmMonth{4}
\ifextended
\else
\received{2022-10-28}
\received[accepted]{2023-02-25}
\fi

\ifextended
\makeatletter
\let\@acmArticle\@empty
\let\@authorsaddresses\@empty
\makeatother
\fi

\usepackage{booktabs}   
\usepackage{caption}
\usepackage{subcaption}
\usepackage{mathpartir}
\usepackage{listings}
\usepackage{listings-rust}
\usepackage{listings-verus}
\usepackage{xspace}
\usepackage{tikz}
\usepackage{colortbl}
\usepackage{siunitx}

\usepackage{pifont}      

\lstdefinestyle{DefaultStyle}
{
  basicstyle=\scriptsize\ttfamily\linespread{0.8}\selectfont,
  breaklines=true,
  tabsize=1,
  breakindent=2em,
  literate={\ \ }{{\ }}1
}

\lstdefinestyle{Verus}%
{ style=DefaultStyle,
, identifierstyle=%
, commentstyle=\color[gray]{0.4}%
, stringstyle=\color[rgb]{0, 0, 0.5}%
, keywordstyle={[1]\bfseries}
, keywordstyle={[2]\color[rgb]{0.75, 0, 0}}
, keywordstyle={[3]\color[rgb]{0, 0.5, 0}}
, keywordstyle={[4]\color[rgb]{0, 0.5, 0}}
, keywordstyle={[5]\color[rgb]{0, 0, 0.75}}
, keywordstyle={[6]\color[rgb]{0, 0, 0.75}}
, columns=spaceflexible%
, keepspaces=true%
, showspaces=false%
, showtabs=false%
, showstringspaces=true%
}%

\lstdefinestyle{VerusLineNos}{
  style=Verus%
, numbers=left%
, firstnumber=auto%
, numberblanklines=true%
, numberstyle=\color{gray}%
, numbersep=5pt%
, xleftmargin=15pt%
}

\newcommand\fstar{F$^\star$\xspace}
\newcommand{\rust}{\cite{the-rust-language,the-rust-programming-language}\xspace}
\newcommand*{\verus}[0]{Verus\xspace}

\newcommand{\codemarker}[1]{\hspace{-2em}\tikz[baseline=(char.base)]{\node[shape=rectangle,draw,inner sep=.7pt] (char) {\scriptsize #1};}}
\newcommand{\inlinecodemarker}[1]{\tikz[baseline=(char.base)]{\node[shape=rectangle,draw,inner sep=.7pt] (char) {\scriptsize #1};}}

\newcommand*\circled[1]{\,\tikz[baseline=(char.base)]{
            \node[shape=circle,draw,inner sep=.7pt] (char) {\scriptsize\texttt{#1}};}\,}
\newcounter{clnumcounter}
\renewcommand{\theclnumcounter}{\Alph{clnumcounter}}
\newcommand{\clnum}{\refstepcounter{clnumcounter}\circled{\theclnumcounter}}
\newcommand{\clref}[1]{\hyperref[#1]{\circled{\ref{#1}}}}

\newcounter{lclnumcounter}
\renewcommand{\thelclnumcounter}{\alph{lclnumcounter}}
\newcommand{\lclnum}{\refstepcounter{lclnumcounter}\circled{\thelclnumcounter}}

\newcommand{\lclref}[1]{\hyperref[#1]{\circled{\ref{#1}}}}

\newif\ifsubmit    
\submittrue

\newcommand{\memoizetotal}{68}
\newcommand{\memoizeexe}{43}
\newcommand{\memoizespec}{23}
\newcommand{\memoizeproof}{2}
\newcommand{\queuefifototal}{477}
\newcommand{\queuefifoexe}{138}
\newcommand{\queuefifospec}{220}
\newcommand{\queuefifoproof}{119}
\newcommand{\allocatorpagestotal}{23}
\newcommand{\allocatorpagesexe}{18}
\newcommand{\allocatorpagesspec}{5}
\newcommand{\allocatorpagesproof}{0}
\newcommand{\invcelltotal}{72}
\newcommand{\invcellexe}{42}
\newcommand{\invcellspec}{21}
\newcommand{\invcellproof}{9}

\newcommand{\pptrtotal}{19}
\newcommand{\pptrexe}{14}
\newcommand{\pptrspec}{0}
\newcommand{\pptrproof}{5}
\newcommand{\internertotal}{196}
\newcommand{\internerexe}{88}
\newcommand{\internerspec}{88}
\newcommand{\internerproof}{20}
\newcommand{\rwlocktotal}{425}
\newcommand{\rwlockexe}{145}
\newcommand{\rwlockspec}{200}
\newcommand{\rwlockproof}{80}
\newcommand{\rctotal}{324}
\newcommand{\rcexe}{97}
\newcommand{\rcspec}{119}
\newcommand{\rcproof}{108}
\newcommand{\doublylinkedxortotal}{385}
\newcommand{\doublylinkedxorexe}{151}
\newcommand{\doublylinkedxorspec}{116}
\newcommand{\doublylinkedxorproof}{118}
\newcommand{\fibototal}{58}
\newcommand{\fiboexe}{22}
\newcommand{\fibospec}{20}
\newcommand{\fiboproof}{16}
\newcommand{\pcelltotal}{18}
\newcommand{\pcellexe}{12}
\newcommand{\pcellspec}{0}
\newcommand{\pcellproof}{6}
\newcommand{\vectortotal}{67}
\newcommand{\vectorexe}{41}
\newcommand{\vectorspec}{22}
\newcommand{\vectorproof}{4}

\newcommand{\timeinvcell}{\SI{2.24}{s}}
\newcommand{\timepcell}{\SI{2.21}{s}}
\newcommand{\timepptr}{\SI{2.2}{s}}
\newcommand{\timerwlock}{\SI{4.44}{s}}
\newcommand{\timedoublylinkedxor}{\SI{5.03}{s}}
\newcommand{\timeinterner}{\SI{3.22}{s}}
\newcommand{\timememoize}{\SI{2.28}{s}}
\newcommand{\timequeuefifo}{\SI{4.58}{s}}
\newcommand{\timefibo}{\SI{2.19}{s}}
\newcommand{\timerc}{\SI{3.51}{s}}
\newcommand{\timevector}{\SI{2.34}{s}}
\newcommand{\timeallocatorpages}{\SI{0.1}{s}}

\begin{document}

\ifextended
\title{\verus: Verifying Rust Programs using Linear Ghost Types (extended version)}
\fi




\author{Andrea Lattuada}
\affiliation{
  \institution{VMware Research}
  \country{Switzerland}
}
\authornote{Research work done mainly at ETH Zurich, Switzerland.}
\email{lattuada@vmware.com}

\author{Travis Hance}
\affiliation{
  \institution{Carnegie Mellon University}
  \country{USA}
}
\email{thance@andrew.cmu.edu}

\author{Chanhee Cho}
\affiliation{
  \institution{Carnegie Mellon University}
  \country{USA}
}
\email{chanheec@andrew.cmu.edu}

\author{Matthias Brun}
\affiliation{
  \institution{ETH Zurich}
  \country{Switzerland}
}
\email{matthias.brun@inf.ethz.ch}

\author{Isitha Subasinghe}
\affiliation{
  \institution{UNSW Sydney}
  \country{Australia}
}
\authornote{Research work done as a student at the University of Melbourne, Australia, and as a research assistant at ETH Zurich.}
\email{i.subasinghe@unsw.edu.au}

\author{Yi Zhou}
\affiliation{
  \institution{Carnegie Mellon University}
  \country{USA}
}
\email{yizhou5@andrew.cmu.edu}

\author{Jon Howell}
\affiliation{
  \institution{VMware Research}
  \country{USA}
}
\email{howell@vmware.com}

\author{Bryan Parno}
\affiliation{
  \institution{Carnegie Mellon University}
  \country{USA}
}
\email{parno@cmu.edu}

\author{Chris Hawblitzel}
\affiliation{
  \institution{Microsoft Research}
  \country{USA}
}
\email{chris.hawblitzel@microsoft.com}

\renewcommand{\shortauthors}{%
A. Lattuada,
T. Hance,
C. Cho,
M. Brun,
I. Subasinghe,
Y. Zhou,
J. Howell,
B. Parno,
C. Hawblitzel}

\begin{abstract}
The Rust programming language provides a powerful type system that checks linearity and borrowing,
allowing code to safely manipulate memory without garbage collection and
making Rust ideal for developing low-level, high-assurance systems.
For such systems, formal verification can be useful
to prove functional correctness properties beyond type safety.
This paper presents Verus, an SMT-based tool for formally verifying Rust programs.
With Verus, programmers express proofs and specifications using the Rust language,
allowing proofs to take advantage of Rust's linear types and borrow checking.
We show how this allows proofs to manipulate linearly typed permissions
that let Rust code safely manipulate memory, pointers, and concurrent resources.
Verus organizes proofs and specifications using a novel mode system that distinguishes
specifications, which are not checked for linearity and borrowing, from executable code and proofs,
which are checked for linearity and borrowing.
We formalize Verus' linearity, borrowing, and modes in a small lambda calculus,
for which we prove type safety and termination of specifications and proofs.
We demonstrate Verus on a series of examples, including pointer-manipulating code
(an xor-based doubly linked list), code with interior mutability, and concurrent code.
\end{abstract}

\ccsdesc[500]{Software and its engineering~Formal software verification}


\keywords{Rust, linear types, systems verification}

\maketitle


\section{Introduction}

The Rust programming language~\rust has brought linear types into the mainstream.
Rust's sophisticated type system incorporates linear types and borrowing,
making it possible to write low-level systems code in a type-safe way
without requiring garbage collection.
This makes Rust an attractive language for developing low-level software
that needs both high performance and high assurance,
and Rust has gained rapid acceptance for system programming over the last
few years~\cite{rustandroid, rustlinux}.

Nevertheless, even type-safe code can still contain bugs
that harm a program's security and reliability.
Furthermore, systems programmers using Rust sometimes resort to unsafe code
(via Rust's \verb`unsafe` keyword)
for programming styles that do not fit into Rust's linearity discipline; e.g.,
it is awkward to encode doubly linked lists in Rust
because the backwards links violate linearity.

Formal verification promises to prove deeper properties about Rust programs,
including properties about low-level code that would otherwise require unsafe Rust features.
Hence, we introduce \verus, an SMT-based tool for verifying Rust code.
SMT-based verification can help
Rust overcome the limitations of Rust's strict typing discipline,
making it possible to safely express low-level code like doubly linked lists
or safe implementations of reader-writer locks for concurrent code.

Just as importantly,
we argue that Rust's linear type system can help make SMT-based verification easier,
bringing the power of substructural logics, like concurrent separation logic~\cite{DBLP:conf/lics/Reynolds02,DBLP:journals/tcs/OHearn07},
to SMT-based reasoning.
In particular, we demonstrate the use of linear ghost permissions
that enable a program to take specific actions on specific resources,
such as writing to a memory location.
Since the permissions are linear, they can track the evolving state of a resource
in the same way that separation logic formulas can track the state of a resource.
Since the permissions are ghost, they exist only during type checking and verification,
and do not impose any overhead on compiled executable code.

To take advantage of Rust's type system for checking linear ghost permissions,
\verus uses Rust to express specifications and proofs,
running Rust's linearity and borrow checking on the proofs.
By using a single language for specifications, proofs, and executable code,
\verus follows in the footsteps of earlier frameworks that combine proofs and programming,
such as Coq~\cite{coq}, Dafny~\cite{DBLP:conf/lpar/Leino10}, \fstar~\cite{mumon}, and Lean~\cite{lean2015}.
However, these systems were designed from scratch for verification,
unlike Rust, which was designed strictly as a programming language.
Therefore, it is important to ensure that the subset of Rust
that \verus allows in proofs and specifications is sound as a proof language.

In particular, we need to make sure that proofs terminate
and that functions used in specifications are pure, mathematical functions,
whereas executable code might contain infinite loops,
be nondeterministic, or have side effects.
In a language like Rust that contains recursive types, higher-order functions,
and type classes (traits), termination can be particularly subtle:
without positivity restrictions on recursive type definitions, for example,
these features together can encode nontermination.
Nevertheless, we want to restrict recursive types only for specifications and proofs,
not for executable code, unlike Coq, Dafny, and Lean,
which restrict all recursive types.

To enforce the distinction between specifications, proofs, and executable code,
\verus introduces a mode system that classifies all code as
specification, proof, or executable,
where specification and proof code are checked for termination.
All three modes of code are type checked;
proof and executable code are checked for linearity and borrowing;
only executable code is compiled to machine code.
We formalize this mode system using a small Rust-inspired lambda calculus,
proving preservation and progress for all well-typed expressions
and termination for all specification and proof expressions (\autoref{sec:formalization}).

\verus currently supports a large set of proof features
and a large subset of ordinary Rust features:
\begin{itemize}
\item Rust's finite range integers (i32, u32, etc.) as well as
  infinite-range integers (int, nat) for specifications and proofs
\item recursive algebraic datatypes (structs and enums),
  including mutual recursion, pattern matching, and pattern match guards
\item mutable variables, while loops, and return statements
\item recursive specification functions and inductive proofs,
  including mutual recursion and lexicographic decreases clauses
\item passing function arguments by borrowing (both \verb`&` and \verb`&mut`)
\item lifetime parameters on structs
\item generics (parametric polymorphism), with support for simple traits (type classes)
\item first-class functions (Rust closures) in specifications
\item modules with public and private definitions
\item preconditions, postconditions, and loop invariants
\item quantifiers (forall, exists, choose) with both automated and manual SMT trigger selection,
  as well as integrated quantifier profiling to diagnose verification performance issues
\item programmer control of SMT performance by selectively hiding and revealing
  specification function definitions,
  including control over recursive function unrolling
\item support for bit-vector reasoning and proof by computation
\item strings and characters
\item a library of types for specifications, including sequences, sets, and maps
\item low-level pointer reasoning
\item concurrency and state machines
\end{itemize}

Verus is also able to handle some situations that require \texttt{unsafe}.
For example, Verus is able to verify the use of raw pointers and unsafe cells, which can be useful
for some low-level pointer reasoning, lock implementations, and interior mutability use-cases.
As we will see, \verus' support for linear ghost state is crucial for this support.
Note though, that Verus does not attempt a full aliasing and provenance model for Rust's pointers;
our simplified model for ``raw pointers'' only handles those that point into the global heap.

Some features are still missing, notably support for separate verification of multiple crates
and functions that return mutable references.
However, we believe that supporting crates and various other Rust features is a matter of engineering,
and supporting functions that return mutable references can follow earlier research (\autoref{sec:related})
by Prusti~\cite{DBLP:conf/nfm/AstrauskasBFGMM22},
RustHorn~\cite{DBLP:conf/esop/0002T020},
Creusot~\cite{denis:hal-03737878},
and Aeneas~\cite{DBLP:journals/pacmpl/HoP22}.

Regardless, this paper will not focus on all the features supported by \verus,
but will instead focus on the most novel contributions:
\begin{enumerate}
\item usage of Rust's linearity and borrow checking in proofs
\item verification of pointer-manipulating Rust code and concurrent Rust code,
  based on a combination of linearity, borrowing, and SMT solving
\item a mode system for enforcing the different properties of specs, proofs,
  and executable code
\item formalization of the mode system, including checking of linearity and borrowing
\end{enumerate}

The remainder of this paper
introduces \verus by example (\autoref{sec:design});
discusses handling unsafe code (\autoref{sec:unsafe});
applies \verus to pointer-based code (including a doubly linked list, \autoref{sec:linear-ghost}),
interior mutability (\autoref{sec:interior-mutability}),
and concurrent code (\autoref{sec:concurrency});
discusses Verus' implementation (\autoref{sec:implementation}),
user experience (\autoref{sec:user-experience}), and limitations (\autoref{sec:limitations});
and presents syntax, semantics, and proofs for a formal lambda calculus
with modes, linearity, and borrowing (\autoref{sec:formalization}).


\section{\verus by example} \label{sec:design}

This section introduces the basic features of \verus
by walking through a simple example that computes Fibonacci numbers,
shown in \autoref{listing:fibonacci}.
The example consists of a set of functions written in Rust.
Each function is annotated with an attribute,
using Rust's \verb`#[...]` attribute syntax,
to indicate whether the function is executable code (\verb`#[exec]`),
proof code (\verb`#[proof]`),
or specification code (\verb`#[spec]`).
We refer to \verb`exec`, \verb`proof`, and \verb`spec` as modes;
\autoref{fig:modes} summarizes the properties of these three modes.

\begin{figure}

\noindent\begin{minipage}{.5\textwidth}

\begin{lstlisting}[language=Verus,style=VerusLineNos]
#[spec] fn fibo(n: nat) -> nat {
    decreases(n); %*\clnum\label{clnum:fibo:fibo-decreases}*
    if n == 0 { 0 }
    else if n == 1 { 1 }
    else { fibo(n - 2) + fibo(n - 1) }
}

#[proof] fn lemma_fibo_is_monotonic(i:nat, j:nat) { 
    requires(i <= j); %*\clnum\label{clnum:fibo:requires-clause}*       %* \codemarker{spec}*
    ensures(fibo(i) <= fibo(j)); %*\clnum\label{clnum:fibo:ensures-clause}*       %* \codemarker{spec}*
    decreases(j - i); %*\clnum\label{clnum:fibo:decreases-clause}*       %* \codemarker{spec}*

    if (i < 2 && j < 2) || i == j {
    } else if i == j - 1 {
        reveal_with_fuel(fibo, 2); %*\clnum\label{clnum:fibo:reveal}*
        lemma_fibo_is_monotonic(i, j - 1); %*\clnum\label{clnum:fibo:lemma-monotonic-induct}*
    } else {
        lemma_fibo_is_monotonic(i, j - 1);
        lemma_fibo_is_monotonic(i, j - 2);
    }
}

#[spec] fn fibo_fits_u64(n: nat) -> bool {
    fibo(n) <= u64::MAX
} 
\end{lstlisting}
\end{minipage}\hfill
\noindent\begin{minipage}{.5\textwidth}

\begin{lstlisting}[language=Verus,style=VerusLineNos,firstnumber=26]
#[exec] fn fibo_impl(n: u64) -> u64 {
    requires(fibo_fits_u64(n)); %*\clnum\label{clnum:fibo:requires-clause-impl}*       %* \codemarker{spec}* 
    ensures(|result: u64| result == fibo(n));%*\clnum\label{clnum:fibo:ensures-clause-impl}*       %* \codemarker{spec}* 

    if n == 0 { return 0; } 
    let mut prev: u64 = 0;
    let mut cur: u64 = 1;
    let mut i: u64 = 1;
    while i < n {
        invariant([ %*\clnum\label{clnum:fibo:invariant-clause}*       %* \codemarker{spec}* 
            0 < i && i <= n,
            fibo_fits_u64(n as nat),
            fibo_fits_u64(i as nat),
            cur == fibo(i),
            prev == fibo(i as nat - 1),
        ]);
        let new_cur = cur + prev;
        prev = cur;
        cur = new_cur;
        assert(prev == fibo(i as nat)); %*\clnum\label{clnum:fibo:fibo-impl-assert}*       %* \codemarker{proof}*
        i = i + 1;
        lemma_fibo_is_monotonic(i, n); %*\clnum\label{clnum:fibo:fibo-impl-lemma}*       %* \codemarker{proof}*
    }
    cur
}
\end{lstlisting}
\end{minipage}
\caption{A proof of correctness of a function computing the n-th
Fibonacci number. We use circled letters, similar to \protect\circled{Z},
to mark points of interest in the code. The markers \protect\inlinecodemarker{spec} and
\protect\inlinecodemarker{proof} indicate specification and proof mode code respectively
when it differs from the mode of the function.}
\label{listing:fibonacci}
\end{figure}

In \verus, specifications and proofs are simply Rust code,
parsed with Rust's parser and checked with Rust's type checker.
This avoids the need for systems programmers to learn a separate verification language,
making verification more accessible and convenient.
It also allows specifications and proofs to take advantage of Rust's features,
such as recursive functions, arithmetic, algebraic datatypes,
pattern matching, modules, closures, traits, etc.
For soundness's sake, \verus places some limits
on the features that specifications and proofs can use.
In particular, specifications must be deterministic,
and recursive \verb`spec` functions and recursive \verb`proof` functions must terminate.

The \verus tool, which extends the Rust compiler,
erases all ghost code (all specifications and proofs) before the code is compiled to machine code.
In the example, \verb`lemma_fibo_is_monotonic`, \verb`fibo`, and \verb`fibo_fits_u64`
are all erased before compilation.
Furthermore, the executable function \verb`fibo_impl` contains small bits of specification and proof
inside its body
(\clref{clnum:fibo:requires-clause-impl},
\clref{clnum:fibo:ensures-clause-impl},
\clref{clnum:fibo:invariant-clause},
\clref{clnum:fibo:fibo-impl-assert},
\clref{clnum:fibo:fibo-impl-lemma}),
and this ghost code is also erased.

\verus encodes preconditions and postconditions as calls to \verus library functions
named \verb`requires` and \verb`ensures`.
Postconditions may refer to the return value;
for this, the ensures function accepts a Rust closure that declares a name for the return value.
(In Rust, first-class functions are called closures and have the syntax ``\verb`|...parameters...| body`''.)
The example uses a postcondition to prove that the executable function \verb`fibo_impl`
computes the same result as the mathematical definition of the $n$\textsuperscript{th} Fibonacci number in the
\verb`fibo` function: the \verb`ensures` clause~\clref{clnum:fibo:ensures-clause-impl}
establishes this postcondition for \verb`fibo_impl`. Because the return value is a bounded 64-bit unsigned
integer, \verb`fibo_impl` can only accept a parameter \verb`n` such that the n-th Fibonacci number fits
in the type of the return value: this is established by the \verb`requires` clause~\clref{clnum:fibo:requires-clause-impl}.
Note that \verus extends Rust's type system with two new integer types,
\verb`int` (mathematical integers $\mathbb{Z}$), and \verb`nat` (natural numbers $\mathbb{N}$),
so that specifications and proofs can talk about arbitrary integers.
Executable code, however, is limited to Rust's finite-width integer types like \verb`u64`,
since \verb`int` and \verb`nat` aren't compilable to machine code.

To help prove the postcondition,
the \verb`fibo_impl` function uses a loop invariant \clref{clnum:fibo:invariant-clause}
containing a list of clauses that must be true before and after each loop iteration.
Given preconditions, postconditions, and loop invariants,
\verus uses standard weakest precondition reasoning~\cite{DBLP:journals/cacm/Dijkstra75}
to generate a verification condition for \verb`fibo_impl`.
It then sends this verification condition to the Z3 SMT solver~\cite{z3}.

In many cases, the SMT solver can prove the verification condition completely automatically.
In other cases, the proof may require reasoning beyond the SMT solver's abilities.
For example, to prove the absence of 64-bit integer overflow,
\verb`fibo_impl` relies on the Fibonacci sequence being monotonic,
which requires an inductive proof that the SMT solver cannot generate automatically.
Instead, the programmer supplies an inductive proof in the form of a recursive \verb`proof` function
(e.g., \verb`lemma_fibo_is_monotonic`).
The programmer can also add explicit assertions\clref{clnum:fibo:fibo-impl-assert} that
serve as hints to the SMT solver.
This style of SMT-based verification with programmer-supplied lemmas and hints
is similar to other verification systems like Boogie~\cite{boogie}, Dafny, and \fstar.

To improve verification performance,
\verus strives to keep the verification condition encoding lightweight,
so that the SMT encoding of specifications is not much larger than the original
specifications written in \verus code.
In particular, calls to \verb`spec` functions are translated directly into calls to SMT functions,
with no additional overhead.
For this reason, \verus \verb`spec` functions are total functions that do not have preconditions and postconditions.
This design is similar to Boogie,
though it differs from Dafny and \fstar.

This design choice has a downside, since
precondition failures can provide the developer with early feedback
to find errors in specification functions and in how they are used.
In order to restore that feedback, \verus introduces \verb`recommends` clauses: soft preconditions
for \verb`spec` functions, which Verus only considers when there is a verification error.
At that point, it performs a separate check for soft preconditions of
\verb`spec` functions that are mentioned in the context of the failure,
and reports failures as warnings for the developer.

Recursive \verb`spec` functions and recursive \verb`proof` functions are valid only if they terminate on all inputs
(otherwise, they could encode unsound circular reasoning).
\verus requires that all such functions contain a \verb`decreases` clause~\clref{clnum:fibo:fibo-decreases}~\clref{clnum:fibo:decreases-clause}
and each recursive call must decrease the expression in the clause. The recursive definition of
the $n$\textsuperscript{th} Fibonacci number~\clref{clnum:fibo:fibo-decreases} in \autoref{listing:fibonacci}
is legal because both recursive calls decrease the expression \verb`n`.
(\verus also imposes positivity restrictions on recursive type definitions to prevent
nontermination, as discussed in \autoref{sec:formalization}.)
The SMT solver may need to unfold definitions of a recursive \verb`spec` function.
As in Dafny and \fstar, \verus uses an integer ``fuel'' to control the number of unfoldings.
The \verb`reveal_with_fuel`~\clref{clnum:fibo:reveal} function controls the fuel level.

\begin{figure}
  \centering
  \small
  \begin{tabular}{r|l|l|l}
    \hline
    ~ & specification mode & proof mode & executable mode \\ \hline
    compiled or ghost & ghost & ghost & compiled \\
    code style & purely functional & mutation allowed & mutation allowed \\
    linearity \& borrowing checking & not checked & checked & checked \\
    can call specification functions & yes & yes & yes \\
    can call proof functions & no & yes & yes \\
    can call executable functions & no & no & yes \\
    determinism & deterministic & nondeterministic & nondeterministic \\
    termination & must terminate & must terminate & nontermination ok \\
    preconditions/postconditions & none & requires/ensures & requires/ensures \\
    \hline
  \end{tabular}
  \caption{Summary of \verus' modes and their properties.}
  \label{fig:modes}
\end{figure}

\subsection{Linearity, Borrowing, Spec Variables, and Proof Variables} \label{sec:design:linearity}

Rust types are linear by default:
unless a type implements the Rust \verb`Copy` trait,
values of the type can only be moved from one variable to another, not copied.
For example, the Rust \verb`Vec<T>` type for vectors is linear.
The following code is illegal in Rust because it attempts to duplicate a \verb`Vec<u64>` value,
returning both copies of the value in a pair:

\begin{lstlisting}[language=Verus]
#[exec] fn f(v: Vec<u64>) -> (Vec<u64>, Vec<u64>) {
    let v1 = v;
    let v2 = v; // illegal, tries to duplicate v
    (v1, v2)
}
\end{lstlisting}

On the other hand, Rust code can duplicate immutable references to values,
as long as the scope of the references is limited.
In Rust terminology, a reference of type \verb`&T` temporarily {\it borrows} from
an owned value of type \verb`T`.
During the borrowing, the original owned value of type \verb`T` is inaccessible.
When the references go out of scope, the original owned value becomes accessible again.
Rust code can also borrow a mutable reference of type \verb`&mut T`;
in contrast to immutable references, mutable references cannot be duplicated.
Rust enforces the property that a value cannot be borrowed both immutably and mutably simultaneously.

Rust contains a sophisticated ``borrow checker'' that checks linearity and borrowing.
Similar to Creusot~\cite{denis:hal-03737878}, \verus trusts the results of Rust's borrow checker,
and does not attempt to recheck these results in the SMT solver,
since this would just slow down the SMT solving.
Because of this, \verus can rely on the properties of borrowing in its SMT encoding.
For example, \verus encodes immutably borrowed references \verb`&T` and owned heap pointers
(\verb`Box<T>`, \verb`Rc<T>`, and \verb`Arc<T>`) simply as values of type \verb`T`,
not as pointers to locations.

\verus specifications, in contrast to ordinary Rust code, are not checked for linearity and borrowing;
specifications can freely copy any value of any type.
This allows specifications to freely talk about linear values,
potentially mentioning a single linear variable multiple times in a precondition or postcondition,
for example.
\verus code can also store nonlinear copies of linear variables inside {\it spec variables},
declared with the attribute \verb`#[spec]`:

\begin{lstlisting}[language=Verus]
#[exec] fn f(v: Vec<u64>) {
    #[spec] let v1 = v; // copies v into spec variable v1
    #[spec] let v2 = v; // copies v into spec variable v2
    assert(v1.len() == v2.len());
}
\end{lstlisting}

Spec variables are similar to ghost variables in Dafny or erased values in \fstar.
However, \verus also supports {\it proof variables}, declared as \verb`#[proof]`,
which do not have a correspondence in Dafny or \fstar.
Proof variables sit midway between \verb`exec` variables and \verb`spec` variables:
like \verb`spec` variables, they are ghost and are not compiled to machine code,
but like \verb`exec` variables, they are checked for linearity.
By default, variables in \verb`exec` functions are \verb`exec`,
while variables in \verb`proof` functions are \verb`spec` unless declared \verb`#[proof]`.
Spec functions can only use \verb`spec` variables, not \verb`proof` variables or \verb`exec` variables.

Since \verb`proof` variables are both linear and ghost,
they can represent abstract linear permissions to perform operations,
which can produce and consume the linear permissions.
The next section describes how \verus exploits this feature to verify
safe low-level pointer manipulation that would, in ordinary Rust, require unsafe code.
(In fact, \verus does not support verification of code marked with Rust's \verb`unsafe` keyword;
instead, its goal is to provide safe replacements for unsafe Rust features,
based on linear ghost permissions and SMT-based verification.)


\subsection{Simplifying Verification Conditions with Linear Types}\label{sec:encoding}

Potential aliasing of variable bindings in the presence of mutation
complicates verification~\cite{on-the-frame-problem}
because it requires explicitly reasoning about memory to determine the potential effects 
of each program statement. Given two bindings \verb`p` and \verb`q`, a predicate \verb`P(p)` about
the data reachable from \verb`p`,
and a statement \verb`S[q]` which mutates one of the memory locations reachable from \verb`q`,
if \verb`P(p)` is true before \verb`S[q]`, then \verb`P(p)` is guaranteed to remain true
after \verb`S[q]` if all the memory locations reachable from \verb`p` and \verb`q` are
disjoint.
Verus relies on the properties enforced by Rust's ``borrow checker'' to avoid explicit memory
reasoning: when encoding \verb`proof` and \verb`exec` function bodies
Verus treats the data associated with a uniquely-owned binding as an immutable value.
Mutable bindings are represented with single static assignment to immutable SMT constants,
one after each mutation.
We discuss Verus' encoding strategy with an example.

\begin{figure}
\makebox[0pt][l]{%
  \raisebox{-\totalheight}[0pt][0pt]{%
    \includegraphics{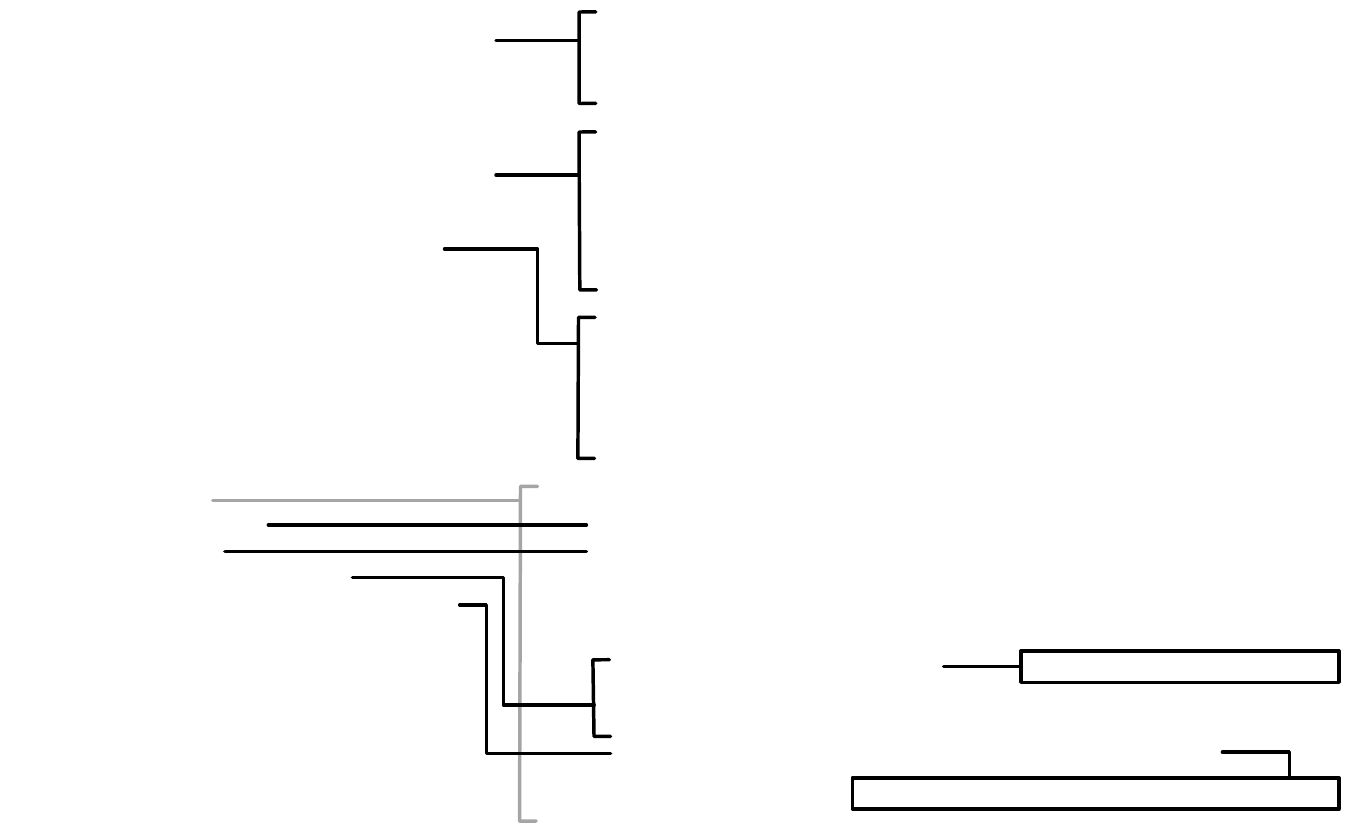}}}%
\makebox[0pt][l]{%
  \raisebox{-\totalheight}[0pt][0pt]{%
  \begin{tikzpicture}[remember picture,overlay]
  \node[outer sep=0pt,inner sep=0pt] at (34em, -19.3em) {\lstinline[style=Verus]|VC: swap\_odd requires|};
          \node[outer sep=0pt,inner sep=0pt] at (31.7em, -22.9em) {\lstinline[style=Verus]|VC: assert(is\_odd(v) && is\_odd(w))|};
  \end{tikzpicture}%
    }}%
\begin{subfigure}[t]{.38\textwidth}
\vspace{.4em}
\begin{lstlisting}[language=Verus,style=VerusLineNos]
#[spec] fn is_odd(n: int) -> bool {
    n % 2 == 1
}

\end{lstlisting}
\vspace{.1em}
\begin{lstlisting}[language=Verus,style=VerusLineNos,firstnumber=5]
fn swap_odd(a: &mut u64, b: &u64) {
    requires((*old(a) as int + *b as %*\\*int) < u64::MAX &&
        is_odd(*b as int));
    ensures(is_odd(*a as int) ==
        !is_odd(*old(a) as int));
    *a = *a + *b;
}

\end{lstlisting}
\vspace{3.3em}
\begin{lstlisting}[language=Verus,style=VerusLineNos,firstnumber=13]
fn main() {
    let mut v = 0;
    let w = 3;
    swap_odd(&mut v, &w);
    assert(is_odd(v) && is_odd(w));
}
\end{lstlisting}
    \caption{Source of example}
    \end{subfigure}
    \hfill
    \begin{subfigure}[t]{.61\textwidth}
    \begin{lstlisting}[style=VerusLineNos]
(declare-fun is_odd.? (Poly) Bool) %*\lclnum\label{lclnum:smt:is-odd-decl}*
(assert (=> (forall ((n@ Poly)) %*\lclnum\label{lclnum:smt:is-odd-axiom}*
  (= (is_odd.? n@) (= (mod (%I n@) 2) 1)) %*\lclnum\label{lclnum:smt:is-odd-defn}*
...
(declare-fun req%swap_odd. (Int Int) Bool) %*\lclnum\label{lclnum:smt:swap_odd-req}*
(assert (forall ((pre%a@ Int) (b@ Int))
  (= (req%swap_odd. pre%a@ b@)
    (and
      (< (+ pre%a@ b@) 18446744073709551615)
      (is_odd.? (I b@))))))
...
(declare-fun ens%swap_odd. (Int Int Int) Bool) %*\lclnum\label{lclnum:smt:swap_odd-ens}*
(assert (forall ((pre%a@ Int) (a@ Int) (b@ Int))
  (= (ens%swap_odd. pre%a@ a@ %*\lclnum\label{lclnum:smt:swap_odd-req-pre}* b@) (and
    (uInv 64 a@) %*\lclnum\label{lclnum:smt:swap_odd-uInv}*
    (= (is_odd.? (I a@)) (not (is_odd.? (I pre%a@))))))))
...
(push)
(declare-const v@0 Int) (declare-const v@1 Int)
(declare-const w@ Int)
(assert (not %*\lclnum\label{lclnum:smt:falsifies}*
  (=> (= v@0 0) (=> (= w@ 3)
    (and
      (req%swap_odd. v@0 w@) %*\lclnum\label{lclnum:smt:main-req-vc}*
      (=> (uInv 64 v@1)
        (=> (ens%swap_odd. v@0 v@1 w@)
          (and (is_odd.? (I v@1)) (is_odd.? (I w@)))) %*\lclnum\label{lclnum:smt:main-assert-vc}*
    ))))))
(pop)
\end{lstlisting}
\caption{SMT-LIB encoding of the verification conditions for \texttt{main}.}
\end{subfigure}
    \caption{A simple example program and relevant parts of its encoding in Z3. The SMTLIB encoding
    has been slightly simplified to aid readability, but without compromising accuracy. In particular,
    we rename the constants, and we elide the patterns chosen for quantifier instantiation in Z3~\cite{e-matching},
    some temporary variables used to optimize the SMT encoding, and some facilities for error reporting.} \label{fig:encoding-example}
\end{figure}

\autoref{fig:encoding-example} shows a simple program, and how Verus encodes it into SMT-LIB~\cite{smtlib}, the input to Z3.
First let us dispatch some boilerplate that clutters the figure.
The functions \verb`%I`, \verb`I` and the sort \verb`Poly` appear often. They are part of the polymorphism encoding machinery of Verus,
which is inspired by Boogie\cite{poly-boogie}. Function \verb`I` is a cast from \verb`Int` to \verb`Poly`, a Z3 sort
representing a polymorphic type, and \verb`%I` is a cast from \verb`Poly` to \verb`Int`. \verb`spec` function arguments
are always \verb`Poly` due to interactions with the Z3's quantifier instantiation.
\verb`uInv 64` \clref{lclnum:smt:swap_odd-uInv} is a typing invariant that restrict the SMT \verb`Int`
type to the range of Rust's \verb`u64` machine type.

In SMT-LIB, functions are defined by constructing axioms (e.g. \clref{lclnum:smt:is-odd-axiom})
that relate their declaration (e.g. \clref{lclnum:smt:is-odd-decl}) to their definition (e.g. \clref{lclnum:smt:is-odd-defn}).
Verus' \verb`spec` is designed to closely match SMT logic, enabling the straightforward encoding of \verb`is_odd`~\clref{lclnum:smt:is-odd-defn}.
Similarly, the SMT functions representing the precondition and postcondition for \verb`swap_odd` (
\verb`req%swap_odd.` \clref{lclnum:smt:swap_odd-req} and \verb`ens%swap_odd.` \clref{lclnum:smt:swap_odd-ens} respectively)
closely match their corresponding \verb`spec`-mode Verus code.
The mutable reference \verb`a: &mut u64` is represented as a pair of constants, \verb`pre%a@` and \verb`a@` \clref{lclnum:smt:swap_odd-req-pre},
respectively the initial and final value. The immutable reference (\verb`b: &u64`) is represented as a single constant \verb`b@`.
Reference types (\verb`&mut` and \verb`&`) do not need special treatment
thanks to the borrow checker's guarantees:
for example the arguments \verb`a` and \verb`b` cannot be aliased because mutable and immutable
references to the same data cannot exist at the same time.

The last SMT-LIB fragment is the encoding of the \verb`main` function and its associated verification conditions.
Like other tools, Verus encodes its proof search as a query to Z3 to find an assignment to constants that
falsifies \clref{lclnum:smt:falsifies} verification conditions: an \verb`unsat` result is a proof that such assignment does not
exist, i.e. verification succeeded.
The constants \verb`v@0` and \verb`v@1` represent the value of binding \verb`v` before and after the call to
\verb`swap_odd`, and \verb`w` represents the immutable binding \verb`w`.
The call to \verb`swap_odd` is encoded modularly with a verification condition to check its precondition \clref{lclnum:smt:main-req-vc}.

Another call to \verb`ens%swap_odd.` introduces its postcondition as an antecedent for all future verification
conditions in the function.
Thanks to Rust's linear type system, there is no need to explicitly model the heap here: the two arguments to \verb`swap_odd`
are guaranteed to point to distinct regions of memory.
Finally, the \verb`assert` is encoded as part of the verification condition \clref{lclnum:smt:main-assert-vc}.


\section{Handling Unsafe Code Safely} \label{sec:unsafe}

Unsafe code can often be a sticking point for users seeking strong guarantees about
their Rust program, as it has the potential to undermine Rust's famous memory-safety guarantees.
In particular, the Rust language guarantees that
a program that does not use \texttt{unsafe} code must unconditionally be memory-safe;
however, when \texttt{unsafe} code is used, the program becomes \emph{conditionally memory-safe};
i.e., the program is only memory-safe if the program obeys certain rules when using
\texttt{unsafe}.

Verus supports a few trusted primitives that are conditionally memory-safe in this manner.
For these cases,
their correctness conditions are encoded as Verus specifications.
Therefore, users can be sure that their code is truly memory-safe (in addition to
\verus' other guarantees) as long as \verus' SMT verification proves that the code upholds
the contracts.

As a simple example, consider the operation of indexing into a vector: one of the most ubiquitous
operations in all of software, yet also one of the most fraught for memory-safety violations.
In Rust, indexing into a vector always performs a bounds check; it is memory-safe because
it will always panic rather than access memory-out-of-bounds.
On the other hand, Rust's \verb`get_unchecked` does not perform any bounds check, and
therefore it an \verb`unsafe` function. That is, \verb`get_unchecked` is conditionally
memory-safe because it is only safe if the user calls the function with a valid index.

We can write this condition as a \verus specification, and provide \verb`get_unchecked`
as a trusted primitive function:

\begin{center}
\begin{lstlisting}[language=Verus,style=VerusLineNos]
fn safe_get_unchecked<V>(v: &Vec<V>, i: usize) -> &V {
    requires 0 <= i && i < v.len()
    ...
}
\end{lstlisting}
\end{center}

In the next sections, we will see some more advanced examples.

\section{Safe Pointer Manipulation with Linear Ghost Types} \label{sec:linear-ghost}

\subsection{Low-Level Pointer Manipulation with Linear Ghost Permissions}\label{sec:ghost:perms}

While Rust's reference types and borrowing rules provide a memory-safe framework for
many use cases, they are sometimes insufficiently expressive and
\emph{raw pointers} may be required.
For example, a doubly-linked list, where each node may be pointed to by two neighbors,
violates the unique-ownership discipline of Rust. Raw pointers are one way to work around
this, although dereferencing raw pointers in Rust requires \verb`unsafe` code.
\verus supports raw (heap) pointers.
The most notable aspect of this support is that,
in order to provide a specification to enforce memory safety, we need to make use of 
linear ghost state.

Specifically, \verus 
introduces a core primitive
\verb`PPtr<T>` (``permissioned pointer'') as a zero-cost alternative
to raw heap pointers, along
with an associated type \verb`PermData<T>` (``permission plus data'') which is to be used in \verb`proof` mode, i.e., they are linear ghost objects as discussed in the previous section.
Calls to the \verb`PPtr<T>` API require ownership of this ghost permission
object in order to dereference the pointer, which prevents data races and other
forms of access that are undefined behavior in Rust's memory model.

However, the \verb`PermData<T>` object does not ``just'' have the role of maintaining memory safety;
it also tracks the data behind the pointer.
Tracking permissions and data this way lets us write proofs in a style similar to that of separation logics.
Specifically, the permissions object has two fields. The first, \verb`perm.view().pptr`,
indicates the pointer that the permission object corresponds to,
and the second, \verb`perm.view().value`,  gives the data behind the pointer.
The value field is an \verb`Option<T>`,
where a value of \verb`Some(v)` means the memory stores \verb`v`,
and a value of \verb`None` indicates that the memory is uninitialized.
(This should not be confused for the \emph{runtime
representation} of an \verb`exec`-mode \verb`Option<T>`, where \verb`None` is a legitimate, initialized value.)

\autoref{fig:pptr:api} shows two key functions from the \verb`PPtr` API: a function to write
through the pointer (\verb`write`) and a function to read through it (\verb`read`).
Both functions require that the permission is actually associated with the 
pointer being dereferenced~\clref{clnum:pptr:write:requiresid}~\clref{clnum:pptr:read:requiresid}
and \verb`read` requires that the memory being read from
is in an initialized state~\clref{clnum:pptr:read:requiresinit}.
Meanwhile, \verb`write`'s postcondition~\clref{clnum:pptr:write:ensuresvalue} says that the updated permission object
contains the written value, while \verb`read`'s postcondition~\clref{clnum:pptr:read:ensuresvalue} says that the returned value
is the value tracked via the permission.
\autoref{fig:pptr:ex} illustrates the usage of \verb`write` and \verb`read`,
together with allocation and deallocation, showing how the permission \verb`value` is updated.

It is crucial that the \verb`proof`-mode object \verb`PermData<T>` 
obeys Rust's ownership rules.
For example, \autoref{fig:pptr} shows how this prevents a use-after-free bug.
When we free the pointer's memory~\clref{clnum:pptr:example:consume},
the \verb`perm` variable is consumed. Thus Rust's linearity checker would report an error 
if the code attempted to read the pointer
again~\clref{clnum:pptr:example:bad-read}, as this produces another use
of \verb`perm`.

\begin{figure}

\begin{subfigure}[c]{.46\textwidth}

\begin{lstlisting}[language=Verus,style=VerusLineNos]
impl<T: Copy> PPtr<T> {
    // Equivalent of `*ptr = v`.
    #[exec] pub fn write(&self,
            #[proof] perm: &mut PermData<V>, v: V) {
        requires(equal(self.id(), old(perm).view().pptr)); %*\clnum\label{clnum:pptr:write:requiresid}*
        ensures([
            equal(perm.view().pptr, self.id()),
            equal(perm.view().value, Option::Some(v)), %*\clnum\label{clnum:pptr:write:ensuresvalue}*
        ]);
        ...
    }

    // Read through the pointer and return the value. Requires the memory to be initialized.
    #[exec] pub fn read(&self,
            #[proof] perm: &PermData<V>) -> V {
        requires([
            equal(self.id(), perm.view().pptr), %*\clnum\label{clnum:pptr:read:requiresid}*
            perm.view().value.is_Some() ]); %*\clnum\label{clnum:pptr:read:requiresinit}*
        ensures(|v: V| equal(Option::Some(v),
            perm.view().value)); %*\clnum\label{clnum:pptr:read:ensuresvalue}*
        ...
    }
}
\end{lstlisting}
\caption{Selected functions from the \texttt{PPtr<T>} API, a core \verus primitive.} \label{fig:pptr:api}
\end{subfigure}
\hfill
\begin{subfigure}[c]{.50\textwidth}

\begin{lstlisting}[language=Verus,style=VerusLineNos]
fn main() {
    // Allocate memory.
    let alloc = PPtr::<u64>::empty();
    // Unpack the return value into the pointer and the (ghost) permission
    let pptr = alloc.0;
    #[proof] let mut perm = alloc.1.0;

    // Initially, pptr points to unitialized memory, and the `perm` proof-object represents that as the value `None`.
    assert(equal(perm.view().pptr, pptr.id()));
    assert(equal(perm.view().value, Option::None));

    // We can write a value through the pptr (thus initializing the memory).
    pptr.write(&mut perm, 5); 

    // Having written the value, this is reflected in the permission object:
    assert(equal(perm.view().value, Option::Some(5)));

    // We can now read it:
    let x = pptr.read(&perm); 
    assert(x == 5);

    // Free the memory:
    pptr.free(perm); %*\clnum\label{clnum:pptr:example:consume}*

    // This would error as `perm` was just consumed 
    // let z = pptr.read(&perm); %*\clnum\label{clnum:pptr:example:bad-read}*
}
\end{lstlisting}
\caption{Example usage of \texttt{PPtr<T>}} \label{fig:pptr:ex}
\end{subfigure}
    \caption{
        \label{fig:pptr}
        The \texttt{PPtr<T>} API and an example usage.
        Though the two functions shown here require \texttt{T: Copy}, this is not a general
        restriction on the \texttt{PPtr} library.
    }
\end{figure}

Finally, observe that the safety of this API depends crucially on our ability to add
preconditions (and validate them via the prover).
For example, we saw that the specification of \verb`write` requires that the permission correspond to
the pointer being written through~\clref{clnum:pptr:write:requiresid}, and without this requirement,
it would be wildly unsound.
Thus, a safe API like this is \emph{not possible to implement in vanilla Rust}:
in order to be unconditionally safe,
the precondition would need to be a run-time check,
which would mean the \verb`PermData` object could not be ghost, and the abstraction
could not be zero-cost.

\subsection{Verified Example: Doubly-Linked List} \label{sec:dlist}
Rust's ownership model typically forces data structures to be acyclic,
unless they use unsafe code.
Here, we illustrate how \verb`PPtr` can be used to verify data structures that have cyclic 
pointer arrangements by verifying a double-ended queue implemented with a doubly-linked list.
Specifically, we use a doubly-linked list to represent a sequence $v_0, v_1, \ldots, v_{n-1}$,
and we implement the four operations
$\{\text{\texttt{push}}, \text{\texttt{pop}}\}
\times
\{\text{\texttt{front}}, \text{\texttt{back}}\}$.
The $i$th node in the list has both a \verb`prev` and \verb`next` pointer alongside
a single element of the sequence, $v_i$. The top level datatype, \verb`DList`, contains
\verb`head` and \verb`tail` pointers, pointing to the first and last nodes, respectively.
The full version (which can be found in our supplementary materials~\cite{verussupplementary})
contains an additional space-saving optimization,
where each node does not store its two pointers separately, but rather, stores their bitwise XOR.

\begin{figure}
    \begin{subfigure}{.48\textwidth}
      \includegraphics[width=\textwidth]{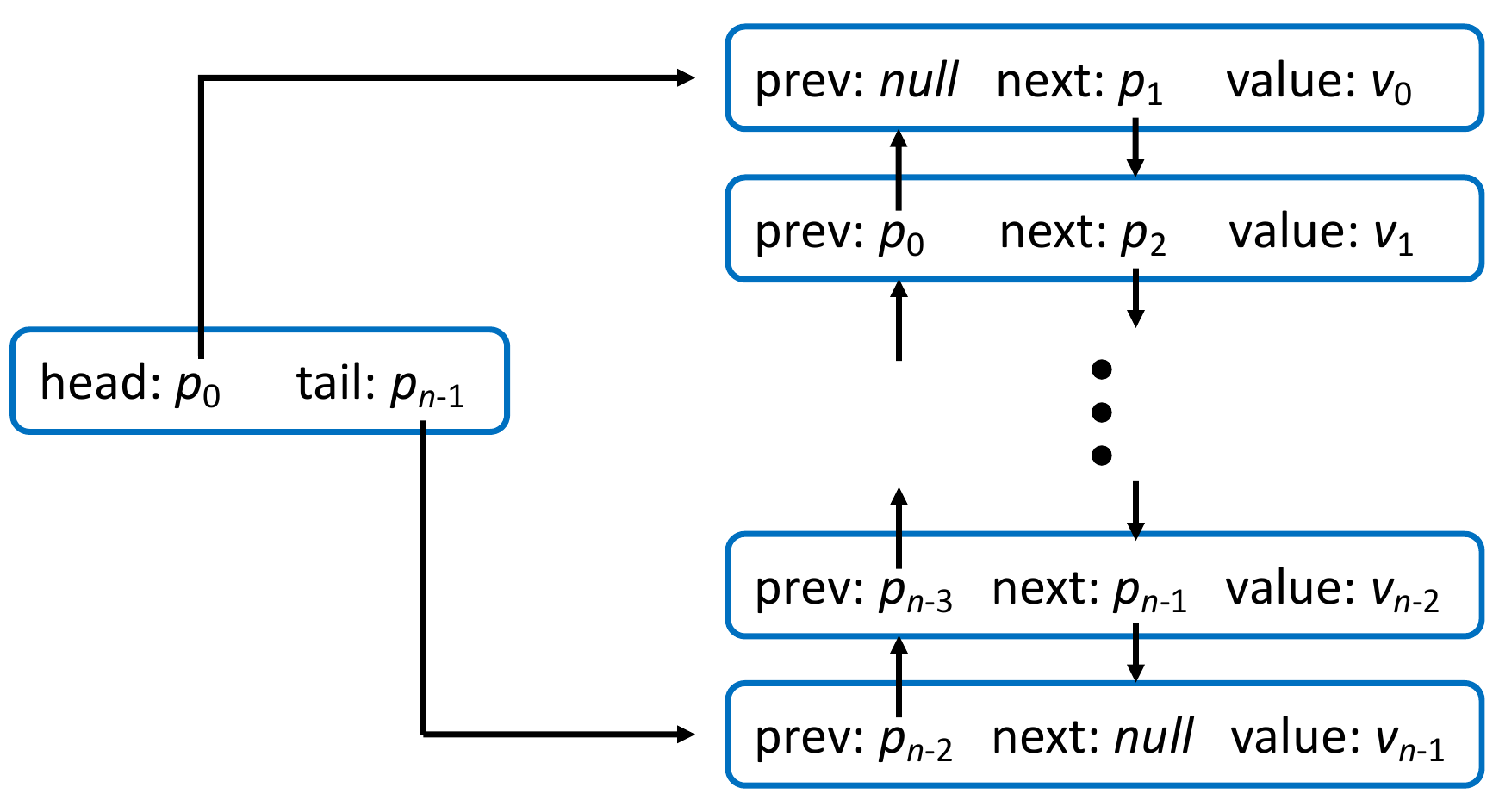}
      \caption{Physical pointer structure of a doubly-linked list.} \label{fig:dlist:phys}
    \end{subfigure}
    \hfill
    \begin{subfigure}{.48\textwidth}
      \includegraphics[width=\textwidth]{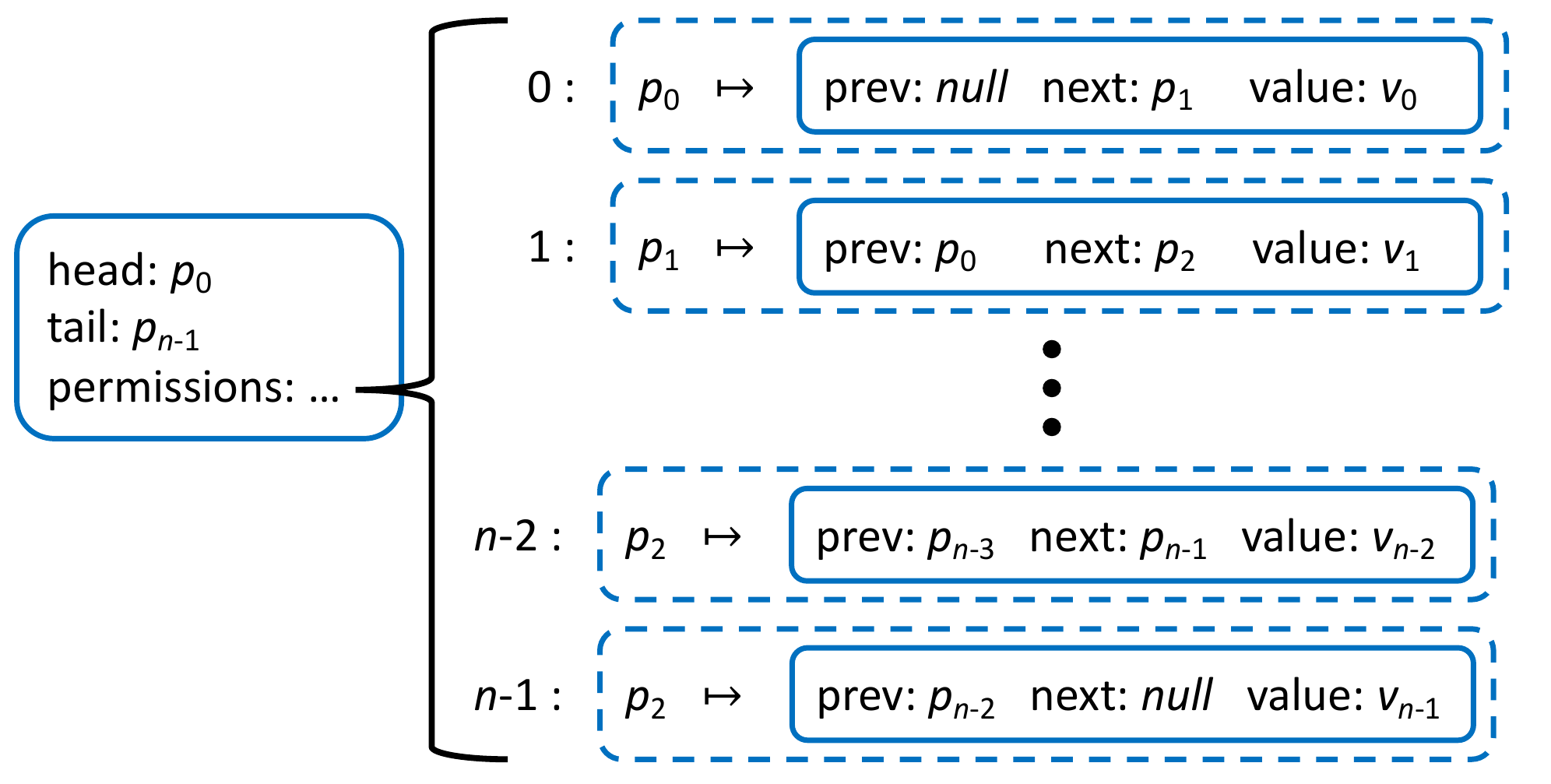}
      \caption{Ownership structure of a \verus doubly-linked list, which includes ghost state.} \label{fig:dlist:ghost}

    \end{subfigure}
    \caption{Doubly-linked lists. The dashed boxes are ghost, \texttt{proof}-mode \texttt{PermData} objects.} \label{fig:dlist}
\end{figure}

\begin{figure}

\begin{subfigure}{.48\textwidth}
\begin{lstlisting}[language=Verus,style=VerusLineNos]
struct Node<V> {
    prev: Option<PPtr<Node<V>>>,
    next: Option<PPtr<Node<V>>>,
    value: V,
}

struct DList<V> {
    #[spec] ptrs: Seq<PPtr<Node<V>>>,
    #[proof] perms: Map<nat, PermData<Node<V>>>,
    #[exec] head: Option<PPtr<Node<V>>>,
    #[exec] tail: Option<PPtr<Node<V>>>,
}

impl<V> DList<V> {
    #[spec] fn view(&self) -> Seq<V> { /* ... */ }

    #[exec] fn new() -> Self {
        ensures(|s: Self| s.well_formed(),
            && s.view().len() == 0); %*\lclnum\label{clnum:dlist:new}*
        /* ... */
    }

    #[exec] fn push_back(&mut self, v: V) {
        requires(old(self).well_formed());
        ensures(self.well_formed() && %*\lclnum\label{clnum:dlist:push-back}*
            equal(self.view(), old(self).view().push(v)));
        /* ... */
    }

    /* push_front, pop_back, pop_front similar */
}

fn main() {
    let mut t = DList::<u32>::new();
    t.push_back(2);   // 2
    t.push_back(3);   // 2, 3
    t.push_front(1);  // 1, 2, 3
    let x = t.pop_back();  // returns 3
    let y = t.pop_front(); // returns 1
    let z = t.pop_front(); // returns 2
    assert(x == 3);
    assert(y == 1);
    assert(z == 2);
}
\end{lstlisting}
\caption{Definition of the \texttt{DList} struct for the doubly-linked list example,
        along with the double-ended queue API, and example usage.} \label{fig:dlist:code:defn}
\end{subfigure}\hfill
\begin{subfigure}{.48\textwidth}
\begin{lstlisting}[language=Verus,style=VerusLineNos]
impl<V> DList<V> {
    #[spec] fn prev_of(&self, i: nat)
            -> Option<PPtr<Node<V>>> {
        if i == 0 {
            None
        } else {
            Some(self.ptrs.index(i as int - 1))
        }
    }

    #[spec] fn next_of(&self, i: nat)
            -> Option<PPtr<Node<V>>> {
        if i + 1 == self.ptrs.len() {
            None
        } else {
            Some(self.ptrs.index(i as int + 1))
        }
    }

    #[spec] fn wf_perm(&self, i: nat) -> bool { %*\lclnum\label{clnum:dlist:wf-perm}*
        self.perms.dom().contains(i) %*\lclnum\label{clnum:dlist:wf-perm-dom}*
        && equal(self.perms.index(i).view().pptr,
            self.ptrs.index(i as int).id()) %*\lclnum\label{clnum:dlist:wf-perm-id}*
        && match self.perms.index(i).view().value {
            Some(node) => %*\lclnum\label{clnum:dlist:wf-perm-prev-next}*
                equal(node.prev, self.prev_of(i)) &&
                equal(node.next, self.next_of(i)),
            None => false,
        }
    }

    #[spec] fn well_formed(&self) -> bool { %*\lclnum\label{clnum:dlist:well-formed}*
        (if self.ptrs.len() != 0 {
            equal(self.head, Some(self.ptrs.index(0))) && %*\lclnum\label{clnum:dlist:well-formed-head-tail}*
            equal(self.tail, Some(self.ptrs.index(self.ptrs.len() as int - 1)))
        } else {
            equal(self.head, None) && %*\lclnum\label{clnum:dlist:well-formed-head-tail-none}*
            equal(self.tail, None)
        })
        && forall(|i: nat| imply(0 <= i && i < self.ptrs.len(), self.wf_perm(i))) %*\lclnum\label{clnum:dlist:well-formed-forall}*
    }
}
\end{lstlisting}
\caption{Definition of \texttt{well\_formed}, used internally by the \texttt{DList}
         implementation to prove correctness of \texttt{push\_back} and others.} \label{fig:dlist:code:wf}
\end{subfigure}
    \caption{Doubly-linked list example} \label{fig:dlist:code}
\end{figure}

\autoref{fig:dlist:phys} shows the physical pointer structure of the list.
However, the diagram does not properly reflect a valid ownership structure
because it shows each node with multiple incoming pointers.
In the \verus implementation, we include an additional field in \verb`DList`:
the ghost \verb`permissions` field, which maintains permissions for \emph{every} node
in the doubly-linked list via a simple ``flattened'' structure, as in \autoref{fig:dlist:ghost}.
Specifically, for each $i \in \{0,\ldots,n-1\}$, we maintain a \verb`PermData` object
that maps pointer $p_i$ to the value it points to: the content of $i$th node, which contains
$v_i$ and the appropriate pointers, \verb`prev` as $p_{i-1}$ and \verb`next` as $p_{i+1}$.
To traverse the doubly-linked list, a user may use \verb`head` to determine $p_0$,
dereference $p_0$ using the $0$th permission object, find $p_1$, and so on.
In other words, we ghostily track the entire state of the list, but to get the same data in \emph{exec-mode}, we need to actually walk the pointers.

\autoref{fig:dlist:code} shows a snippet of the API. The spec-mode \verb`view()` function provides an abstraction
of the list as a simple sequence $v_0, v_1, \ldots, v_{n-1}$.
The specifications of the exec-mode API functions are all given in terms of this abstraction.
For example, the postcondition of \verb`DList::new()` says that the list represents the empty sequence,
while the postcondition of \verb`DList::push_front(v)` says that \verb`v` is appended to the end of the sequence.
The remaining three API functions (\verb`push_back`, \verb`pop_front`, and \verb`pop_back`) are similar.

The exec implementations are too involved to show here, so instead we show the definition of
the spec-mode predicate \verb`well_formed`~\lclref{clnum:dlist:well-formed},
i.e., the invariant that holds on \verb`DList<T>` and which each operation must preserve.
This definition says that~\lclref{clnum:dlist:well-formed-head-tail} the \verb`head` and \verb`tail` pointers are the first and last, respectively
(unless the list is empty, in which case~\lclref{clnum:dlist:well-formed-head-tail-none} they are both \verb`None`).
Finally, the \verb`forall`~\lclref{clnum:dlist:well-formed-forall} says that for each $0 \le i < n$, \verb`wf_perm(i)` holds; i.e.,
the $i$th permission is correct.
The definition of \verb`wf_perm(i)`~\lclref{clnum:dlist:wf-perm} says that
the permission is in our \verb`perms` map~\lclref{clnum:dlist:wf-perm-dom},
the permission corresponds to $p_i$~\lclref{clnum:dlist:wf-perm-id},
and that the \verb`prev` and \verb`next` fields of the node have the correct values~\lclref{clnum:dlist:wf-perm-prev-next}.

\begin{figure}
\noindent\begin{minipage}{.48\textwidth}
\begin{lstlisting}[language=Verus,style=VerusLineNos]
struct InvCell<T> { /* ... */ }

impl InvCell<T> {
    // Well-formedness of the InvCell
    #[spec] pub fn wf(&self) -> bool;

    // Boolean predicate indicating the values allowed to be stored.
    #[spec] pub fn inv(&self, val: T) -> bool;

    // Construct a new InvCell, with initial value `val` and invariant given by `f`.
    #[exec] pub fn new(val: T, #[spec] f: impl Fn(T) -> bool) -> Self {
        requires(f(val));
        ensures(|cell: Self| cell.wf() && forall(|t: T| f(t) == cell.inv(t)));
        /* ... */
    }
\end{lstlisting}
\end{minipage}
\hfill\begin{minipage}{.48\textwidth}
\begin{lstlisting}[language=Verus,style=VerusLineNos,firstnumber=17]
    // Write to the cell and return the old value.
    #[exec] pub fn replace(&self, val: T) -> T {
        requires(self.wf() && self.inv(val)); %*\clnum\label{clnum:invcell:replace:pre}*
        ensures(|old_val: T| self.inv(old_val));
        /* ... */
    }

    // Read the current value of the cell.
    #[exec] pub fn get(&self) -> T
        where T: Copy
    {
        requires(self.wf());
        ensures(|val: T| self.inv(val));  %*\clnum\label{clnum:invcell:get:post}*
        /* ... */
    }
}
\end{lstlisting}
\end{minipage}
    \caption{
        \label{fig:inv-cell-api}
        API and specification for \texttt{InvCell<T>}.
    }
\end{figure}

\section{Supporting Interior Mutability} \label{sec:interior-mutability}

Interior mutability is a Rust pattern in which the contents (the ``interior'') of a datatype \verb`X` may be modified even when it is shared via a reference type \verb`&X`.
Since \verb`&` is supposed to be an ``immutable'' reference, interior mutability appears to be at odds with the core tenets of Rust's type system, and in fact
interior mutability is only sound when restricted appropriately.
Rust's standard library provides a handful of types with interior mutability, e.g., \verb`Cell`, \verb`RefCell`, \verb`RwLock`, each of which provides a different set of restrictions
and characteristics. For example, \verb`Cell` may not be shared across threads, while \verb`RwLock` is thread-safe but incurs all the costs of being a lock.
The most flexible Rust datatype supporting interior mutability is the \verb`UnsafeCell`, upon which the aforementioned types are implemented. 
Since \verb`UnsafeCell` has no restrictions, it is---as the name implies---\emph{not} safe in general, and implementations that use it must take great care.

Providing safe and correct versions of such types in \verus is challenging, 
since in our SMT encoding, values of type \verb`&T` are always treated as immutable.
Therefore, to handle any \verb`Cell`-like datatypes, our SMT representation of \verb`&Cell<T>` cannot include an encoding of its mutable ``interior'' \verb`T`.
How, then, are we able to verify programs that require reasoning about this interior?

There are two broad classes of strategies a \verus developer can use:
\begin{enumerate}
  \item Use linear ghost state to represent the contents of a cell, similar to the way linear ghost state represents the value pointed to by a pointer.
  \item Avoid ``keeping track of'' the interior value entirely. Instead, when the interior value is read, model the result as being effectively nondeterministic,
      potentially using invariants to restrict the set of values that can be stored in the interior.
\end{enumerate}
\verus provides primitives and additional verified libraries supporting both styles, which the user can mix-and-match as needed.

The first strategy is the one used by our primitive \verb`PCell` (``permissioned \verb`Cell`'').
In the same way that \verb`PPtr` is our safe alternative to Rust's raw pointers,
\verb`PCell` is our safe alternative to \verb`UnsafeCell`.
\verb`PCell` uses a ghost permission mechanism with a similar API and specification to \verb`PPtr`, allowing us to track the interior value on the permission object.

The second strategy is exemplified by the type \verb`InvCell` of Verus' standard
library (\autoref{fig:inv-cell-api}), which provides a \verb`Cell`-like interface and allows the user to specify an invariant as a boolean predicate on values.
Whenever they write to the cell, they must prove the written value satisfies the invariant~\clref{clnum:invcell:replace:pre},
and when they read from it, they obtain an arbitrary value that they can \emph{assume} satisfies it~\clref{clnum:invcell:get:post}.
We first illustrate how this can be useful in our next section, and then we
discuss how \verb`InvCell` is itself verified in terms of lower-level invariant primitives.

\subsection{Verified Example: Memoized Function Calls} \label{sec:memoize}

\begin{figure}

\noindent\begin{minipage}{.5\textwidth}

\begin{lstlisting}[language=Verus,style=VerusLineNos]
#[spec] fn expected_result() -> u64 { /* ... */ }

#[exec] fn computation() -> u64 {
    ensures(|res: u64| res == expected_result()); %*\lclnum\label{clnum:comp:post}*
    /* ... */
}

#[spec] fn cell_value_inv(v: Option<u64>) -> bool {
    equal(v, Option::Some(expected_result()))
    || equal(v, Option::None) %*\lclnum\label{clnum:memo:cell-value-inv}*
}

#[spec] fn cell_is_valid(
        cell: InvCell<Option<u64>>) -> bool {
    cell.wf()
    && forall(|v| (#[trigger] cell.inv(v) ==
        cell_value_inv(v)))
}

#[exec] fn init_cell() -> InvCell<Option<u64>> {
    ensures(|c| cell_is_valid(c));
    InvCell::new(Option::None,
        |v: Option<u64>| cell_value_inv(v)) %*\lclnum\label{clnum:memo:init}*
}
\end{lstlisting}
\end{minipage}\hfill
\noindent\begin{minipage}{.5\textwidth}

\begin{lstlisting}[language=Verus,style=VerusLineNos]
#[exec] fn memoized_computation(
        cell: &InvCell<Option<u64>>) -> u64 {
    requires(cell_is_valid(*cell));
    ensures(|res: u64| res == expected_result());

    match cell.get() {
        Option::Some(res) => res, %*\lclnum\label{clnum:memo:case-none}*
        Option::None => {
            let res = computation();
            cell.replace(Option::Some(res));
            res %*\lclnum\label{clnum:memo:case-some}*
        }
    }
}

struct Client<'a> {
    cell: &'a InvCell<Option<u64>>,
}

fn main() {
    let c = init_cell();
    let client1 = Client { cell: &c };
    let client2 = Client { cell: &c };
    let x = memoized_computation(&client1.cell);
    let y = memoized_computation(&client2.cell);
    assert(x == y);
}
\end{lstlisting}
\end{minipage}

    \caption{
        \label{fig:memoize}
        Memoization example built on top of \texttt{InvCell}. 
    }
\end{figure}

\emph{Memoization} is an optimization technique whereby a user saves time
by storing the result of a computation the first time it is invoked; on future invocations,
they use the stored value. Here, we show how to memoize a function call
\texttt{computation()}.
In \autoref{fig:memoize} we use a function that takes 0 arguments for simplicity, so there is only a single value to memoize;
in our supplementary materials~\cite{verussupplementary}, we provide a slightly more complex example that memoizes
a single-argument function \texttt{computation(i)}.

To set up the problem, we assume that \verb`computation()` has a postcondition~\lclref{clnum:comp:post} ensuring
that its result is equal to some desired (spec-mode) value, \verb`expected_result()`.
(This is similar to the setup of \verb`fibo_impl` and \verb`fibo` from earlier.)
The aim is to construct a function \verb`memoized_computation` that also returns
\verb`expected_result()`.
To keep the problem interesting, we also insist that it be possible to share the 
``result store'' across potentially many clients.
As such, we need to use a shared reference type \verb`&`; however,
a given update invocation might need to update the result store, which requires
mutability. Therefore, we need to use some form of interior mutability.

In our approach, we use
an \verb`InvCell` with a simple invariant on the data held by the cell.
When initializing the cell~\lclref{clnum:memo:init}, we specify the data invariant as a (spec-mode) boolean predicate on the interior values;
here, we set it to the function \verb`cell_value_inv`, defined at \lclref{clnum:memo:cell-value-inv}.
The resulting property of the cell is expressed as in \verb`cell_is_valid`.
This definition says that the value is valid if and only if the stored value is either
\verb`None` (not yet computed) or contains the correct answer.

To implement \verb`memoized_computation`, we first read from the cell; if the value we get is
\verb`Some`, then we return the value immediately~\lclref{clnum:memo:case-none} (as we can assume it satisfies the invariant
we just specified).
Otherwise, we perform the computation, store it in the cell,
and return it~\lclref{clnum:memo:case-some}.

Finally, in \verb`main`, we show that we can create multiple ``clients,'' sharing a
reference to the cell, and use them to call \verb`memoized_computation`.

\subsection{Invariant Primitives and \texttt{InvCell} Verification} \label{sec:invcell}

Just as Rust's standard library implements \verb`Cell` via \verb`UnsafeCell`,
in \verus we can implement and verify \verb`InvCell` via our 
\verb`UnsafeCell`-equivalent, \verb`PCell`.
To do this, though, we first need to introduce our \emph{invariant primitives}.

To see what these are for, consider what happens when we
try to implement \verb`InvCell<T>` using \verb`UnsafeCell<T>`.
From the API, we know that we need to be able to write
even when we only have access to a shared reference \verb`&InvCell<T>`,
but writing to the underlying \verb`UnsafeCell<T>` requires exclusive ownership
of the \verb`PermData<T>` object.

Once again, we run into this problem of trying to gain exclusive ownership of something
that is shared. However, we have pushed the problem one layer down---to the \emph{ghost}
layer, and this is where \verus introduces its trusted invariant primitives to escape
the problem.

The two primitives are called \verb`LocalInvariant<G>` and
\verb`AtomicInvariant<G>`.\footnote{Both these types also have additional type parameters used to specify the invariant as a boolean predicate on \texttt{G}.}
Each one allows the user to store a (ghost) object \verb`G`; each one allows the user to perform
a ghost operation called \emph{opening the invariant}, where they obtain temporary, exclusive
ownership over the \verb`G`. For example, this snippet shows how the implementation of
\verb`InvCell<T>::replace` temporarily gains access to the \verb`PermData<T>` object:

\begin{lstlisting}[language=Verus,style=VerusLineNos]
impl InvCell<T> {
    pub fn replace(&self, val: T) -> T {
        requires(self.wf() && self.inv(val));
        ensures(|old_val| self.inv(old_val));

        let r;

        // Opens the invariant `&self.perm_inv` which has type `LocalInvariant<PermData<T>>`.
        // Opening the invariant is a ghost operation, and it binds to the ghost variable `perm`.
        open_local_invariant!(&self.perm_inv => perm => {
            // The code inside, however, is executable. This is where we actually perform
            // the write, using ownership of the ghost `perm` object, of type `PermData<T>`.
            r = self.pcell.replace(&mut perm, val);
        });

        r // Return the old value.
    }
}
\end{lstlisting}

The difference between the two primitives is that \verb`LocalInvariant<G>` is restricted for use
on a single thread: it does not implement \verb`Send` or \verb`Sync`, the traits Rust uses to mark thread-safety.
\verb`AtomicInvariant<G>` is thread-safe, and it does implement these traits: however,
this comes with an additional requirement, that the invariant may only be opened
for atomic operations.
Since \verb`InvCell` (like Rust's \verb`Cell`) is for single-threaded use,
we use \verb`LocalInvariant<V>` here. 

The reader might wonder what happens if we attempt to nest calls to the invariant-opening
operation, \verb`open_local_invariant`. It would certainly be unsound if we could open the same
\verb`LocalInvariant<G>` object twice, and obtain double-ownership of the `T`.
Indeed, \verus generates extra verification conditions to disallow such things
by tracking which invariants are ``open'' at a given time. These verification conditions
are designed to be lightweight, and they have no impact on our SMT generation
for cases outside of those which use the low-level invariant APIs.

\section{Concurrency, User-Defined Linear Ghost State, and Atomics} \label{sec:concurrency}

Rust's memory safety and ownership discipline allows our verification methodology to be sound in the presence of multi-threading.
However, verifying low-level code with fine-grained concurrency still requires additional techniques.

One key such technique is \emph{user-defined ghost state}: just as \verus provides
\verb`PermData` to track memory ownership,
the user can define their own ghost state to track elements of a custom concurrent protocol.
For defining ghost state, we primarily use a technique of prior work~\cite{ironsync-tr},
which suggests viewing user-defined ghost state as a ``localized transition system.''
In a localized transition system,
the user defines state transitions that can be expressed in terms of
thread-local views of the global program state, and then proves inductive invariants
on the resulting state transition system.

The result of this construction is a collection of \verb`proof`-mode (ownership-checked)  ghost types
representing components of the system state, along with an API for performing operations that manipulate
the ghost objects (constructing them, dropping them, or modifying them).
These operations might require certain properties to hold of the ghost state,
which can be proved from the inductive invariants of the transition system.
Finally, the programmer can manipulate these objects like any other ghost object, e.g., putting them
inside cells, invariant objects (\autoref{sec:invcell}), locks, atomics, or other mechanisms.

For example, we use this technique to verify a FIFO queue using a ring buffer with
atomic head and tail pointers.
At a very high level, we do this by first defining ghost state to represent the evolution
of the FIFO state. This state includes both the head and tail pointers, and as a result,
\verus gives us access to ghost objects that represent the head and tail,
and we then associate these ghost objects with atomic memory using an \verb`AtomicInvariant`.

For example, one of the transitions defined in this system (out of four total)
is called \verb`consume_start`.
Its corresponding API function has the following type signature:
\begin{lstlisting}[language=Verus,style=VerusLineNos]
#[proof] pub fn consume_start(
    #[proof] &self, #[proof] tail: &Fifo::tail<T>, #[proof] consumer: &mut Fifo::consumer<T>
  ) -> PermData<T> { /* auto-generated by Verus ghost state machinery */ }
\end{lstlisting}
The \verb`self` object, here, is a (ghost) metadata object that gives access to the API.
The interesting parameters are the \verb`tail`, a user ghost state object that
represents the value of the tail pointer,
and \verb`consumer`, which represents the thread-local state of the consumer thread.
Intuitively, this signature requires two things:
first, that the client ``prove'' that they are the consumer
by exhibiting the ghost state thread in order to perform the action.
Second, that they access the tail pointer while performing the action.
If the value of the tail pointers indicates that a message is waiting to be received,
then the consumer thread obtains the permission to access
a cell of the ring buffer, from which it can read a message, and which it relinquishes
at the end of the ``consume'' operation.
The validity of the operation (i.e., its ability to return this particular ghost object)
is encoded in the correctness conditions of the transition system,
and \verus requires the user to prove that these conditions hold from the inductive invariants.

Though user ghost state is usually intended for concurrent code, it is sometimes useful
for single-threaded code as well.
The supplementary materials~\cite{verussupplementary} include the following collection of examples for both
single-threaded and multi-threaded code, all with user-defined ghost state:
\begin{itemize}
  \item A concurrent FIFO queue based on a ring buffer, with head and tail pointers manipulated by atomics, as described above.
  \item A string interner that returns an identifier and ghost state, allowing the user to reason about the identifier
      as if it were the originally interned value.
  \item A thread-safe reader-writer lock, also implemented with atomics, which allows the user to specify an invariant on the protected data, in a similar fashion to \verb`InvCell`.
  \item A (non-thread-safe) reference-counted pointer, similar to Rust's \verb`Rc` (though without weak-pointers),
      which uses a \verb`PPtr` for the heap allocation and a \verb`PCell` for the reference counter.
\end{itemize}


\section{Implementation} \label{sec:implementation}

We forked the Rust compiler to introduce additional hooks and typechecking rules.
We then implemented \verus as a separate ``driver'' that links against the Rust compiler.
Both our fork and \verus are open source (\url{https://github.com/verus-lang/verus}) and in use by various verification projects.
We are working with the Rust compiler developers to 
extend Rust with additional language features, such as support
for ghost code, to better integrate \verus with Rust. 

In \autoref{fig:impl:examples} we list programs and examples we have verified using \verus.
For each, we report the number of lines of \verb`spec`, \verb`proof`,
and \verb`exec` code, the time to verify the example, and interesting \verus features they employ. 
The code snippets used in the figures in the paper are extracts from these examples,
which are available in full in the supplementary material~\cite{verussupplementary}.

\begin{table}
\caption{Example programs}
\begin{tabular}{l|l|l|l|l|l|p{4cm}}
    \hline
                           & \multicolumn{4}{l|}{sloc} & verif. &  \\
    Example                & \verb`spec` & \verb`proof` & \verb`exec` & total & time & \verus features \\ \hline
    \arrayrulecolor{gray}
    Allocator pages        & \allocatorpagesspec & \allocatorpagesproof & \allocatorpagesexe & \allocatorpagestotal & \timeallocatorpages & linearity \\ \hline
    XOR doubly-linked list & \doublylinkedxorspec & \doublylinkedxorproof & \doublylinkedxorexe & \doublylinkedxortotal & \timedoublylinkedxor & permissions \\ \hline
    Fibonacci              & \fibospec & \fiboproof & \fiboexe & \fibototal & \timefibo & \\ \hline
    Vector                 & \vectorspec & \vectorproof & \vectorexe & \vectortotal & \timevector & linearity \\ \hline
    Interner               & \internerspec & \internerproof & \internerexe & \internertotal & \timeinterner & user ghost state \\ \hline
    Memoization            & \memoizespec & \memoizeproof & \memoizeexe & \memoizetotal & \timememoize & interior mutability \\ \hline
    \verb`PCell` example usage           & \pcellspec & \pcellproof & \pcellexe & \pcelltotal & \timepcell & permissions \\ \hline
    \verb`PPtr` example usage            & \pptrspec & \pptrproof & \pptrexe & \pptrtotal & \timepptr &  permissions \\ \hline
    \verb`InvCell`         & \invcellspec & \invcellproof & \invcellexe & \invcelltotal & \timeinvcell &  permissions, \texttt{LocalInvariant} \\ \hline
    FIFO queue             & \queuefifospec & \queuefifoproof & \queuefifoexe & \queuefifototal & \timequeuefifo & permissions, atomics, \phantom{xyz} user ghost state \\ \hline
    \verus \verb`Rc`       & \rcspec & \rcproof & \rcexe & \rctotal & \timerc & permissions, cells, \phantom{padding} user ghost state \\ \hline
    \verus \verb`RwLock`   & \rwlockspec & \rwlockproof & \rwlockexe & \rwlocktotal & \timerwlock & permissions, user ghost state \\ \hline
\end{tabular}
\label{fig:impl:examples}
\end{table}

\section{User experience and error reporting} \label{sec:user-experience}

We discuss the Verus user experience by example. Suppose the user starts by defining an \verb`Account` struct and
an \verb`exec` function to transfer funds between accounts.

\begin{lstlisting}[language=Verus,style=VerusLineNos,firstnumber=6]
pub struct Account { pub balance: u64 }

pub fn transfer_funds(orig: &mut Account, dest: &mut Account, amount: u64) {
  requires([ old(orig).balance >= amount, old(dest).balance as nat + amount < u64::MAX ]);
  ensures([ dest.balance == old(dest).balance + amount, orig.balance == old(orig).balance - amount ]);
  orig.balance = orig.balance - amount;
  dest.balance = dest.balance + amount;
}
\end{lstlisting}

This function verifies, because Rust's type system ensures that \verb`orig` and \verb`dest`
are not aliased. In fact, if the user accidentally aliased the two arguments when calling \verb`transfer_funds`,
\begin{lstlisting}[language=Verus,style=VerusLineNos,firstnumber=15]
fn main() {
  let mut acct1 = Account { balance: 20_000 };
  transfer_funds(&mut acct1, &mut acct1, 20_000);
  assert(acct1.balance == 10_000);
}
\end{lstlisting}
\noindent the user would quickly get an error from the Rust borrow checker, and Verus would not
attempt to invoke Z3 to verify the invalid program, thereby allowing the user to quickly iterate by fixing
the issue and re-running Verus.
\begin{lstlisting}[style=Verus,literate=]
error[E0499]: cannot borrow `acct1` as mutable more than once at a time
  --> account.rs:17:32
   |
17 |     transfer_funds(&mut acct1, &mut acct1, 20_000);
   |     -------------- ----------  ^^^^^^^^^^ second mutable borrow occurs here
   |     |              |
   |     |              first mutable borrow occurs here
   |     first borrow later used by call
\end{lstlisting}

If one wrote similar code in Dafny, using a \verb`class` (a reference type) to
represent the \verb`Account`, they would declare the \verb`transfer_funds` method as
\lstinline[style=Verus]|method TransferFunds(orig: Accnt, dest: Accnt, amnt: nat)| with
similar preconditions and postconditions. Dafny would report that the postconditions
cannot be verified, which can be misleading to the developer, who has to determine that such
a failure is due to potential aliasing of \verb`orig` and \verb`dest`; the developer
would then need to add a framing condition to \verb`transfer_funds`,
\verb`requires orig != dest`.

In response to the Rust borrow checker failure above, the user may try and fix the \verb`main` function,

\begin{lstlisting}[language=Verus,style=VerusLineNos,firstnumber=15]
fn main() {
  let mut acct1 = Account { balance: 10_000 }; let mut acct2 = Account { balance: 20_000 };
  #[spec] let total_balance = acct1.balance + acct2.balance;
  transfer_funds(&mut acct1, &mut acct2, 20_000);
  assert(total_balance == acct1.balance + acct2.balance);
}
\end{lstlisting}

\noindent but inadvertently introduce a logical error, which Verus reports with precise pointers to the
offending code, and the relevant context (in this case, the failing precondition on the definition
of \verb`transfer_funds`):

\begin{lstlisting}[style=Verus,literate=]
error: precondition not satisfied
  --> account.rs:18:5
   |
9  |   requires([ old(orig).balance >= amount, old(dest).balance as nat + amount < u64::MAX ]);
   |              --------------------------- failed precondition
...
18 |     transfer_funds(&mut acct1, &mut acct2, 20_000);
   |     ^^^^^^^^^^^^^^^^^^^^^^^^^^^^^^^^^^^^^^^^^^^^^^
\end{lstlisting}

Adjusting the transferred amount to \verb`10_000` would result in successful verification upon re-running Verus.

This user experience is conceptually similar to that of the Viper separation logic engine~\cite{DBLP:conf/vmcai/0001SS16}
and the VeriFast C and Java verification tool~\cite{VeriFast}, with the distinction that both of these
tools use a separate substructural logic to reason about memory permissions:
the user has to explicitly manipulate permissions and separation logic predicates for the program data.
In Rust's linear type system, memory-reasoning permissions are implicitly associated with data ownership,
and manipulated by moving values, or taking references.

\section{Limitations} \label{sec:limitations}

Verus currently only supports mutable borrows (\verb`&mut`) of data passed as arguments to a function call:
mutable references in return values and explicit borrows on the right-hand side of assignments are not
supported.
We believe that adding more complete support with an approach similar to Creusot's is mainly a
matter of syntax, interface design, and engineering.

Unlike tools that re-encode ownership properties (e.g., with separation logic in Viper~\cite{DBLP:conf/vmcai/0001SS16}),
Verus relies on borrow-checking rules and hence cannot reason about traditional Rust unsafe code.
This may limit its applicability in applications that heavily rely on unsafe,
e.g., when direct memory manipulation is required to communicate with memory-mapped devices.
\autoref{sec:unsafe}, \autoref{sec:linear-ghost}, and \autoref{sec:dlist} discuss
encapsulations that alleviate the need for unsafe in certain contexts.

Verus is closely tied to Rust's type system,
which is more limited in some ways than the dependent type systems of Coq and F*.
This may preclude some more sophisticated styles of structuring proofs that are supported by Coq and F*.
While the limitation on mutable borrows will be lifted in the near future, limitations tied to Rust's type
system are imposed by Verus's design choices.

\newcommand{\sep}{\;\mid\;}

\newcommand{\integer}{i}
\newcommand{\structname}{\textrm{S}}
\newcommand{\usage}{u}
\newcommand{\mode}{m}
\newcommand{\modeu}{\mu}
\newcommand{\modu}[2]{#1\: #2}
\newcommand{\linear}{\textrm{linear}}
\newcommand{\shared}{\textrm{shared}}
\newcommand{\spec}{\textrm{spec}}
\newcommand{\prf}{\textrm{proof}}
\newcommand{\exec}{\textrm{exec}}
\newcommand{\lifetime}{\textrm{L}}
\newcommand{\callability}{\textrm{O}}
\newcommand{\datadecl}{\textrm{d}}
\newcommand{\datadecls}{\textrm{D}}
\newcommand{\static}{\textrm{static}}
\newcommand{\restricted}{\textrm{restricted}}
\newcommand{\once}{\textrm{Once}}
\newcommand{\many}{\textrm{Many}}

\newcommand{\val}{v}
\newcommand{\expr}{e}
\newcommand{\utyp}[2]{#1\: #2}
\newcommand{\typ}{\tau}
\newcommand{\typint}{\textrm{int}}
\newcommand{\typunit}{\textrm{Unit}}
\newcommand{\typnever}{\textrm{Never}}
\newcommand{\typpermi}[1]{\textrm{permission}(\integer \mapsto #1)}
\newcommand{\typopt}[1]{\textrm{Option}(#1)}
\newcommand{\typfun}[7]{\textrm{Fn}^{#1}_{#2\: #3}\: #4\:#5\rightarrow #6\:#7}

\newcommand{\exprunit}{()}
\newcommand{\exprbottom}{\bot}
\newcommand{\exprnone}[1]{\textrm{None}(#1)}
\newcommand{\exprsome}[2]{\textrm{Some}(#1: #2)}
\newcommand{\exprstruct}[2]{#1(#2)}
\newcommand{\exprfun}[7]{\lambda^{#1}_{#2\: #3}\:#4\!:\! #5\: #6.\: #7}
\newcommand{\exprdefault}[1]{\textrm{default}(#1)}
\newcommand{\exprcrashnever}[1]{\textrm{crash\_never}(#1)}
\newcommand{\exprhdata}{\textrm{hdata}()}
\newcommand{\exprhread}{\textrm{hread}()}
\newcommand{\exprhwrite}[1]{\textrm{hwrite}(#1)}
\newcommand{\exprpermi}[1]{\textrm{permission}(\integer \mapsto #1)}
\newcommand{\exprpdata}[1]{\textrm{pdata}(#1)}
\newcommand{\exprpreadi}[1]{\textrm{pread}(\integer @ #1)}
\newcommand{\exprpwritei}[2]{\textrm{pwrite}(\integer := #1 @ #2)}
\newcommand{\exprdrop}[1]{\textrm{drop}(#1)}
\newcommand{\exprcopy}[1]{\textrm{copy}(#1)}
\newcommand{\exprseq}[2]{#1;\, #2}
\newcommand{\exprlet}[4]{\textrm{let}\: #1\: #2 = {#3}\:\textrm{in}\:{#4}}
\newcommand{\exprifsome}[4]{\textrm{if}\:\textrm{let}\:\textrm{Some}(#1) = #2\: \textrm{then}\: #3\: \textrm{else}\: #4}
\newcommand{\exprletstruct}[4]{\textrm{let}\: #1(#2) = {#3}\:\textrm{in}\:{#4}}
\newcommand{\exprapp}[2]{#1\: #2}
\newcommand{\declstruct}[2]{#1\mapsto(#2)}
\newcommand{\structfield}[2]{#1\:#2}

\newcommand{\heapval}{h}
\newcommand{\heaptyp}{\textrm{H}}
\newcommand{\penv}{\textrm{P}}
\newcommand{\xenv}{\Gamma}
\newcommand{\lstrict}{\textrm{strict}}
\newcommand{\llax}{\textrm{lax}}

\ifextended
  \newcommand{\laxity}{\varsigma}
\fi

\ifextended
  \newcommand{\heapexp}[2]{(#1, #2)}
\fi

\newcommand{\modeless}{\sqsubseteq}
\newcommand{\modejoin}{\sqcup}

\newcommand{\turnstile}{\mbox{\Large $\vdash$}}
\newcommand{\sturnstile}{\mbox{\normalsize $\vdash$}}
\newcommand{\djudgment}[1]{\turnstile\; #1}
\newcommand{\ljudgment}[3]{#1\: \turnstile\; #2 : #3}
\newcommand{\cjudgment}[3]{#1\,;\, #2\: \turnstile\; #3 : \textrm{Copy}}
\newcommand{\tjudgment}[2]{#1\: \turnstile\; #2}
\newcommand{\trjudgment}[4]{#1\,;\, #2\,;\, #3\: \turnstile\; #4}

\ifextended
  \newcommand{\ejudgment}[9]{#1\,;\, #2\,;\, #3\,;\, #4\,;\, #5\: \turnstile_{\!#6}\; #7 : #8\: #9}
\fi

\ifextended
  \newcommand{\sjudgment}[9]{#1\,;\, #2\,;\, #3\,;\, #4\,;\, #5\: \turnstile\; (#6, #7) : #8\: #9}
\fi

\newcommand{\defjudgment}[3]{#1\: \turnstile\; #2\;\textrm{defaults\_to}\;#3}

\newcommand{\envextend}[3]{#1,\:#2\mapsto #3}
\newcommand{\envjoin}[2]{#1,#2}
\newcommand{\envmerge}[2]{#1\:\#\:#2}
\newcommand{\envmergen}[2]{#1\:\#\:\ldots\:\#\:#2}
\newcommand{\domain}[1]{\textrm{domain}(#1)}
\newcommand{\subst}[3]{#1 [ #2 := #3 ]}
\newcommand{\msubst}[5]{#1 [ #2 := #3, \ldots, #4 := #5 ]}
\newcommand{\ectx}{E}
\newcommand{\ehole}{[\cdot]}
\newcommand{\efill}[1]{\ectx[#1]}
\newcommand{\step}{\rightarrow}
\newcommand{\fullstep}{\longrightarrow}
\newcommand{\versions}[2]{\textrm{versions}_#1(#2)}

\newcommand{\envn}{!}
\newcommand{\envl}{\text{!`}}
\newcommand{\envlinear}[1]{\textrm{linear}(#1)}
\newcommand{\envshared}[1]{\textrm{shared}(#1)}
\newcommand{\envspec}[1]{\textrm{spec}(#1)}
\newcommand{\envlax}[1]{\textrm{lax}(#1)}

\newcommand{\modeof}[1]{\textrm{mode\_of}(#1)}
\newcommand{\lifetimeof}[1]{\textrm{lifetime\_of}(#1)}
\newcommand{\islinear}[1]{\textrm{is\_linear}(#1)}
\newcommand{\isstatic}[1]{\textrm{is\_static}(#1)}
\newcommand{\isunrestricted}[2]{\textrm{is\_unrestricted}(#1, #2)}
\newcommand{\nonspecfunctionmodes}[4]{\textrm{non\_spec\_function\_modes}(#1, #2, #3, #4)}
\newcommand{\functionbodycontext}[7]{\textrm{function\_body\_context}(#1, #2, #3, #4, #5, #6, #7)}

\newcommand{\defaultsto}[2]{#1\: \textrm{defaults\_to}\: #2}


\section{Formalization} \label{sec:formalization}

The previous sections introduced Verus concepts by example.
This section presents a small formal lambda calculus to make the concepts
from the previous sections more precise.
The goal of this lambda calculus is to serve as a model
to demonstrate particular features and their type safety.
We do not attempt to capture all of the semantics of Rust and Verus,
since formalizing Rust semantics is by itself a large and
challenging problem~\cite{pearce-rust-toplas2021, DBLP:journals/corr/abs-1903-00982, DBLP:journals/pacmpl/0002JKD18}.
Instead, we focus on a small set of topics that are novel to Verus
and are particularly relevant for type safety:

\begin{itemize}
\item \verb`spec`, \verb`proof`, and \verb`exec` functions
\item \verb`spec`, \verb`proof`, and \verb`exec` variables,
  showing how \verb`exec` and \verb`proof` variables are treated linearly,
  while \verb`spec` variables can capture nonlinear snapshots of data
  from \verb`exec` and \verb`proof` variables
\item \verb`spec`, \verb`proof`, and \verb`exec` annotations on datatype fields
\item linear ghost permissions, with read-only borrowing
\item ensuring termination of \verb`spec` and \verb`proof` code,
  particularly in the presence of mutation, recursive types,
  and higher-order features like traits or first-class functions
\ifextended
\item default values in \verb`spec` code,
  particularly for types are are uninhabited in \verb`proof` code and \verb`exec` code
\fi
\end{itemize}

Since this is already a sizable list of topics,
we aggressively minimize other features in our model language.
First, we omit concurrency entirely.
Second, we omit preconditions, postconditions, and verification condition generation,
focusing instead on type safety and termination.
(We believe that verification condition generation could be added
in a style similar to the formalization of Linear Dafny~\cite{DBLP:journals/pacmpl/LiLZCHPH22}.)
Third, our lambda calculus is a mostly-functional language that manipulates values,
rather than an imperative language that mutates values stored in locations.
(This contrasts with more detailed formalizations of Rust centered on
locations~\cite{pearce-rust-toplas2021, DBLP:journals/corr/abs-1903-00982}.)
\ifextended
The model language does, however, include two forms of mutation.
First, the language supports load and store operations that are controlled by linear ghost permissions.
Second, the lambda calculus contains a tiny nonlinear mutable heap
(actually, just a single nonlinear mutable heap location),
for the purpose of demonstrating the type safety of higher-order code in the presence of mutation.
\fi

Since our model language is based on values rather than locations,
it lacks Rust's distinction between a value (e.g. of type \verb`int`)
and a reference to that value (e.g. of type \verb`&int` or \verb`&mut int`).
Nevertheless, we still want to capture some notion of borrowing in order to demonstrate
borrowed linear ghost permissions.
For this, we associate $\linear$ and $\shared$ {\it usages}
with variable typings and expression typings.
The usage ``$\shared$'' represents immutable borrowing (as in \verb`&int`),
which we use for reading permissions.
For simplicity, we omit mutable borrowing,
instead annotating permissions with ``$\linear$''.

We build these usages into the mode system,
in a style similar to Linear Dafny~\cite{DBLP:journals/pacmpl/LiLZCHPH22}
(which in turn built on earlier work by Wadler's ``let!'' feature~\cite{DBLP:conf/ifip2/Wadler90}
and Cogent's purely functional support for borrowing~\cite{cogent}).
We define a mode $\mode$ to be $\spec$, $\prf$, or $\exec$ (see \autoref{fig:formal:syntax}),
with a reflexive, transitive ordering $\exec \modeless \prf \modeless \spec$
and a least upper bound $\mode_1 \modejoin \mode_2$ that
is the least $\mode$ such that
$\mode_1 \modeless \mode$ and
$\mode_2 \modeless \mode$.
We then associate a usage $\usage$ with proof and exec modes,
since proof and exec variables can be linear or borrowed:
\[
\modeu ::= \spec \sep \modu{\prf}{\usage} \sep \modu{\exec}{\usage}
\]
The environment
$\xenv ::= \{x_1 \mapsto \utyp{\modeu_1}{\typ_1}, \ldots, x_n \mapsto \utyp{\modeu_n}{\typ_n}\}$
tracks the mode, usage, and type of each variable.
We refer to a binding $x \mapsto \utyp{\modu{\mode}{\linear}}{\typ}$ as a linear binding,
and we refer to $x \mapsto \utyp{\modu{\mode}{\shared}}{\typ}$
and $x \mapsto \utyp{\spec}{\typ}$ as nonlinear bindings.
We write $\envl\xenv$ to extract just the linear bindings from $\xenv$
and we write $\envn\xenv$ to extract just the nonlinear bindings from $\xenv$
(see \autoref{fig:formal:typingdefs}).

We write $\xenv_1, \xenv_2$ to concatenate two environments together.
For writing typing rules, though,
we often want to split environments in a more sophisticated way then simple concatenation.
In particular, we want to split linear bindings between subexpressions
while sharing nonlinear bindings among subexpressions.
For this, we write $\xenv = \envmerge{\xenv_1}{\xenv_2}$.
For example, in the typing rule for adding two integers (see \autoref{fig:formal:typing}),
the left subexpression gets environment $\xenv_1$ and the right
subexpression gets $\xenv_2$:
\[
\inferrule{
  \ejudgment{\datadecls}{\heaptyp}{\penv_1}{\xenv_1}{\mode}{\laxity}{\expr_1}{\modeu}{\typint}
  \\
  \ejudgment{\datadecls}{\heaptyp}{\penv_2}{\xenv_2}{\mode}{\laxity}{\expr_2}{\modeu}{\typint}
}{
  \ejudgment{\datadecls}{\heaptyp}{\envmerge{\penv_1}{\penv_2}}{\envmerge{\xenv_1}{\xenv_2}}{\mode}{\laxity}{\expr_1 + \expr_2}{\modeu}{\typint}
}
\]
(The other environments can be ignored for now;
\autoref{sec:termination} discusses $\datadecls$ and $\heaptyp$,
and \autoref{sec:permissions} discusses $\penv$ and $\mode$.)

When $\xenv = \envmerge{\xenv_1}{\xenv_2}$,
all nonlinear bindings in $\xenv$ appear in both $\xenv_1$ and $\xenv_2$.
Linear bindings, however are more subtle:
a linear binding in $\xenv$ appears as-is in one of the environments ($\xenv_1$ or $\xenv_2$),
and is demoted to mode $\spec$ in the other environment.
Thus, the environment that didn't get the linear binding can still talk about the variable
in specifications.
For example, if $\xenv$ has a linear binding for $x_2$ and we split $\xenv$
into $\xenv = \envmerge{\xenv_1}{\xenv_2}$,
and $\xenv_1$ receives the linear binding for $x_2$,
then $\xenv_2$ will receive a $\spec$ binding for $x_2$:

$\xenv = \{x_1 \mapsto \utyp{\modu{\exec}{\shared}}{\typ},\: x_2 \mapsto \utyp{\modu{\exec}{\linear}}{\typ}\}$

$\xenv_1 = \{x_1 \mapsto \utyp{\modu{\exec}{\shared}}{\typ},\: x_2 \mapsto \utyp{\modu{\exec}{\linear}}{\typ}\}$

$\xenv_2 = \{x_1 \mapsto \utyp{\modu{\exec}{\shared}}{\typ},\: x_2 \mapsto \utyp{\spec}{\typ}\}$

(See \autoref{fig:formal:typingdefs} for a formal definition of $\envmerge{\xenv_1}{\xenv_2}$.)

Following Linear Dafny's formalization,
our model language allows borrowing by temporarily viewing linear variables as shared
within a lexical scope.
For example, in the sequencing expression $\exprseq{\expr_1}{\expr_2}$
the first expression $\expr_1$ can view a portion of the environment $\xenv_b$ as shared,
and these variables then revert to linear in $\expr_2$:
\[
\inferrule{
  \ejudgment{\datadecls}{\heaptyp}{\penv_1, \envshared{\penv_b}}{\xenv_1, \envshared{\xenv_b}}{\mode}{\laxity}{\expr_1}{\modeu_1}{\typunit}
  \\
  \ejudgment{\datadecls}{\heaptyp}{\penv_2, \envlinear{\penv_b}}{\xenv_2, \envlinear{\xenv_b}}{\mode}{\laxity}{\expr_2}{\modeu_2}{\typ_2}
}{
  \ejudgment{\datadecls}{\heaptyp}{(\envmerge{\penv_1}{\penv_2}), \envlinear{\penv_b}}{(\envmerge{\xenv_1}{\xenv_2}), \envlinear{\xenv_b}}{\mode}{\laxity}{\exprseq{\expr_1}{\expr_2}}{\modeu_2}{\typ_2}
}
\]
Here, the notation $\envlinear{\xenv_b}$ means $\xenv_b$ with all proof/exec bindings made linear,
$\envshared{\xenv_b}$ means $\xenv_b$ with all proof/exec bindings made shared.
(For more detail on this style of borrowing, which was inspired by
Wadler's ``let!'' feature~\cite{DBLP:conf/ifip2/Wadler90}, see~\cite{DBLP:journals/pacmpl/LiLZCHPH22}.)

For simplicity and clarity,
the model language implements a linear type system that prohibits discarding linear resources;
for example, it disallows discarding permissions,
and the only way to deallocate a linear \verb`struct` is to deconstruct it with pattern matching.
(Rust behaves more like an affine type system, allowing dropping of any value.)
Rust includes a \verb`Copy` trait,
implemented by simple types like \verb`bool` and \verb`u64`,
for types that are inherently nonlinear and may be freely copied.
Our model language also includes a judgment $\cjudgment{\datadecls}{\mode}{\typ}$
\ifextended
(see \autoref{fig:formal:additional} for the formal definition)
\fi
to indicate that a type $\typ$ may be copied or dropped,
although, for simplicity, the copies and drops are explicit,
using the expressions $\exprcopy{\expr}$ and $\exprdrop{\expr}$.

Even though the linear type system prohibits copying and dropping linear resources,
it allows arbitrary implicit copying and dropping in specifications.
It defines $\cjudgment{\datadecls}{\spec}{\typ}$ to be true of all types,
so that in $\spec$ mode, code can always use $\exprcopy{\expr}$ and $\exprdrop{\expr}$.
Furthermore, $\spec$ variables can be implicitly copied when splitting
variables among subexpressions using $\envmerge{\xenv_1}{\xenv_2}$.
In particular, since environment splitting creates $\spec$ copies of linear variable bindings,
$\spec$ variables can be used to capture immutable snapshots of mutable linear resources.
For example if variable $x_l$ is bound linearly,
the expression ``$\exprlet{\spec}{x_s}{x_l}{\expr}$''
can make a $\spec$ copy $x_s$ of the linear variable $x_l$ without consuming $x_l$.
Here, $\expr$ can continue to use $x_l$ linearly while simultaneously
keeping the immutable snapshot $x_s$.
This allows specifications to talk about the past state (old snapshots) of linear resources
as well as the current state,
which is useful for specifications that relate old states to new states.

\subsection{Permissions} \label{sec:permissions}

\autoref{sec:ghost:perms} described how linear ghost permissions
allow safe manipulation of low-level pointers.
To model permissions and pointers,
the model language contains a $\typpermi{\typ}$ type representing permission
to read or write a value to pointer $\integer$,
which, for simplicity, is simply an integer constant.
There are three operations on pointers and permissions:

\begin{itemize}
\item $\exprpreadi{\expr_p}$ reads the value stored at pointer $\integer$,
  based on the access granted by permission $\expr_p$
\item $\exprpwritei{\expr_v}{\expr_p}$ writes a new value $\expr_v$ to pointer $\integer$,
  based on the access granted by permission $\expr_p$
\item $\exprpdata{\expr_p}$ takes a $\spec$-mode snapshot of the value currently stored at pointer $\integer$,
  based on the access granted by a $\spec$-mode copy of the permission $\expr_p$
\end{itemize}

\autoref{fig:formal:typingdefs} shows the typing rules for these operations.
Each operation uses the same permission,
but with a different mode.
Writing requires linear access to the permission,
so that no aliased views of the permission can have a stale view of the permission.
Reading, on the other hand, can be performed on borrowed permissions with mode $\shared$.
Finally, $\exprpdata{\expr_p}$ uses the permission with mode $\spec$,
allowing use in specifications.
For simplicity, we omit operations to deallocate permissions or allocate new permissions;
we assume that all permissions are passed in to a program when the program starts and
returned at the end of the program.
However, the typing rule for $\exprpwritei{\expr_v}{\expr_p}$ allows the program to
change the type of a permission, effectively reallocating the memory for a new type.
Thus, the linear handling of permissions is crucial;
if the permissions were not linear, a program could use a stale permission to read
a value memory from memory with an out-of-date type, subverting type safety.
Our type safety theorem (\autoref{sec:soundness}) ensures that this cannot happen.

While $\exprpdata{\expr_p}$ is a ghost-only operation,
$\exprpreadi{\expr_p}$ and $\exprpwritei{\expr_v}{\expr_p}$ perform run-time actions that,
in an implementation, would be compiled to machine code.
Since proofs and specifications are ghost code,
they are not allowed to perform $\exprpreadi{\expr_p}$ and $\exprpwritei{\expr_v}{\expr_p}$ operations.
To enforce this, the typing rules include an access level $\mode$
that limits what operations the code is allowed to perform.
Many operations (such as integer addition) can be performed in any mode,
but $\exprpreadi{\expr_p}$ and $\exprpwritei{\expr_v}{\expr_p}$ can only be performed in $\exec$ mode:
\[
\inferrule{
  \ldots
}{
  \ejudgment{\datadecls}{\heaptyp}{\envmerge{\penv_1}{\penv_2}}{\envmerge{\xenv_1}{\xenv_2}}{\exec}{\laxity}{\exprpwritei{\expr_v}{\expr_p}}{\modu{\prf}{\linear}}{\typ}
}
\]
The bodies of $\exec$ functions are type-checked with access level $\exec$,
and can perform run-time reads and writes,
while the bodies of $\prf$ and $\spec$ functions are type-checked with access level
$\prf$ and $\spec$, and therefore cannot perform run-time reads and writes.

The formal semantics of our mode language use a value $\exprpermi{\val}$
to represent the storage location pointed to by pointer $\integer$,
holding contents $\val$.
Notice that this storage location is just a value,
so it may get passed around from expression to expression,
in and out of functions,
although the type system's linearity ensures that there will never be two inconsistent linear copies
of a permission for the same pointer.
There may, however, be many $\spec$ copies of the permission floating around that
contain snapshots old permission state,
and the code is allowed to execute $\exprpdata{x_s}$ on these snapshots to obtain
the old contents as $\spec$ values.
In fact, it's important that $\exprpdata{x_s}$ return the contents associated with
the snapshot $x_s$,
rather than the most up-to-date value,
because specifications must be deterministic:
they cannot produce different values just because the state has changed.

In order to prove the type safety of our model language,
we have to prove that well-typedness is preserved, including the well-typedness of permission values.
For this, we use an environment $\penv$ that keeps track of whether each storage location $\integer$
is currently linear or borrowed ($\shared$).
We also need to type-check the snapshotted $\spec$ copies of permissions.
For this, we provide a special ``dead-end'' rule that allows stale copies of a permission
to persist as $\spec$-only (see \autoref{fig:formal:typing2});
this is sound because the $\spec$-only permission cannot be coerced
back to a $\shared$ or $\linear$ permission for run-time reads and writes
(hence our description of the $\spec$ copy as a ``dead-end'').

\subsection{Functions and Lifetimes}

Since our model language is a lambda calculus,
it supports functions.
This models both first-order functions,
as shown in the examples in previous section,
and higher-order features.
For example, Verus supports first-class functions in specifications.
Verus also supports simple traits with methods taking a \verb`self` argument;
these simple traits can encode first-class functions.

First-class functions in Rust (called closures in Rust terminology)
are considerably more complicated
than simply-typed lambda calculus functions, though.
First, Rust distinguishes between \verb`Fn`,
which represents functions that may be called many times,
and \verb`FnOnce`,
which represents functions that can only be called once.
(There is also \verb`FnMut`, which we do not model.)
\verb`FnOnce` functions may capture linear variables,
while \verb`Fn` functions cannot.
We define a {\it callability} $\callability ::= \once \sep \many$
to represent this distinction.

Rust first-class functions also have lifetimes associated with them,
so that a function cannot outlive the variables that it captures,
even if these variables are nonlinear (e.g., variables of type \verb`&int`).
Rust lifetimes are quite sophisticated, including parameterization over lifetime variables,
but, for simplicity,
our model language contains just two hard-coded lifetimes
$\lifetime ::= \static \sep \restricted$.
The lifetime $\static$ means that a function may be passed around freely,
because it does not capture any $\shared$ variables,
while the lifetime $\restricted$ means that a function may have captured $\shared$ variables,
and therefore the function cannot be returned past the nearest enclosing borrowing scope.
(Note that this rather strict limitation is only for the model language;
the actual \verus implementation allows Rust's more sophisticated lifetime variables.)
The definition $\textrm{function\_body\_context}$ (see \autoref{fig:formal:typingdefs})
specifies exactly which variables may be captured by the body of a function definition
for each of the four combinations of $\callability$ and $\lifetime$.

Finally, Verus adds yet another dimension to functions: a mode $\mode$
that represents a function being a $\spec$ function, $\prf$ function, or $\exec$ function.
The typing rule for function calls $\exprapp{\expr_f}{\expr_a}$ (\autoref{fig:formal:typing2})
require that the function $\expr_f$'s mode be accessible
according to the current access level,
which means $\mode \modeless \mode_f$ if the current access level is $\mode$
and $\expr_f$ is a function of mode $\mode_f$.

With all of these configuration options,
we can write function definitions $\exprfun{\mode}{\callability}{\lifetime}{x}{\modeu}{\typ}{\expr}$
of type $\typfun{\mode}{\callability}{\lifetime}{\modeu_1}{\typ_1}{\modeu_2}{\typ_2}$.
\autoref{fig:formal:typing2} shows two main rules for assigning function types to function definitions,
one for non-spec functions and one for spec functions.
The latter allows spec functions to capture snapshots of shared variables without worrying about lifetimes;
it does not allow direct capturing of linear variables (since this would effectively discard the linear variable),
but programs can always capture a linear variable indirectly by splitting off
a $\spec$ copy of the linear variable from the surrounding environment and capturing the $\spec$ copy.

The language also allows $\spec$ snapshots of non-spec functions.
Just as snapshots of permissions required a dead-end rule,
functions also require dead-end rules, which bring a slightly annoying technicality.
We could write very simple dead-end rules that just ignore the function's body completely,
and this would be sound,
since a non-spec function snapshotted as a $\spec$ value can never be called,
so the body doesn't matter.
However, our proof of termination in \autoref{sec:soundness} is based on a translation
of our model language into the calculus of inductive constructions (CIC, the logic used by Coq),
and for this translation we need to retain enough of the body to form a well-typed CIC term.
For this, we need to relax the linearity checking in order for the retained body to remain well-typed
in the model language.
\ifextended
Therefore, we parameterize all of the typing rules with a flag $\laxity ::= \lstrict \sep \llax$ that enables ($\lstrict$) or disables
($\llax$) linearity checking, and use $\llax$ for the dead-end function rules.
Note, however, that $\laxity$ is just for assisting the CIC translation,
and does not correspond to anything in the real Verus implementation of type checking and mode checking;
Verus always uses the $\lstrict$ rules.
\fi

\ifextended
\subsubsection{Default values} \label{sec:default}

In Verus, $\spec$ functions are total and correspond closely to SMT total functions,
for the sake of enabling a direct, efficient translation from Verus into SMT queries.
As in Z3, Boogie, and Creusot,
but unlike Dafny, \fstar, and Coq,
Verus $\spec$ functions always return some value of their output type, for all possible inputs.
For example, the division function is uninterpreted when dividing by zero;
it returns some integer, but we can't know which integer.
A Verus function that wraps a division operation is itself well-formed
and returns whatever arbitrary integer the division returns:

\begin{lstlisting}[language=Verus]
#[spec]
fn my_div(i: int, j: int) -> int { i / j }
\end{lstlisting}

On the other hand, non-spec functions are partial:
Verus prohibits division by zero, for example, in $\prf$ and $\exec$ functions.
(If the code above were declared \verb`#[proof]` or \verb`#[exec]`,
it would need a precondition specifying that $j \neq 0$.)
Therefore, it is important to ensure that $\spec$ values don't leak back into non-spec code.
This is especially important for spec functions that return an uninhabited type,
such as Rust's ``Never'' type (written as ``!'' in Rust syntax);
it would be unsound for a value of the Never type to appear as an $\exec$ value.
(For a real-world example of values of uninhabited type causing unsoundness in Dafny,
see~\cite{dafny:unsoundness}.)

To model this, we include the type $\typnever$ in our language,
along with a special default value $\exprbottom$ that has this type in specifications
(and {\it only} in specifications).
To demonstrate that the value doesn't leak into $\prf$ variables or $\exec$ variables,
we include an expression $\exprcrashnever{\expr}$, usable in $\prf$ and $\exec$ modes,
that crashes (fails to step) when given a value of type $\typnever$.
Our type safety proof ensures that this crash never happens.

Building on $\exprbottom$, we can define default values of all well-formed types.
(The details are found in the definition of ``$\defaultsto{\typ}{\val}$'' in \autoref{fig:formal:additional}.)
\fi

\subsection{Termination} \label{sec:termination}

When Verus code is compiled,
all ghost code is erased (not compiled to machine code).
This erasure is sound only if the ghost code always terminates with no side effects.
The access level described in \autoref{sec:permissions} enforces the absence of side effects.
Enforcing termination, though, is more delicate because
mutation and recursive types can often encode nontermination
when combined with higher-order features,
like traits and first-class functions.

To see how this can happen, consider the following two examples, written in OCaml.
The first example creates a mutable reference that holds a function of type \verb`unit -> unit`.
It then stores a new function into the mutable reference
The new function recursively calls itself by reading itself from the reference,
causing an infinite loop:

\begin{lstlisting}
let r: (unit -> unit) ref = ref (fun () -> ()) in
r := (fun () -> !r ());
!r ()
\end{lstlisting}

The second example passes a function to itself as an argument,
using a recursive type \verb`R` to encapsulate the function in a well-typed way.
The function then calls its argument, which means it calls itself,
causing an infinite loop:

\begin{lstlisting}
type r = | R of (r -> unit)
let f (R x) = x (R x) in f (R f)
\end{lstlisting}

\noindent(Note: this can also be encoded directly in Rust as follows:
\begin{lstlisting}[language=Verus]
        trait T { fn f(&self); }
        fn rec<A: T>(x: &A) { x.f(); }
        struct S {}
        impl T for S { fn f(&self) { rec(self); } }
        fn foo() { let s = S {}; s.f(); }
\end{lstlisting}
although this is more complicated.)

Neither of these examples would be caught by \verb`decreases` clauses,
because there are no explicit recursively-defined functions in the code.

\ifextended
To demonstrate that Verus can correctly prohibit these sources of nontermination,
we include two more features in our model language.
First, we add a small heap that consists of a single value $\heapval = \val$
stored in a single heap location, having type $\heaptyp = \typ$.
We also add operations $\exprhread$, $\exprhwrite{\expr}$, and $\exprhdata$
to read, write, and snapshot the heap data.
Using this, we can express nontermination with the following heap value $\heapval$:
\[
\heapval =
  \exprfun{\exec}{\many}{\static}{x}{\modu{\exec}{\linear}}{\typunit}
    {\exprapp{(\exprhread)}{x}}
\]
As in the first OCaml example, this function reads itself from the heap
and then calls itself, causing an infinite loop.
In fact, our model language allows this nontermination for $\exec$ functions.
It prohibits it for $\spec$ and $\prf$ functions,
though, because $\spec$ and $\prf$ functions don't have a sufficent access level $\mode$
to call $\exprhread$, which requires access level $\exec$.
More subtly, they don't have sufficient access to call $\exprhdata$ either,
which could be another source of nontermination ---
unlike the $\exprpdata{\expr_p}$ operation, $\exprhdata$ requires access level $\exec$,
not access level $\spec$.
(This access-level restriction also ensures that specifications are deterministic,
since $\exprhdata$ reads directly from the global heap instead of from a snapshotted permission value.)

Note that Verus currently uses linear permissions rather than a heap,
but other languages like Dafny and \fstar use heaps,
and Linear Dafny uses regions, which have similar properties to heaps.
We include both the heap and linear permissions in the model language
to highlight the distinction between the rules for the two approaches.
In particular, we found the difference between
the $\exprpdata{\expr_p}$ rules and the $\exprhdata$ rules surprising and worth formalizing.

The second feature we add is recursive type definitions
in the form of recursive structs.
\fi
A recursive struct declaration
$\datadecl ::= \declstruct{\structname}{\structfield{\mode_1}{\typ_1},\ldots,\structfield{\mode_n}{\typ_n}}$
declares a struct $\structname$ with $n$ fields,
each having a mode and a type.
When constructing or destructing structs,
the typing rules join the mode of the fields with the mode of the overall struct value
using the $\modejoin$ operator (see \autoref{fig:formal:typing2}).
This joining ensures, for example, that when reading fields from a $\spec$ snapshot
of an $\exec$ struct value, the result will have mode $\spec$ even if the field mode
is $\prf$ or $\exec$.

\ifextended
To complement recursive structs, the language also includes an option type $\typopt{\typ}$,
allowing interesting recursive types like lists and trees.
The rules for options are straightforward, with one nuance:
the expression for matching on options ($\exprifsome{x}{\expr_1}{\expr_2}{\expr_3}$)
restricts the access level for $\expr_2$ and $\expr_3$ depending on the mode of the option
(see \autoref{fig:formal:typing}).
This prohibits, for example, $\exec$ code from testing whether a $\spec$ option
is Some or None and causing a side effect that depends on the test result.
It does allow $\prf$ functions to test $\spec$ options, though, which is sound and useful in proofs.
\fi

The rules for well-formed struct declarations
allow recursive structs (although, for simplicity, they disallow mutual recursion).
These rules enforce a standard ``strict positivity'' restriction
(used by Coq, Lean, \fstar, and Dafny).
However, they only require strict positivity in $\spec$ and $\prf$ function types;
non-positive uses are allowed in $\exec$ function types
(unlike in Coq, Lean, and Dafny, where all function types are restricted).

\ifextended
The main judgment for well-formed types has the form
$\trjudgment{\datadecls}{\datadecls_r}{\datadecls_p}{\typ}$ (see \autoref{fig:formal:typing2}).
The environment $\datadecls$ contains all declarations before the current struct
that we're checking,
and $\datadecls_r$ contains the current struct if it is legal
to use recursively in the current position.
The global environment $\datadecls$ is well formed ($\djudgment{\datadecls}$)
if the empty environment is well formed ($\djudgment{\varnothing}$) and
each subsequent struct declaration is well formed:
\[
\inferrule{
  \djudgment{\datadecls}
  \\
  \datadecl = \declstruct{\structname}{\structfield{\mode_1}{\typ_1},\ldots,\structfield{\mode_n}{\typ_n}}
  \\
  \forall 1\le i \le n,\;(\trjudgment{\datadecls}{\varnothing}{\datadecl}{\typ_i}\;\;\textrm{and}\;\;\ljudgment{\mode_i}{\typ_i}{\static})
}{
  \djudgment{\datadecls, \datadecl}
}
\]
For example, the rule for $\spec$ function types makes the struct available
in the return type, but not in the argument type ($\datadecls_r$ is set to $\varnothing$
when checking the argument type):

\[
\inferrule{
  \trjudgment{\datadecls}{\varnothing}{\varnothing}{\typ_1}
  \\
  \trjudgment{\datadecls}{\datadecls_r}{\datadecls_p}{\typ_2}
}{
  \trjudgment{\datadecls}{\datadecls_r}{\datadecls_p}
    {\typfun{\spec}{\many}{\static}{\spec}{\typ_1}{\spec}{\typ_2}}
}
\]

By contrast, the rule for $\exec$ function types allows recursion in the argument type,
so that executable code can still express nontermination through recursive types.
(Note: the two separate environments $\datadecls_r$ and $\datadecls_p$
are used because the rules actually enforce two separate properties:
first, strict positivity and second, that recursive structs have default values.
For the second property, declarations start in $\datadecls_p$ and then shift to $\datadecls_r$.)
\fi

\subsection{Semantics and Type Safety} \label{sec:soundness}

\ifextended
\autoref{fig:formal:evaluation}
\fi
defines evaluation rules
$\heapexp{\heapval}{\expr} \fullstep \heapexp{\heapval'}{\expr'}$
\ifextended
for a heap and expression to take a single step to a new heap and expression.
\fi
Based on this and the typing rules,
we have proven type preservation,
progress, and ghost-code termination:

\begin{itemize}
\item Preservation:
if $\djudgment\datadecls$
and $\sjudgment{\datadecls}{\heaptyp}{\penv}{\xenv}{\mode}{\heapval}{\expr}{\modeu}{\typ}$
and $\heapexp{\heapval}{\expr} \fullstep \heapexp{\heapval'}{\expr'}$,\newline
then $\sjudgment{\datadecls}{\heaptyp}{\penv}{\xenv}{\mode}{\heapval'}{\expr'}{\modeu}{\typ}$
\item Progress:
if $\djudgment\datadecls$
and $\sjudgment{\datadecls}{\heaptyp}{\penv}{\varnothing}{\mode}{\heapval}{\expr}{\modeu}{\typ}$
and $\expr$ is not a value,\newline
then there is some $\heapexp{\heapval'}{\expr'}$ such that $\heapexp{\heapval}{\expr} \fullstep \heapexp{\heapval'}{\expr'}$.
\item Termination:
if $\mode\in\{\spec, \prf\}$
and $\djudgment\datadecls$
and $\sjudgment{\datadecls}{\heaptyp}{\penv}{\varnothing}{\mode}{\heapval_0}{\expr_0}{\modeu}{\typ}$
then there is no infinite evaluation sequence
$\heapexp{\heapval_0}{\expr_0} \fullstep \heapexp{\heapval_1}{\expr_1} \fullstep \heapexp{\heapval_2}{\expr_2} \fullstep \heapexp{\heapval_3}{\expr_3} \fullstep \ldots$
\end{itemize}

The supplementary material~\cite{verussupplementary} contains proofs of these theorems.
The preservation and progress proofs are straightforward.
The termination proof works by translating the declarations, types, and expressions
into CIC (calculus of inductive constructions) declarations and terms,
and then proving that the CIC declarations and terms are well-typed
and proving a simulation between the CIC reduction steps and the $\heapexp{\heapval}{\expr}$ evaluation steps.
The translation to CIC is fairly simple:
$\spec$ and $\prf$ function types are translated into corresponding
CIC function types, while $\exec$ function types are simply translated into the unit type,
$\exec$ functions are erased completely (translated into the unit value),
\ifextended
$\exprpermi{\val}$ is translated into $\val$, and the heap is erased completely.
\fi

\begin{figure}
\begin{tabular}{lcrl}
variable        & $ x $ & & \\
integer         & $ \integer $ & ::= & $ \ldots, -2, -1, 0, 1, 2, \ldots $ \\
struct name     & $ \structname $ & & \\
usage           & $ \usage $ & ::= & $ \linear \sep \shared $ \\
mode            & $ \mode $ & ::= & $ \spec \sep \prf \sep \exec$ \\
mode + usage\!\!  & $ \modeu $ & ::= & $ \spec \sep \modu{\prf}{\usage} \sep \modu{\exec}{\usage}$ \\
callability     & $ \callability $ & ::= & $ \once \sep \many $ \\
lifetime        & $ \lifetime $ & ::= & $ \static \sep \restricted $ \\
type            & $ \typ $ & ::= & $ \typint \sep \typunit
\ifextended
                             \sep \typnever
\fi
                             \sep \typpermi{\typ} \sep \typopt{\typ} $ \\
                &           & $ \mid $ & $ \structname \sep \typfun{\mode}{\callability}{\lifetime}{\modeu_1}{\typ_1}{\modeu_2}{\typ_2} $ \\
value           & $ \val $ & ::= & $ \integer \sep \exprunit
\ifextended
                             \sep \exprbottom
\fi
                             \sep \exprpermi{v} \sep \exprnone{\typ} \sep \exprsome{\val}{\typ} $ \\
                &           & $ \mid $ & $ \exprstruct{\structname}{\val_1,\ldots,\val_n} \sep \exprfun{\mode}{\callability}{\lifetime}{x}{\modeu}{\typ}{\expr} $ \\
expression      & $ \expr $ & ::= & $ x \sep \integer \sep \expr_1 + \expr_2 \sep \exprunit
\ifextended
                             \sep \exprbottom \sep \exprdefault{\typ} \sep \exprcrashnever{\expr}
\fi
                             $ \\
\ifextended
                &           & $ \mid $ & $ \exprhdata \sep \exprhread \sep \exprhwrite{\expr} $ \\
\fi
                &           & $ \mid $ & $ \exprpermi{v} \sep \exprpdata{\expr_p} \sep \exprpreadi{\expr_p} \sep \exprpwritei{\expr_v}{\expr_p} $ \\
                &           & $ \mid $ & $ \exprdrop{\expr} \sep \exprcopy{\expr} \sep \exprseq{\expr_1}{\expr_2} \sep \exprlet{\mode}{x}{\expr_1}{\expr_2} $ \\
                &           & $ \mid $ & $ \exprnone{\typ} \sep \exprsome{\expr}{\typ} \sep \exprifsome{x}{\expr_1}{\expr_2}{\expr_3} $ \\
                &           & $ \mid $ & $ \exprstruct{\structname}{\expr_1,\ldots,\expr_n} \sep \exprletstruct{\structname}{x_1,\ldots,x_n}{\expr_1}{\expr_2} $ \\
                &           & $ \mid $ & $ \exprfun{\mode}{\callability}{\lifetime}{x}{\modeu}{\typ}{\expr} \sep \exprapp{\expr_1}{\expr_2} $ \\
datatype decl   & $ \datadecl $ & ::= & $ \declstruct{\structname}{\structfield{\mode_1}{\typ_1},\ldots,\structfield{\mode_n}{\typ_n}} $ \\
datatype decls\!\!  & $ \datadecls $ & ::= & $ \datadecl_1,\ldots,\datadecl_n $ \\
\ifextended
heap value      & $ \heapval $ & ::= & $\val$ \\
heap type       & $ \heaptyp $ & ::= & $\typ$ \\
\fi
permission env\!\!\!\!\!\!  & $ \penv $ & ::= & $ \{\integer_1 \mapsto \usage_1, \ldots, \integer_n \mapsto \usage_n\} $ \\
variable env    & $ \xenv $ & ::= & $ \{x_1 \mapsto \utyp{\modeu_1}{\typ_1}, \ldots, x_n \mapsto \utyp{\modeu_n}{\typ_n}\} $ \\
\ifextended
lax checking    & $ \laxity $ & ::= & $ \lstrict \sep \llax $ \\
\fi
\end{tabular}
\caption{Formal Model Language Syntax}
\label{fig:formal:syntax}
\end{figure}

\begin{figure}
\begin{flushleft}
{\small

$\modeof{\spec} = \spec \quad\quad \modeof{\utyp{\mode}{\usage}} = \mode$

$\islinear{\modeu} = \textrm{true}$ iff $\modeu = \modu{\mode}{\linear}$

\begin{tabular}{ll}
$\envn\penv = \{\integer \mapsto \shared \;|\; \integer \mapsto \shared \in \penv\}$
&
$\envn\xenv = \{x \mapsto \utyp{\modeu}{\typ} \;|\; x \mapsto \utyp{\modeu}{\typ} \in \xenv \wedge \neg \islinear{\modeu}\}$
\\
$\envl\penv =  \{\integer \mapsto \linear \;|\; \integer \mapsto \linear \in \penv\}$
&
$\envl\xenv =  \{x \mapsto \utyp{\modeu}{\typ} \;|\; x \mapsto \utyp{\modeu}{\typ} \in \xenv \wedge \islinear{\modeu}\}$
\end{tabular}

\begin{tabular}{ll}
$\envlinear{\penv} = \{\integer \mapsto \linear \;|\; \integer \mapsto \usage \in \penv\}$
&
$\envlinear{\xenv} = \{x \mapsto \utyp{(\modu{\mode}{\linear})}{\typ} \;|\; x \mapsto \utyp{(\modu{\mode}{\usage})}{\typ} \in \xenv\}$
\\
$\envshared{\penv} = \{\integer \mapsto \shared \;|\; \integer \mapsto \usage \in \penv\}$
&
$\envshared{\xenv} = \{x \mapsto \utyp{(\modu{\mode}{\shared})}{\typ} \;|\; x \mapsto \utyp{(\modu{\mode}{\usage})}{\typ} \in \xenv\}$
\\
$ $
&
$\envspec{\xenv} = \{x \mapsto \utyp{\spec}{\typ} \;|\; x \mapsto \utyp{\modeu}{\typ} \in \xenv\}$
\end{tabular}

$\penv = \envmerge{\penv_1}{\penv_2}$ iff $\envl\penv = \envl\penv_1, \envl\penv_2$ and $\envn\penv = \envn\penv_1 = \envn\penv_2$

$\xenv = \envmerge{\xenv_1}{\xenv_2}$ iff $\envl\xenv = \envl\xenv_1, \envl\xenv_2$ and $(\envn\xenv, \envspec{\envl\xenv)} = (\envn\xenv_1, \envspec{\envl\xenv_1)} = (\envn\xenv_2, \envspec{\envl\xenv_2})$

Define $\isunrestricted{\modeu}{\typ}$ to mean: $\modeu \neq (\modu{\mode}{\shared})$
and $\ljudgment{\modeof{\modeu}}{\typ}{\static}$

Define $\isstatic{\xenv}$ to mean: for all $x\mapsto\modu{\modeu}{\typ}\in\xenv$, $\lifetimeof{\typ} = \static$

Define $\nonspecfunctionmodes{m_f}{\modeu_x}{\modeu_b}{\typ_b}$ to mean:
$\isunrestricted{\modeu_b}{\typ_b}$ and
$\mode_f \neq spec$ and
$\mode_f \modeless \modeof{\modeu_x}$ and
$\mode_f \modeless \modeof{\modeu_b}$

Define $\functionbodycontext{\callability}{\lifetime}{\penv}{\xenv}{\penv_b}{\xenv_b}{\usage}$ to mean:
\begin{itemize}
\item if $\callability = \once$ and $\lifetime = \restricted$ then $\penv_b = \penv$ and $\xenv_b = \xenv$
\item if $\callability = \many$ and $\lifetime = \restricted$ then $\penv = \envn\penv$ and $\xenv = \envn\xenv$ and $\penv_b = \penv$ and $\xenv_b = \xenv$
\item if $\callability = \many$ and $\lifetime = \static$ then $\penv = \envn\penv$ and $\penv_b = \varnothing$ and $\xenv = \envn\xenv$ and $\xenv_b = \envspec{\xenv}$
\item if $\callability = \once$ and $\lifetime = \static$ then $\penv_b = \envl\penv$ and $\xenv_b = \envl\xenv, \envspec{\envn\xenv}$ and $\isstatic{\envl\xenv}$
\item if $\callability = \once$ then $\usage = \linear$
\end{itemize}

Define $\lifetimeof{\typ}$ to be:
\begin{itemize}
  \item $\lifetimeof{\typfun{\mode}{\callability}{\lifetime}{\modeu_1}{\typ_1}{\modeu_2}{\typ_2}} = \lifetime$
  \item $\lifetimeof{\typopt{\typ}} = \lifetimeof{\typ}$
  \item $\lifetimeof{\typ} = \static$ for all other $\typ$
\end{itemize}
}
\end{flushleft}
\vspace{-2mm}
\caption{Notation and definitions for type checking}
\label{fig:formal:typingdefs}
\end{figure}

\begin{figure}
\begin{flushleft}{\bf Well-typed expression (main rules)} $\ejudgment{\datadecls}{\heaptyp}{\penv}{\xenv}{\mode}{\laxity}{\expr}{\modeu}{\typ}$\end{flushleft}
{\small

$
\inferrule{
  \mode \modeless \modeof{\modeu_x}
}{
  \ejudgment{\datadecls}{\heaptyp}{\envn\penv}{\envn\xenv, x\mapsto\utyp{\modeu_x}{\typ_x}}{\mode}{\laxity}{x}{\modeu_x}{\typ_x}
}
$
\;\;\;\;\;\;\;\;
$
  \ejudgment{\datadecls}{\heaptyp}{\envn\penv}{\envn\xenv, x\mapsto\utyp{\modu{\mode_x}{\shared}}{\typ_x}}{\mode}{\laxity}{x}{\spec}{\typ_x}
$
\medskip\;\;\;\;\;\;\;\;
$
  \ejudgment{\datadecls}{\heaptyp}{\envn\penv}{\envn\xenv}{\mode}{\laxity}{\integer}{\modeu}{\typint}
$
\medskip\;\;\;\;\;\;\;\;
\inferrule{
  \ejudgment{\datadecls}{\heaptyp}{\penv_1}{\xenv_1}{\mode}{\laxity}{\expr_1}{\modeu}{\typint}
  \\
  \ejudgment{\datadecls}{\heaptyp}{\penv_2}{\xenv_2}{\mode}{\laxity}{\expr_2}{\modeu}{\typint}
}{
  \ejudgment{\datadecls}{\heaptyp}{\envmerge{\penv_1}{\penv_2}}{\envmerge{\xenv_1}{\xenv_2}}{\mode}{\laxity}{\expr_1 + \expr_2}{\modeu}{\typint}
}
\medskip\;\;\;\;\;\;\;\;
$
  \ejudgment{\datadecls}{\heaptyp}{\envn\penv}{\envn\xenv}{\mode}{\laxity}{\exprunit}{\modeu}{\typunit}
$
\medskip\;\;\;\;\;\;\;\;
\ifextended
$
  \ejudgment{\datadecls}{\heaptyp}{\envn\penv}{\envn\xenv}{\mode}{\laxity}{\exprbottom}{\spec}{\typnever}
$
\medskip\;\;\;\;\;\;\;\;
\inferrule{
  \tjudgment{\datadecls}{\typ}
}{
  \ejudgment{\datadecls}{\heaptyp}{\envn\penv}{\envn\xenv}{\mode}{\laxity}{\exprdefault{\typ}}{\spec}{\typ}
}
\medskip\;\;\;\;\;\;\;\;
\inferrule{
  \ejudgment{\datadecls}{\heaptyp}{\penv}{\xenv}{\mode}{\laxity}{\expr}{\modeu}{\typnever}
  \\
  \modeu \neq \spec
}{
  \ejudgment{\datadecls}{\heaptyp}{\penv}{\xenv}{\mode}{\laxity}{\exprcrashnever{\expr}}{\modeu}{\typunit}
}
\medskip\;\;\;\;\;\;\;\;
$
  \ejudgment{\datadecls}{\heaptyp}{\envn\penv}{\envn\xenv}{\exec}{\laxity}{\exprhdata}{\spec}{\heaptyp}
$
\medskip\;\;\;\;\;\;\;\;
$
  \ejudgment{\datadecls}{\heaptyp}{\envn\penv}{\envn\xenv}{\exec}{\laxity}{\exprhread}{\modu{\exec}{\linear}}{\heaptyp}
$
\medskip\;\;\;\;\;\;\;\;
\inferrule{
  \ejudgment{\datadecls}{\heaptyp}{\penv}{\xenv}{\exec}{\laxity}{\expr}{\modu{\exec}{\linear}}{\heaptyp}
}{
  \ejudgment{\datadecls}{\heaptyp}{\penv}{\xenv}{\exec}{\laxity}{\exprhwrite{\expr}}{\modeu}{\typunit}
}
\medskip\;\;\;\;\;\;\;\;
\fi
\inferrule{
  \ejudgment{\datadecls}{\heaptyp}{\envn\penv}{\envn\xenv}{\mode}{\laxity}{\val}{\modu{\exec}{\linear}}{\typ}
  \\
  \cjudgment{\datadecls}{\exec}{\typ}
}{
  \ejudgment{\datadecls}{\heaptyp}{\envn\penv, \integer\mapsto\usage}{\envn\xenv}{\mode}{\laxity}{\exprpermi{\val}}{\modu{\prf}{\usage}}{\typpermi{\typ}}
}
\medskip\;\;\;\;\;\;\;\;
\inferrule{
  \ejudgment{\datadecls}{\heaptyp}{\penv}{\xenv}{\mode}{\laxity}{\expr}{\spec}{\typpermi{\typ}}
}{
  \ejudgment{\datadecls}{\heaptyp}{\penv}{\xenv}{\mode}{\laxity}{\exprpdata{\expr}}{\spec}{\typ}
}
\medskip\;\;\;\;\;\;\;\;
\inferrule{
  \ejudgment{\datadecls}{\heaptyp}{\penv}{\xenv}{\exec}{\laxity}{\expr_p}{\modu{\prf}{\shared}}{\typpermi{\typ}}
}{
  \ejudgment{\datadecls}{\heaptyp}{\penv}{\xenv}{\exec}{\laxity}{\exprpreadi{\expr_p}}{\modu{\exec}{\shared}}{\typ}
}
\medskip\;\;\;\;\;\;\;\;
\inferrule{
  \ejudgment{\datadecls}{\heaptyp}{\penv_1}{\xenv_1}{\exec}{\laxity}{\expr_v}{\modu{\exec}{\linear}}{\typ'}
  \\
  \ejudgment{\datadecls}{\heaptyp}{\penv_2}{\xenv_2}{\exec}{\laxity}{\expr_p}{\modu{\prf}{\linear}}{\typpermi{\typ}}
  \\
  \cjudgment{\datadecls}{\exec}{\typ'}
}{
  \ejudgment{\datadecls}{\heaptyp}{\envmerge{\penv_1}{\penv_2}}{\envmerge{\xenv_1}{\xenv_2}}{\exec}{\laxity}{\exprpwritei{\expr_v}{\expr_p}}{\modu{\prf}{\linear}}{\typ'}
}
\medskip\;\;\;\;\;\;\;\;
\inferrule{
  \ejudgment{\datadecls}{\heaptyp}{\penv}{\xenv}{\mode}{\laxity}{\expr}{\modu{\mode_e}{\linear}}{\typ}
  \\
  \cjudgment{\datadecls}{\mode_e}{\typ}
}{
  \ejudgment{\datadecls}{\heaptyp}{\penv}{\xenv}{\mode}{\laxity}{\exprdrop{\expr}}{\modu{\mode_e}{\shared}}{\typ}
}
\medskip\;\;\;\;\;\;\;\;
\inferrule{
  \ejudgment{\datadecls}{\heaptyp}{\penv}{\xenv}{\mode}{\laxity}{\expr}{\modu{\mode_e}{\shared}}{\typ}
  \\
  \cjudgment{\datadecls}{\mode_e}{\typ}
}{
  \ejudgment{\datadecls}{\heaptyp}{\penv}{\xenv}{\mode}{\laxity}{\exprcopy{\expr}}{\modu{\mode_e}{\linear}}{\typ}
}
\medskip\;\;\;\;\;\;\;\;
\inferrule{
  \ejudgment{\datadecls}{\heaptyp}{\penv_1, \envshared{\penv_b}}{\xenv_1, \envshared{\xenv_b}}{\mode}{\laxity}{\expr_1}{\modeu_1}{\typunit}
  \\
  \ejudgment{\datadecls}{\heaptyp}{\penv_2, \envlinear{\penv_b}}{\xenv_2, \envlinear{\xenv_b}}{\mode}{\laxity}{\expr_2}{\modeu_2}{\typ_2}
}{
  \ejudgment{\datadecls}{\heaptyp}{(\envmerge{\penv_1}{\penv_2}), \envlinear{\penv_b}}{(\envmerge{\xenv_1}{\xenv_2}), \envlinear{\xenv_b}}{\mode}{\laxity}{\exprseq{\expr_1}{\expr_2}}{\modeu_2}{\typ_2}
}
\medskip\;\;\;\;\;\;\;\;
\inferrule{
  \ejudgment{\datadecls}{\heaptyp}{\penv_1, \envshared{\penv_b}}{\xenv_1, \envshared{\xenv_b}}{\mode}{\laxity}{\expr_1}{\modeu_1}{\typ_1}
  \\
  \ejudgment{\datadecls}{\heaptyp}{\penv_2, \envlinear{\penv_b}}{\xenv_2, \envlinear{\xenv_b}, x\mapsto\modu{\modeu_1}{\typ_1}}{\mode}{\laxity}{\expr_2}{\modeu_2}{\typ_2}
  \\
  \isunrestricted{\modeu_1}{\typ_1}\:\textrm{or}\:(\penv_b = \varnothing\:\textrm{and}\:\xenv_b = \varnothing)
  \\
  \ljudgment{\modeof{\modeu_2}}{\typ_2}{\static}
  \\
  \mode_1 = \modeof{\modeu_1}
  \\
  \mode \modeless \mode_1
}{
  \ejudgment{\datadecls}{\heaptyp}{(\envmerge{\penv_1}{\penv_2}), \envlinear{\penv_b}}{(\envmerge{\xenv_1}{\xenv_2}), \envlinear{\xenv_b}}{\mode}{\laxity}
    {\exprlet{\mode_1}{x}{\expr_1}{\expr_2}}{\modeu_2}{\typ_2}
}
\medskip\;\;\;\;\;\;\;\;
\inferrule{
  \tjudgment{\datadecls}{\typ}
}{
  \ejudgment{\datadecls}{\heaptyp}{\envn\penv}{\envn\xenv}{\mode}{\laxity}{\exprnone{\typ}}{\modeu}{\typopt{\typ}}
}
\medskip\;\;\;\;\;\;\;\;
\inferrule{
  \ejudgment{\datadecls}{\heaptyp}{\penv}{\xenv}{\mode}{\laxity}{\expr}{\modeu}{\typ}
}{
  \ejudgment{\datadecls}{\heaptyp}{\penv}{\xenv}{\mode}{\laxity}{\exprsome{\expr}{\typ}}{\modeu}{\typopt{\typ}}
}
\medskip\;\;\;\;\;\;\;\;
\inferrule{
  \ejudgment{\datadecls}{\heaptyp}{\penv_1}{\xenv_1}{\mode}{\laxity}{\expr_1}{\modeu_1}{\typopt{\typ_1}}
  \\
  \ejudgment{\datadecls}{\heaptyp}{\penv_b}{\xenv_b, x\mapsto{\modu{\modeu_1}{\typ_1}}}{\mode_b}{\laxity}{\expr_2}{\modeu_b}{\typ_b}
  \\
  \ejudgment{\datadecls}{\heaptyp}{\penv_b}{\xenv_b}{\mode_b}{\laxity}{\expr_3}{\modeu_b}{\typ_b}
  \\
  \mode \modeless \mode_b
  \\
  (\modeof{\modeu_1} \modeless \mode_b)\:\textrm{or}\:(\modeof{\modeu_1} = \spec\:\textrm{and}\:\mode_b = \prf)
}{
  \ejudgment{\datadecls}{\heaptyp}{\envmerge{\penv_1}{\penv_b}}{\envmerge{\xenv_1}{\xenv_b}}{\mode}{\laxity}{\exprifsome{x}{\expr_1}{\expr_2}{\expr_3}}{\modeu_b}{\typ_b}
}
}
\caption{Type Checking Rules}
\label{fig:formal:typing}
\end{figure}

\begin{figure}
\begin{flushleft}{\bf Well-typed expression (main rules, continued)}\end{flushleft}
{\small
$
\inferrule{
  \datadecls = \ldots, \declstruct{\structname}{\utyp{\mode_1}{\typ_1},\ldots,\utyp{\mode_n}{\typ_n}}, \ldots
  \\
  \forall 1\le i \le n,\;\ejudgment{\datadecls}{\heaptyp}{\penv_i}{\xenv_i}{\mode}{\laxity}{\expr_i}{(\mode_i\modejoin\modeu)}{\typ_i}
}{
  \ejudgment{\datadecls}{\heaptyp}{\envmergen{\penv_1}{\penv_n}}{\envmergen{\xenv_1}{\xenv_n}}{\mode}{\laxity}{\exprstruct{\structname}{\expr_1,\ldots,\expr_n}}{\modeu}{\structname}
}
$
\medskip\;\;\;\;\;\;\;\;
\inferrule{
  \datadecls = \ldots, \declstruct{\structname}{\utyp{\mode_1}{\typ_1},\ldots,\utyp{\mode_n}{\typ_n}}, \ldots
  \\
  \ejudgment{\datadecls}{\heaptyp}{\penv_0}{\xenv_0}{\mode}{\laxity}{\expr_0}{\modeu_0}{\structname}
  \\
  \ejudgment{\datadecls}{\heaptyp}{\penv_b}{\xenv_b, x_1\mapsto\utyp{(\mode_1\modejoin\modeu_0)}{\typ_1}, \ldots, x_n\mapsto\utyp{(\mode_n\modejoin\modeu_0)}{\typ_n}}{\mode}{\laxity}{\expr_b}{\modeu_b}{\typ_b}
  \\
  \ljudgment{\modeof{\modeu_b}}{\typ_b}{\static}
}{
  \ejudgment{\datadecls}{\heaptyp}{\envmerge{\penv_0}{\penv_b}}{\envmerge{\xenv_0}{\xenv_b}}{\mode}{\laxity}{\exprletstruct{\structname}{x_1,\ldots,x_n}{\expr_0}{\expr_b}}{\modeu_b}{\typ_b}
}
\medskip\;\;\;\;\;\;\;\;
\inferrule{
  \ejudgment{\datadecls}{\heaptyp}{\penv_b}{\xenv_b, x\mapsto \utyp{\modeu_x}{\typ_x}}{\mode_f}{\laxity}{\expr_b}{\modeu_b}{\typ_b}
  \\
  \tjudgment{\datadecls}{\typ_x}
  \\
  \functionbodycontext{\callability}{\lifetime}{\penv}{\xenv}{\penv_b}{\xenv_b}{\usage}
  \\
  \nonspecfunctionmodes{\mode_f}{\modeu_x}{\modeu_b}{\typ_b}
}{
  \ejudgment{\datadecls}{\heaptyp}{\penv}{\xenv}{\mode}{\laxity}{(\exprfun{\mode_f}{\callability}{\lifetime}{x}{\modeu_x}{\typ_x}{\expr_b})}{\utyp{\mode_f}{\usage}}{(\typfun{\mode_f}{\callability}{\lifetime}{\modeu_x}{\typ_x}{\modeu_b}{\typ_b})}
}
\medskip\;\;\;\;\;\;\;\;
\inferrule{
  \ejudgment{\datadecls}{\heaptyp}{\envn\penv}{\envn\xenv, x\mapsto \utyp{\spec}{\typ_x}}{\spec}{\laxity}{\expr_b}{\spec}{\typ_b}
  \\
  \tjudgment{\datadecls}{\typ_x}
}{
  \ejudgment{\datadecls}{\heaptyp}{\envn\penv}{\envn\xenv}{\mode}{\laxity}{(\exprfun{\spec}{\many}{\static}{x}{\spec}{\typ_x}{\expr_b})}{\spec}{(\typfun{\spec}{\many}{\static}{\spec}{\typ_x}{\spec}{\typ_b})}
}
\medskip\;\;\;\;\;\;\;\;
\inferrule{
  \callability = \once\Longrightarrow \islinear{\modeu_1}
  \\
  \ejudgment{\datadecls}{\heaptyp}{\penv_1}{\xenv_1}{\mode}{\laxity}{\expr_f}{\modeu_1}{(\typfun{\mode_f}{\callability}{\lifetime}{\modeu_a}{\typ_a}{\modeu_b}{\typ_b})}
  \\
  \ejudgment{\datadecls}{\heaptyp}{\penv_2}{\xenv_2}{\mode}{\laxity}{\expr_a}{\modeu_a}{\typ_a}
  \\
  \modeof{\modeu_1} \modeless \mode_f
  \\
  \mode \modeless \mode_f
}{
  \ejudgment{\datadecls}{\heaptyp}{\envmerge{\penv_1}{\penv_2}}{\envmerge{\xenv_1}{\xenv_2}}{\mode}{\laxity}{\exprapp{\expr_f}{\expr_a}}{\modeu_b}{\typ_b}
}
\ifextended
\medskip\;\;\;\;\;\;\;\;
\inferrule{
  \ejudgment{\datadecls}{\heaptyp}{\penv}{\xenv}{\mode}{\lstrict}{\expr}{\modeu}{\typ}
  \\
  \ejudgment{\datadecls}{\heaptyp}{\varnothing}{\envspec{\xenv}}{\exec}{\lstrict}{\heapval}{\modu{\exec}{\linear}}{\heaptyp}
  \\
  \cjudgment{\datadecls}{\exec}{\heaptyp}
  \\
  \lifetimeof{\heaptyp} = \static
}{
  \sjudgment{\datadecls}{\heaptyp}{\penv}{\xenv}{\mode}{\heapval}{\expr}{\modeu}{\typ}
}
\fi
}

\begin{flushleft}{\bf Well-typed expression (dead-end rules)} $\ejudgment{\datadecls}{\heaptyp}{\penv}{\xenv}{\mode}{\laxity}{\expr}{\modeu}{\typ}$\end{flushleft}
{\small
$
\inferrule{
  \envspec{\xenv} = \envspec{\xenv'}
  \\
  \modeof{\modeu} = \modeof{\modeu'}
  \\
  \ejudgment{\datadecls}{\heaptyp}{\penv'}{\xenv'}{\mode}{\llax}{\expr}{\modeu'}{\typ}
}{
  \ejudgment{\datadecls}{\heaptyp}{\penv}{\xenv}{\mode}{\llax}{\expr}{\modeu}{\typ}
}
$
\medskip\;\;\;\;\;\;\;\;
\inferrule{
  \ejudgment{\datadecls}{\heaptyp}{\envn\penv}{\envn\xenv}{\mode}{\laxity}{\val}{\modeu}{\typ}
}{
  \ejudgment{\datadecls}{\heaptyp}{\envn\penv}{\envn\xenv}{\mode}{\laxity}{\exprpermi{\val}}{\spec}{\typpermi{\typ}}
}
\medskip\;\;\;\;\;\;\;\;
\inferrule{
  \ejudgment{\datadecls}{\heaptyp}{\envn\penv}{\envn\xenv, x\mapsto \utyp{\modeu_x}{\typ_x}}{\mode_f}{\llax}{\expr_b}{\modeu_b}{\typ_b}
  \\
  \tjudgment{\datadecls}{\typ_x}
  \\
  \nonspecfunctionmodes{\mode_f}{\modeu_x}{\modeu_b}{\typ_b}
}{
  \ejudgment{\datadecls}{\heaptyp}{\envn\penv}{\envn\xenv}{\mode}{\laxity}{(\exprfun{\mode_f}{\callability}{\lifetime}{x}{\modeu_x}{\typ_x}{\expr_b})}{\spec}{(\typfun{\mode_f}{\callability}{\lifetime}{\modeu_x}{\typ_x}{\modeu_b}{\typ_b})}
}
\medskip\;\;\;\;\;\;\;\;
\inferrule{
  \ejudgment{\datadecls}{\heaptyp}{\envn\penv}{\envn\xenv, x\mapsto \utyp{\modeu_x}{\typ_x}}{\mode_f}{\llax}{\expr_b}{\modeu_b}{\typ_b}
  \\
  \tjudgment{\datadecls}{\typ_x}
  \\
  \nonspecfunctionmodes{\mode_f}{\modeu_x}{\modeu_b}{\typ_b}
}{
  \ejudgment{\datadecls}{\heaptyp}{\envn\penv}{\envn\xenv}{\mode}{\laxity}{(\exprfun{\mode_f}{\once}{\lifetime}{x}{\modeu_x}{\typ_x}{\expr_b})}{\modu{\mode_f}{\shared}}{(\typfun{\mode_f}{\once}{\lifetime}{\modeu_x}{\typ_x}{\modeu_b}{\typ_b})}
}
}

\ifextended

\begin{flushleft}{\bf Well-formed types} $\trjudgment{\datadecls}{\datadecls_r}{\datadecls_p}{\typ}$ \end{flushleft}
{\small
$
\trjudgment{\datadecls}{\datadecls_r}{\datadecls_p}{\typint}
$
\;\;\;\;\;\;\;\;
$
\trjudgment{\datadecls}{\datadecls_r}{\datadecls_p}{\typunit}
$
\;\;\;\;\;\;\;\;
$
\trjudgment{\datadecls}{\datadecls_r}{\datadecls_p}{\typnever}
$
\;\;\;\;\;\;\;\;
$
\inferrule{
  \trjudgment{\datadecls}{\datadecls_r}{\datadecls_p}{\typ}
}{
  \trjudgment{\datadecls}{\datadecls_r}{\datadecls_p}{\typpermi{\typ}}
}
$
\;\;\;\;\;\;\;\;
\inferrule{
  \trjudgment{\datadecls}{\datadecls_r, \datadecls_p}{\varnothing}{\typ}
}{
  \trjudgment{\datadecls}{\datadecls_r}{\datadecls_p}{\typopt{\typ}}
}
\medskip\;\;\;\;\;\;\;\;
\inferrule{
  \datadecls, \datadecls_r = \ldots, \declstruct{\structname}{\ldots}, \ldots
}{
  \trjudgment{\datadecls}{\datadecls_r}{\datadecls_p}{\structname}
}
\medskip\;\;\;\;\;\;\;\;
\inferrule{
  \trjudgment{\datadecls}{\datadecls_r, \datadecls_p}{\varnothing}{\typ_1}
  \\
  \trjudgment{\datadecls}{\datadecls_r, \datadecls_p}{\varnothing}{\typ_2}
  \\
  \isunrestricted{\modeu_2}{\typ_2}
}{
  \trjudgment{\datadecls}{\datadecls_r}{\datadecls_p}{\typfun{\exec}{\callability}{\lifetime}{\modeu_1}{\typ_1}{\modeu_2}{\typ_2}}
}
\medskip\;\;\;\;\;\;\;\;
\inferrule{
  \prf \modeless \modeof{\modeu_1}
  \\
  \prf \modeless \modeof{\modeu_2}
  \\
  \trjudgment{\datadecls}{\varnothing}{\varnothing}{\typ_1}
  \\
  \trjudgment{\datadecls}{\datadecls_r}{\datadecls_p}{\typ_2}
  \\
  \isunrestricted{\modeu_2}{\typ_2}
}{
  \trjudgment{\datadecls}{\datadecls_r}{\datadecls_p}{\typfun{\prf}{\callability}{\lifetime}{\modeu_1}{\typ_1}{\modeu_2}{\typ_2}}
}
\medskip\;\;\;\;\;\;\;\;
\inferrule{
  \trjudgment{\datadecls}{\varnothing}{\varnothing}{\typ_1}
  \\
  \trjudgment{\datadecls}{\datadecls_r}{\datadecls_p}{\typ_2}
}{
  \trjudgment{\datadecls}{\datadecls_r}{\datadecls_p}{\typfun{\spec}{\many}{\static}{\spec}{\typ_1}{\spec}{\typ_2}}
}

}
\fi

\caption{Type Checking Rules, continued}
\label{fig:formal:typing2}
\end{figure}

\ifextended
\begin{figure}
\begin{flushleft}{\bf Evaluation context} $\efill{\expr}$\end{flushleft}
{\small

\begin{tabular}{lrrl}
& $ \ectx $ & ::= & $
  \ehole
  \sep \ectx_1 + \expr_2
  \sep \val_1 + \ectx_2
  \sep \exprcrashnever{\ectx}
  \sep \exprhwrite{\ectx}
  \sep \exprpdata{\ectx}
  $ \\
  & & $ \mid $ & $
  \exprpreadi{\ectx}
  \sep \exprpwritei{\ectx}{\expr_p}
  \sep \exprpwritei{\val}{\ectx}
  \sep \exprdrop{\ectx}
  \sep \exprcopy{\ectx}
  $ \\
  & & $ \mid $ & $
  \exprseq{\ectx_1}{\expr_2}
  \sep \exprlet{\mode}{x}{\ectx_1}{\expr_2}
  \sep \exprsome{\ectx}{\typ}
  \sep \exprifsome{x}{\ectx}{\expr_2}{\expr_3}
  $ \\
  & & $ \mid $ & $
  \exprstruct{\structname}{\val_1,\ldots,\val_i,\ectx_j,\expr_k,\ldots,\expr_n}
  \sep \exprletstruct{\structname}{x_1,\ldots,x_n}{\ectx_0}{\expr_b}
  \sep \exprapp{\ectx_1}{\expr_2}
  \sep \exprapp{\val_1}{\ectx_2}
  $ \\
\end{tabular}

}
\begin{flushleft}{\bf Evaluation rules} $(\heapval, \expr) \fullstep (\heapval', \expr')$ (implicitly in a context $\datadecls$)\end{flushleft}
{\small

\[
\inferrule{
  \expr\step\expr'
}{
  (\heapval, \expr) \step (\heapval, \expr')
}
\;\;\;\;\;\;\;\;
\inferrule{
  (\heapval, \expr) \step (\heapval', \expr')
}{
  (\heapval, \efill{\expr})\fullstep(\heapval', \efill{\expr'})
}
\]

\[
\inferrule{
  i_3 = \textrm{sum of } i_1, i_2
}{
  i_1 + i_2 \step i_3
}
\;\;\;\;\;\;\;\;
\inferrule{
  \defjudgment{\datadecls}{\typ}{\val}
}{
  \exprdefault{\typ} \step \val
}
\]

\[
(\heapval, \exprhdata) \step (\heapval, \heapval)
\;\;\;\;\;\;\;\;
(\heapval, \exprhread) \step (\heapval, \heapval)
\;\;\;\;\;\;\;\;
(\heapval, \exprhwrite{\val}) \step (\val, \exprunit)
\]

\[
\exprpdata{\exprpermi{\val}} \step \val
\;\;\;\;\;\;\;\;
\exprpreadi{\exprpermi{\val}} \step \val
\]

\[
\exprpwritei{\val'}{\exprpermi{\val}} \step \exprpermi{\val'}
\]

\[
\exprdrop{\val} \step ()
\;\;\;\;\;\;\;\;
\exprcopy{\val} \step \val
\;\;\;\;\;\;\;\;
\exprseq{\exprunit}{\expr_2} \step \expr_2
\;\;\;\;\;\;\;\;
\exprlet{\mode}{x}{\val_1}{\expr_2} \step \subst{\expr_2}{x}{\val_1}
\]

\[
\exprifsome{x}{\exprnone{\typ}}{\expr_2}{\expr_3} \step \expr_3
\]

\[
\exprifsome{x}{\exprsome{\val}{\typ}}{\expr_2}{\expr_3} \step \subst{\expr_2}{x}{\val}
\]

\[
\exprletstruct{\structname}{x_1,\ldots,x_n}{\exprstruct{\structname}{\val_1,\ldots,\val_n}}{\expr_b}
\step \msubst{\expr_b}{x_1}{\val_1}{x_n}{\val_n}
\]

\[
\exprapp{(\exprfun{\mode}{\callability}{\lifetime}{x}{\modeu_x}{\typ_x}{\expr_b})}{\val_x} \step \subst{\expr_b}{x}{\val_x}
\]

$ $

(Note: $\exprcrashnever{\exprbottom}$ does not step.  By not stepping, it ``crashes''.)

}

\caption{Evaluation Rules}
\label{fig:formal:evaluation}
\end{figure}

\begin{figure}
\begin{flushleft}{\bf Default values} $\defjudgment{\datadecls}{\typ}{\val}$\end{flushleft}
{\small

\[
\defjudgment{\datadecls}{\typint}{0}
\;\;\;\;\;\;\;\;
\defjudgment{\datadecls}{\typunit}{\exprunit}
\;\;\;\;\;\;\;\;
\defjudgment{\datadecls}{\typnever}{\exprbottom}
\]

\[
\inferrule{
  \defjudgment{\datadecls}{\typ}{\val}
}{
  \defjudgment{\datadecls}{\typpermi{\typ}}{\exprpermi{\val}}
}
\;\;\;\;\;\;\;\;
\defjudgment{\datadecls}{\typopt{\typ}}{\exprnone{\typ}}
\]

\[
\inferrule{
  \datadecls = \ldots, \declstruct{\structname}{\utyp{\mode_1}{\typ_1},\ldots,\utyp{\mode_n}{\typ_n}}, \ldots
  \\
  \defjudgment{\datadecls}{\typ_1}{\val_1}
  \\
  \ldots
  \\
  \defjudgment{\datadecls}{\typ_n}{\val_n}
}{
  \defjudgment{\datadecls}{\structname}{\exprstruct{\structname}{\val_1,\ldots,\val_n}}
}
\]

\[
\defjudgment{\datadecls}
  {(\typfun{\mode}{\callability}{\lifetime}{\modeu_1}{\typ_1}{\modeu_2}{\typ_2})}
  {(\exprfun{\mode}{\callability}{\lifetime}{x}{\modeu_1}{\typ_1}{\exprdefault{\typ_2}})}
\]

}
\begin{flushleft}{\bf Copyable types} $\cjudgment{\datadecls}{\mode}{\typ}$\end{flushleft}
{\small

\[
\cjudgment{\datadecls}{\spec}{\typ}
\;\;\;\;\;\;\;\;
\cjudgment{\datadecls}{\mode}{\typint}
\;\;\;\;\;\;\;\;
\cjudgment{\datadecls}{\mode}{\typunit}
\;\;\;\;\;\;\;\;
\cjudgment{\datadecls}{\mode}{\typnever}
\]

\[
\inferrule{
  \datadecls = \ldots, \declstruct{\structname}{\utyp{\mode_1}{\typ_1},\ldots,\utyp{\mode_n}{\typ_n}}, \ldots
  \\
  \cjudgment{\datadecls}{\mode_1}{\typ_1}
  \\
  \ldots
  \\
  \cjudgment{\datadecls}{\mode_n}{\typ_n}
}{
  \cjudgment{\datadecls}{\mode}{\structname}
}
\]

\[
\inferrule{
  \cjudgment{\datadecls}{\mode}{\typ}
}{
  \cjudgment{\datadecls}{\mode}{\typopt{\typ}}
}
\;\;\;\;\;\;\;\;
  \cjudgment{\datadecls}{\mode}{(\typfun{\mode_f}{\many}{\lifetime}{\modeu_1}{\typ_1}{\modeu_2}{\typ_2})}
\]

}
\caption{Additional Rules}
\label{fig:formal:additional}
\end{figure}

\fi


\section{Related work}\label{sec:related}

Many tools for verifying Rust code exist.
As far as we know, no other tool leverages Rust's borrow checker to enforce linear ghost permissions.
However, in other dimensions, there is significant overlap between Verus and other projects.

Creusot~\cite{denis:hal-03737878} may be the closest tool to Verus,
since it uses Rust code to express specifications and proofs,
based on a macro named Pearlite.
Creusot functions can be annotated as \verb`#[logic]` or \verb`#[predicate]`
to indicate that the functions are ghost.
These are similar to Verus' \verb`spec` functions,
in that they are not checked for linearity and borrowing
(``Pearlite formulas are type-checked by the front-end of the Rust compiler,
but they are not borrow checked'').
Creusot does not have ghost code that is checked for linearity and borrowing,
the way Verus' \verb`proof` functions and \verb`proof` variables are.
Verus' SMT-LIB encoding is conceptually similar to the one produced 
by Creusot~\cite{denis:hal-03737878} via the Why3~\cite{boogie11why3} prover,
which requires an intermediate step: in Creusot the Rust code is first
lowered into Why3's MLCFG (an ML with labelled blocks and gotos),
and then Why3 encodes verification conditions for the backend solvers.

Prusti~\cite{DBLP:journals/pacmpl/Astrauskas0PS19,DBLP:conf/nfm/AstrauskasBFGMM22}
verifies Rust code by translating it into the Viper separation logic engine~\cite{DBLP:conf/vmcai/0001SS16},
effectively reverifying ownership properties enforced by Rust's borrow checker.
This relatively heavyweight encoding creates larger formulas for an SMT solver,
but can be used for Rust \verb`unsafe` code that subverts Rust's borrow checking rules.
By contrast, \verus relies on the memory safety enforced by Rust's borrow checker,
obviating the need to use separation logic ubiquitously---instead,
the user can selectively apply separation logic-style techniques
(based on linear ghost permissions) only for the tricky cases that require them.

Aeneas~\cite{DBLP:journals/pacmpl/HoP22} verifies Rust code by translating it into a purely functional
representation in \fstar.
In this style of verification,
programmers develop a proof about the functional representation of executable Rust code,
which is quite different from Verus' Hoare-logic style,
where the programmer annotates the Rust code with preconditions, postconditions, and loop invariants.



\citet{DBLP:journals/pacmpl/YanovskiDJD21} propose a datatype called \verb`GhostCell`, which separates
data from permission in a manner similar to our \verb`PCell` and \verb`PermData`.
The main difference is that \verb`GhostCell` employs a polymorphic type trick to enforce that a permission may
only be used with the cells to which it corresponds, while \verb`PCell` uses a \verb`requires` clause to enforce this,
which is more flexible and allows permissions to depend on data that is not statically determined during type-checking.
Furthermore, while \verb`GhostCell` is used to enforce memory safety, to our knowledge, it has not been used to show functional correctness properties.

RustBelt~\cite{DBLP:journals/pacmpl/0002JKD18} is a verification framework that establishes
a semantic model for type safety in Rust: it allows a user to verify \verb`unsafe` code
with safe APIs, i.e., prove that any well-typed, \verb`unsafe`-free Rust program using the API
will be memory safe.
This makes it complementary to \verus, which relies on that memory safety, and indeed, it might be possible to use RustBelt to verify \verus' memory primitives (\verb`PPtr` and \verb`PCell`) and their specifications.
RustBelt can also handle atomics with relaxed memory ordering~\cite{DBLP:journals/pacmpl/DangJKD20}, which \verus does not support.
RustBelt is implemented in Coq, and thus proofs are written via tactics rather than by SMT.

RustBelt has also been used as part of RustHornBelt~\cite{DBLP:conf/pldi/MatsushitaDJD22}, which validates RustHorn~\cite{DBLP:conf/esop/0002T020},
the encoding used by Creusot.
However, RustHornBelt still requires that \verb`unsafe` code be proved correct in Coq,
while \verus provides safe, zero-cost alternatives to commonly used \verb`unsafe` Rust features via its linear ghost state.
Specifically, \verus provides \verb`PPtr` for raw pointers and \verb`PCell` for \verb`UnsafeCell`, so that users can write code (which would otherwise need those
\verb`unsafe` features) within \verus itself.

Note though that while \verus supports some \verb`unsafe` use-cases, including raw pointers,
our specification for pointers
is very simple, only handling pointers that point into heap allocations from the global
memory allocator. A complete pointer model for Rust would support pointers to the
stack variables, cell interiors, struct fields, references, and so on, as well as handle
thorny issues such as pointer provenance.
By comparison, Stacked Borrows~\cite{StackedBorrows} is a promising operational semantics 
for Rust memory accesses that aims to handle all these concepts.

Separation logic~\cite{DBLP:conf/lics/Reynolds02, DBLP:journals/tcs/OHearn07, Iris31}
was one inspiration for our linear ghost permissions,
although the techniques used in Verus and separation logic are quite different.
In separation logic, a permission is part of the logic rather than a program-level value,
and two permissions are combined together using separating conjunction.
In Verus, permissions are values and two permissions are combined together by placing them in a datatype.
Thus, in Verus, programmers manipulate permissions directly as data,
which can require extra programmer effort,
but makes generating verification conditions for an SMT solver much easier,
since SMT solvers handle classical logic, not separation logic.

Another inspiration for linear ghost permissions was earlier work on
using linearity in type systems to manage changing
state~\cite{crary1999capcalculus, smith2000aliastypes, zhu2005atsviews, morrisett2005l3}
Alias Types~\cite{smith2000aliastypes}, for example, tracks a set of constraints on the memory state,
and these constraints change linearly as the memory state evolves.
ATS~\cite{zhu2005atsviews} combines this idea, in the form of ``stateful views'',
with reasoning about integer arithmetic via a simple dependent type system.
Most similar to our approach is L3~\cite{morrisett2005l3},
which treats ``capabilities'' (permissions) as first-class linear ghost values,
as in Verus.  L3 uses type variables (specifically, location variables)
to connect the capabilities to pointers, whereas
Verus uses SMT solving, which avoids the burden on the programmer of
universally quantifying or existentially quantifying over location variables.
The combination of SMT solving and Rust's automated borrow checking means that
ideas from ATS and L3 are now not only possible within a mainstream language,
but convenient.

Dafny~\cite{DBLP:conf/lpar/Leino10} and \fstar~\cite{mumon} support ghost code and ghost variables.
\fstar uses an effect system to distinguish ghost functions from executable functions,
and has an \verb`erased` type to represent ghost data.
\fstar does not have a linear type system,
although the \fstar Steel system~\cite{DBLP:journals/pacmpl/FromherzRSGMMR21}
supports separation logic reasoning.
Dafny supports \verb`ghost` annotations on variables,
similar to Verus \verb`spec` variables,
and Dafny supports \verb`lemma`s, similar to Verus \verb`proof` functions.
Linear Dafny~\cite{DBLP:journals/pacmpl/LiLZCHPH22} extends Dafny with linear types and borrowing,
although the linearity and borrowing is less sophisticated than in Rust (for example,
Linear Dafny lacks lifetime variables).


\section{Conclusions}\label{sec:conclusions}

By taking advantage of Rust's linearity and borrow checking,
Verus can express linear ghost permissions that aid the verification
of tricky, low-level and/or concurrent code.
This allows Verus to safely express code that would be unsafe in ordinary Rust,
and to prove strong correctness guarantees about the code.
Even for more straightforward code,
Rust's type safety and control over aliasing makes verification considerably easier,
allowing \verus' generation of verification conditions to treat Rust code more as functional code
than as imperative code.
In other words, we've found that one of the most valuable tools for verifying Rust code is Rust itself.
So we conclude with a simple slogan for Verus' style of verification:
ask not what verification can do for Rust --- ask what Rust can do for verification.


\begin{acks}                            
  The authors would like to thank Jay Bosamiya, Nikhil Swamy, Guido Martinez,
  and the anonymous reviewers
  for their help and suggestions on the paper.
  Work at CMU was supported, in part, by a gift from VMware, 
  the Alfred P.\ Sloan Foundation, the Intel Corporation,
  and the Future Enterprise Security initiative at Carnegie Mellon CyLab (FutureEnterprise@CyLab).
  At ETH Zurich Andrea Lattuada was supported, in part, by a gift from VMware.
\end{acks}

\bibliography{bib}


\begin{thebibliography}{44}


\ifx \showCODEN    \undefined \def \showCODEN     #1{\unskip}     \fi
\ifx \showDOI      \undefined \def \showDOI       #1{#1}\fi
\ifx \showISBNx    \undefined \def \showISBNx     #1{\unskip}     \fi
\ifx \showISBNxiii \undefined \def \showISBNxiii  #1{\unskip}     \fi
\ifx \showISSN     \undefined \def \showISSN      #1{\unskip}     \fi
\ifx \showLCCN     \undefined \def \showLCCN      #1{\unskip}     \fi
\ifx \shownote     \undefined \def \shownote      #1{#1}          \fi
\ifx \showarticletitle \undefined \def \showarticletitle #1{#1}   \fi
\ifx \showURL      \undefined \def \showURL       {\relax}        \fi
\providecommand\bibfield[2]{#2}
\providecommand\bibinfo[2]{#2}
\providecommand\natexlab[1]{#1}
\providecommand\showeprint[2][]{arXiv:#2}

\bibitem[Amani et~al\mbox{.}(2016)]%
        {cogent}
\bibfield{author}{\bibinfo{person}{Sidney Amani}, \bibinfo{person}{Alex Hixon},
  \bibinfo{person}{Zilin Chen}, \bibinfo{person}{Christine Rizkallah},
  \bibinfo{person}{Peter Chubb}, \bibinfo{person}{Liam O'Connor},
  \bibinfo{person}{Joel Beeren}, \bibinfo{person}{Yutaka Nagashima},
  \bibinfo{person}{Japheth Lim}, \bibinfo{person}{Thomas Sewell},
  \bibinfo{person}{Joseph Tuong}, \bibinfo{person}{Gabriele Keller},
  \bibinfo{person}{Toby Murray}, \bibinfo{person}{Gerwin Klein}, {and}
  \bibinfo{person}{Gernot Heiser}.} \bibinfo{year}{2016}\natexlab{}.
\newblock \showarticletitle{Cogent: Verifying High-Assurance File System
  Implementations}. In \bibinfo{booktitle}{\emph{Proceedings of the ACM
  Conference on Architectural Support for Programming Languages and Operating
  Systems (ASPLOS)}}.
\newblock
\urldef\tempurl%
\url{https://doi.org/10.1145/2872362.2872404}
\showDOI{\tempurl}


\bibitem[Astrauskas et~al\mbox{.}(2022)]%
        {DBLP:conf/nfm/AstrauskasBFGMM22}
\bibfield{author}{\bibinfo{person}{Vytautas Astrauskas}, \bibinfo{person}{Aurel
  B{\'{\i}}l{\'{y}}}, \bibinfo{person}{Jon{\'{a}}s Fiala},
  \bibinfo{person}{Zachary Grannan}, \bibinfo{person}{Christoph Matheja},
  \bibinfo{person}{Peter M{\"{u}}ller}, \bibinfo{person}{Federico Poli}, {and}
  \bibinfo{person}{Alexander~J. Summers}.} \bibinfo{year}{2022}\natexlab{}.
\newblock \showarticletitle{The {Prusti} Project: Formal Verification for
  {Rust}}. In \bibinfo{booktitle}{\emph{{NASA} Formal Methods - 14th
  International Symposium, {NFM} 2022, Pasadena, CA, USA, May 24-27, 2022,
  Proceedings}} \emph{(\bibinfo{series}{LNCS}, Vol.~\bibinfo{volume}{13260})}.
  \bibinfo{publisher}{Springer}, \bibinfo{pages}{88--108}.
\newblock
\urldef\tempurl%
\url{https://doi.org/10.1007/978-3-031-06773-0\_5}
\showDOI{\tempurl}


\bibitem[Astrauskas et~al\mbox{.}(2019)]%
        {DBLP:journals/pacmpl/Astrauskas0PS19}
\bibfield{author}{\bibinfo{person}{Vytautas Astrauskas}, \bibinfo{person}{Peter
  M{\"{u}}ller}, \bibinfo{person}{Federico Poli}, {and}
  \bibinfo{person}{Alexander~J. Summers}.} \bibinfo{year}{2019}\natexlab{}.
\newblock \showarticletitle{Leveraging {Rust} Types for Modular Specification
  and Verification}.
\newblock \bibinfo{journal}{\emph{Proc. {ACM} Program. Lang.}}
  \bibinfo{volume}{3}, \bibinfo{number}{{OOPSLA}} (\bibinfo{year}{2019}),
  \bibinfo{pages}{147:1--147:30}.
\newblock
\urldef\tempurl%
\url{https://doi.org/10.1145/3360573}
\showDOI{\tempurl}


\bibitem[Barnett et~al\mbox{.}(2005)]%
        {boogie}
\bibfield{author}{\bibinfo{person}{Michael Barnett},
  \bibinfo{person}{Bor{-}Yuh~Evan Chang}, \bibinfo{person}{Robert DeLine},
  \bibinfo{person}{Bart Jacobs}, {and} \bibinfo{person}{K.~Rustan~M. Leino}.}
  \bibinfo{year}{2005}\natexlab{}.
\newblock \showarticletitle{Boogie: {A} Modular Reusable Verifier for
  Object-Oriented Programs}. In \bibinfo{booktitle}{\emph{Formal Methods for
  Components and Objects, 4th International Symposium, {FMCO} 2005, Amsterdam,
  The Netherlands, November 1-4, 2005, Revised Lectures}}
  \emph{(\bibinfo{series}{LNCS}, Vol.~\bibinfo{volume}{4111})}.
  \bibinfo{publisher}{Springer}, \bibinfo{pages}{364--387}.
\newblock
\urldef\tempurl%
\url{https://doi.org/10.1007/11804192\_17}
\showDOI{\tempurl}


\bibitem[Barrett et~al\mbox{.}(2010)]%
        {smtlib}
\bibfield{author}{\bibinfo{person}{Clark Barrett}, \bibinfo{person}{Aaron
  Stump}, {and} \bibinfo{person}{Cesare Tinelli}.}
  \bibinfo{year}{2010}\natexlab{}.
\newblock \showarticletitle{{The SMT-LIB Standard: Version 2.0}}. In
  \bibinfo{booktitle}{\emph{Proceedings of the 8th International Workshop on
  Satisfiability Modulo Theories (Edinburgh, UK)}},
  \bibfield{editor}{\bibinfo{person}{A.~Gupta} {and}
  \bibinfo{person}{D.~Kroening}} (Eds.).
\newblock


\bibitem[Bobot et~al\mbox{.}(2011)]%
        {boogie11why3}
\bibfield{author}{\bibinfo{person}{Fran\c{c}ois Bobot},
  \bibinfo{person}{Jean-Christophe Filli\^atre}, \bibinfo{person}{Claude
  March\'e}, {and} \bibinfo{person}{Andrei Paskevich}.}
  \bibinfo{year}{2011}\natexlab{}.
\newblock \showarticletitle{Why3: Shepherd Your Herd of Provers}. In
  \bibinfo{booktitle}{\emph{Boogie 2011: First International Workshop on
  Intermediate Verification Languages}}. \bibinfo{address}{Wroc\l{}aw, Poland},
  \bibinfo{pages}{53--64}.
\newblock
\newblock
\shownote{\url{https://hal.inria.fr/hal-00790310}}.


\bibitem[Borgida et~al\mbox{.}(1995)]%
        {on-the-frame-problem}
\bibfield{author}{\bibinfo{person}{Alexander Borgida}, \bibinfo{person}{John
  Mylopoulos}, {and} \bibinfo{person}{Raymond Reiter}.}
  \bibinfo{year}{1995}\natexlab{}.
\newblock \showarticletitle{On the Frame Problem in Procedure Specifications}.
\newblock \bibinfo{journal}{\emph{{IEEE} Trans. Software Eng.}}
  \bibinfo{volume}{21}, \bibinfo{number}{10} (\bibinfo{year}{1995}),
  \bibinfo{pages}{785--798}.
\newblock
\urldef\tempurl%
\url{https://doi.org/10.1109/32.469460}
\showDOI{\tempurl}


\bibitem[Cohen and Leino(2020)]%
        {dafny:unsoundness}
\bibfield{author}{\bibinfo{person}{Ernie Cohen} {and} \bibinfo{person}{Rustan
  Leino}.} \bibinfo{year}{2020}\natexlab{}.
\newblock \bibinfo{title}{Dafny issue 851: unsoundness: dafny seems to assume
  tuple and inductive datatypes are inhabited}.
\newblock
\newblock
\urldef\tempurl%
\url{https://github.com/dafny-lang/dafny/issues/851}
\showURL{%
\tempurl}


\bibitem[{Coq Development Team}(2022)]%
        {coq}
\bibfield{author}{\bibinfo{person}{{Coq Development Team}}.}
  \bibinfo{year}{2022}\natexlab{}.
\newblock \bibinfo{title}{The {C}oq {P}roof {A}ssistant}.
\newblock \bibinfo{howpublished}{\url{https://coq.inria.fr/}}.
\newblock


\bibitem[Crary et~al\mbox{.}(1999)]%
        {crary1999capcalculus}
\bibfield{author}{\bibinfo{person}{Karl Crary}, \bibinfo{person}{David Walker},
  {and} \bibinfo{person}{Greg Morrisett}.} \bibinfo{year}{1999}\natexlab{}.
\newblock \showarticletitle{Typed Memory Management in a Calculus of
  Capabilities}. In \bibinfo{booktitle}{\emph{Proceedings of the 26th ACM
  SIGPLAN-SIGACT Symposium on Principles of Programming Languages}}
  \emph{(\bibinfo{series}{POPL '99})}.
\newblock
\urldef\tempurl%
\url{https://doi.org/10.1145/292540.292564}
\showDOI{\tempurl}


\bibitem[Dang et~al\mbox{.}(2020)]%
        {DBLP:journals/pacmpl/DangJKD20}
\bibfield{author}{\bibinfo{person}{Hoang{-}Hai Dang},
  \bibinfo{person}{Jacques{-}Henri Jourdan}, \bibinfo{person}{Jan{-}Oliver
  Kaiser}, {and} \bibinfo{person}{Derek Dreyer}.}
  \bibinfo{year}{2020}\natexlab{}.
\newblock \showarticletitle{RustBelt meets relaxed memory}.
\newblock \bibinfo{journal}{\emph{Proc. {ACM} Program. Lang.}}
  \bibinfo{volume}{4}, \bibinfo{number}{{POPL}} (\bibinfo{year}{2020}),
  \bibinfo{pages}{34:1--34:29}.
\newblock
\urldef\tempurl%
\url{https://doi.org/10.1145/3371102}
\showDOI{\tempurl}


\bibitem[de~Moura et~al\mbox{.}(2015)]%
        {lean2015}
\bibfield{author}{\bibinfo{person}{Leonardo de Moura}, \bibinfo{person}{Soonho
  Kong}, \bibinfo{person}{Jeremy Avigad}, \bibinfo{person}{Floris van Doorn},
  {and} \bibinfo{person}{Jakob von Raumer}.} \bibinfo{year}{2015}\natexlab{}.
\newblock \showarticletitle{The {Lean} Theorem Prover}. In
  \bibinfo{booktitle}{\emph{Proceedings of the Conference on Automated
  Deduction (CADE)}}.
\newblock


\bibitem[de~Moura and Bj{\o}rner(2007)]%
        {e-matching}
\bibfield{author}{\bibinfo{person}{Leonardo~Mendon{\c{c}}a de Moura} {and}
  \bibinfo{person}{Nikolaj~S. Bj{\o}rner}.} \bibinfo{year}{2007}\natexlab{}.
\newblock \showarticletitle{Efficient E-Matching for {SMT} Solvers}. In
  \bibinfo{booktitle}{\emph{Automated Deduction - CADE-21, 21st International
  Conference on Automated Deduction, Bremen, Germany, July 17-20, 2007,
  Proceedings}} \emph{(\bibinfo{series}{Lecture Notes in Computer Science},
  Vol.~\bibinfo{volume}{4603})}, \bibfield{editor}{\bibinfo{person}{Frank
  Pfenning}} (Ed.). \bibinfo{publisher}{Springer}, \bibinfo{pages}{183--198}.
\newblock
\urldef\tempurl%
\url{https://doi.org/10.1007/978-3-540-73595-3\_13}
\showDOI{\tempurl}


\bibitem[de~Moura and Bj{\o}rner(2008)]%
        {z3}
\bibfield{author}{\bibinfo{person}{Leonardo~Mendon{\c{c}}a de Moura} {and}
  \bibinfo{person}{Nikolaj~S. Bj{\o}rner}.} \bibinfo{year}{2008}\natexlab{}.
\newblock \showarticletitle{{Z3:} An Efficient {SMT} Solver}. In
  \bibinfo{booktitle}{\emph{Tools and Algorithms for the Construction and
  Analysis of Systems, 14th International Conference, {TACAS} 2008, Held as
  Part of the Joint European Conferences on Theory and Practice of Software,
  {ETAPS} 2008, Budapest, Hungary, March 29-April 6, 2008. Proceedings}}
  \emph{(\bibinfo{series}{LNCS}, Vol.~\bibinfo{volume}{4963})}.
  \bibinfo{publisher}{Springer}, \bibinfo{pages}{337--340}.
\newblock
\urldef\tempurl%
\url{https://doi.org/10.1007/978-3-540-78800-3\_24}
\showDOI{\tempurl}


\bibitem[Denis et~al\mbox{.}(2022)]%
        {denis:hal-03737878}
\bibfield{author}{\bibinfo{person}{Xavier Denis},
  \bibinfo{person}{Jacques-Henri Jourdan}, {and} \bibinfo{person}{Claude
  March{\'e}}.} \bibinfo{year}{2022}\natexlab{}.
\newblock \showarticletitle{Creusot: A Foundry for the Deductive Verication of
  {Rust} Programs}. In \bibinfo{booktitle}{\emph{{Proceedings of ICFEM 2022 -
  International Conference on Formal Engineering Methods}}}
  \emph{(\bibinfo{series}{Lecture Notes in Computer Science})}.
  \bibinfo{publisher}{{Springer Verlag}}, \bibinfo{address}{Madrid, Spain}.
\newblock
\urldef\tempurl%
\url{https://doi.org/10.1007/978-3-031-17244-1_6}
\showDOI{\tempurl}


\bibitem[Dijkstra(1975)]%
        {DBLP:journals/cacm/Dijkstra75}
\bibfield{author}{\bibinfo{person}{Edsger~W. Dijkstra}.}
  \bibinfo{year}{1975}\natexlab{}.
\newblock \showarticletitle{Guarded Commands, Nondeterminacy and Formal
  Derivation of Programs}.
\newblock \bibinfo{journal}{\emph{Commun. {ACM}}} \bibinfo{volume}{18},
  \bibinfo{number}{8} (\bibinfo{year}{1975}), \bibinfo{pages}{453--457}.
\newblock
\urldef\tempurl%
\url{https://doi.org/10.1145/360933.360975}
\showDOI{\tempurl}


\bibitem[Fromherz et~al\mbox{.}(2021)]%
        {DBLP:journals/pacmpl/FromherzRSGMMR21}
\bibfield{author}{\bibinfo{person}{Aymeric Fromherz}, \bibinfo{person}{Aseem
  Rastogi}, \bibinfo{person}{Nikhil Swamy}, \bibinfo{person}{Sydney Gibson},
  \bibinfo{person}{Guido Mart{\'{\i}}nez}, \bibinfo{person}{Denis Merigoux},
  {and} \bibinfo{person}{Tahina Ramananandro}.}
  \bibinfo{year}{2021}\natexlab{}.
\newblock \showarticletitle{Steel: proof-oriented programming in a dependently
  typed concurrent separation logic}.
\newblock \bibinfo{journal}{\emph{Proc. {ACM} Program. Lang.}}
  \bibinfo{volume}{5}, \bibinfo{number}{{ICFP}} (\bibinfo{year}{2021}),
  \bibinfo{pages}{1--30}.
\newblock
\urldef\tempurl%
\url{https://doi.org/10.1145/3473590}
\showDOI{\tempurl}


\bibitem[{Google Security Blog}(2021)]%
        {rustandroid}
\bibfield{author}{\bibinfo{person}{{Google Security Blog}}.}
  \bibinfo{year}{2021}\natexlab{}.
\newblock \bibinfo{title}{Rust in the Android platform}.
\newblock
\newblock
\urldef\tempurl%
\url{https://security.googleblog.com/2021/04/rust-in-android-platform.html}
\showURL{%
\tempurl}


\bibitem[Hance et~al\mbox{.}(2022)]%
        {ironsync-tr}
\bibfield{author}{\bibinfo{person}{Travis Hance}, \bibinfo{person}{Yi Zhou},
  \bibinfo{person}{Andrea Lattuada}, \bibinfo{person}{Reto Achermann},
  \bibinfo{person}{Alex Conway}, \bibinfo{person}{Ryan Stutsman},
  \bibinfo{person}{Gerd Zellweger}, \bibinfo{person}{Chris Hawblitzel},
  \bibinfo{person}{Jon Howell}, {and} \bibinfo{person}{Bryan Parno}.}
  \bibinfo{year}{2022}\natexlab{}.
\newblock \bibinfo{booktitle}{\emph{Sharding the State Machine: Automated
  Modular Reasoning for Complex Concurrent Systems}}.
\newblock \bibinfo{type}{{T}echnical {R}eport} CMU-CyLab-22-003.
  \bibinfo{institution}{CyLab, Carnegie Mellon University}.
\newblock


\bibitem[Ho and Protzenko(2022)]%
        {DBLP:journals/pacmpl/HoP22}
\bibfield{author}{\bibinfo{person}{Son Ho} {and} \bibinfo{person}{Jonathan
  Protzenko}.} \bibinfo{year}{2022}\natexlab{}.
\newblock \showarticletitle{Aeneas: Rust Verification by Functional
  Translation}.
\newblock \bibinfo{journal}{\emph{Proc. {ACM} Program. Lang.}}
  \bibinfo{volume}{6}, \bibinfo{number}{{ICFP}} (\bibinfo{year}{2022}),
  \bibinfo{pages}{711--741}.
\newblock
\urldef\tempurl%
\url{https://doi.org/10.1145/3547647}
\showDOI{\tempurl}


\bibitem[Jacobs et~al\mbox{.}(2011)]%
        {VeriFast}
\bibfield{author}{\bibinfo{person}{Bart Jacobs}, \bibinfo{person}{Jan Smans},
  \bibinfo{person}{Pieter Philippaerts}, \bibinfo{person}{Fr{\'{e}}d{\'{e}}ric
  Vogels}, \bibinfo{person}{Willem Penninckx}, {and} \bibinfo{person}{Frank
  Piessens}.} \bibinfo{year}{2011}\natexlab{}.
\newblock \showarticletitle{VeriFast: {A} Powerful, Sound, Predictable, Fast
  Verifier for {C} and Java}. In \bibinfo{booktitle}{\emph{{NASA} Formal
  Methods - Third International Symposium, {NFM} 2011, Pasadena, CA, USA, April
  18-20, 2011. Proceedings}} \emph{(\bibinfo{series}{Lecture Notes in Computer
  Science}, Vol.~\bibinfo{volume}{6617})},
  \bibfield{editor}{\bibinfo{person}{Mihaela~Gheorghiu Bobaru},
  \bibinfo{person}{Klaus Havelund}, \bibinfo{person}{Gerard~J. Holzmann}, {and}
  \bibinfo{person}{Rajeev Joshi}} (Eds.). \bibinfo{publisher}{Springer},
  \bibinfo{pages}{41--55}.
\newblock
\urldef\tempurl%
\url{https://doi.org/10.1007/978-3-642-20398-5\_4}
\showDOI{\tempurl}


\bibitem[Jung et~al\mbox{.}(2019)]%
        {StackedBorrows}
\bibfield{author}{\bibinfo{person}{Ralf Jung}, \bibinfo{person}{Hoang-Hai
  Dang}, \bibinfo{person}{Jeehoon Kang}, {and} \bibinfo{person}{Derek Dreyer}.}
  \bibinfo{year}{2019}\natexlab{}.
\newblock \showarticletitle{Stacked Borrows: An Aliasing Model for {Rust}}.
\newblock \bibinfo{journal}{\emph{Proc. ACM Program. Lang.}}
  \bibinfo{volume}{4}, \bibinfo{number}{POPL}, Article \bibinfo{articleno}{41}
  (\bibinfo{date}{dec} \bibinfo{year}{2019}), \bibinfo{numpages}{32}~pages.
\newblock
\urldef\tempurl%
\url{https://doi.org/10.1145/3371109}
\showDOI{\tempurl}


\bibitem[Jung et~al\mbox{.}(2018a)]%
        {DBLP:journals/pacmpl/0002JKD18}
\bibfield{author}{\bibinfo{person}{Ralf Jung}, \bibinfo{person}{Jacques{-}Henri
  Jourdan}, \bibinfo{person}{Robbert Krebbers}, {and} \bibinfo{person}{Derek
  Dreyer}.} \bibinfo{year}{2018}\natexlab{a}.
\newblock \showarticletitle{{RustBelt}: Securing the Foundations of the {Rust}
  Programming Language}.
\newblock \bibinfo{journal}{\emph{Proc. {ACM} Program. Lang.}}
  \bibinfo{volume}{2}, \bibinfo{number}{{POPL}} (\bibinfo{year}{2018}),
  \bibinfo{pages}{66:1--66:34}.
\newblock
\urldef\tempurl%
\url{https://doi.org/10.1145/3158154}
\showDOI{\tempurl}


\bibitem[Jung et~al\mbox{.}(2018b)]%
        {Iris31}
\bibfield{author}{\bibinfo{person}{Ralf Jung}, \bibinfo{person}{Robbert
  Krebbers}, \bibinfo{person}{Jacques-Henri Jourdan}, \bibinfo{person}{Ales
  Bizjak}, \bibinfo{person}{Lars Birkedal}, {and} \bibinfo{person}{Derek
  Dreyer}.} \bibinfo{year}{2018}\natexlab{b}.
\newblock \showarticletitle{Iris from the ground up: A modular foundation for
  higher-order concurrent separation logic}.
\newblock \bibinfo{journal}{\emph{Journal of Functional Programming}}
  (\bibinfo{year}{2018}).
\newblock
\urldef\tempurl%
\url{https://doi.org/10.1017/S0956796818000151}
\showDOI{\tempurl}


\bibitem[Klabnik and Nichols(2018)]%
        {the-rust-programming-language}
\bibfield{author}{\bibinfo{person}{Steve Klabnik} {and} \bibinfo{person}{Carol
  Nichols}.} \bibinfo{year}{2018}\natexlab{}.
\newblock \bibinfo{booktitle}{\emph{The Rust Programming Language}}.
\newblock \bibinfo{publisher}{No Starch Press}, \bibinfo{address}{USA}.
\newblock
\showISBNx{1593278284}


\bibitem[Lattuada et~al\mbox{.}(2023)]%
        {verussupplementary}
\bibfield{author}{\bibinfo{person}{Andrea Lattuada}, \bibinfo{person}{Travis
  Hance}, \bibinfo{person}{Chanhee Cho}, \bibinfo{person}{Matthias Brun},
  \bibinfo{person}{Isitha Subasinghe}, \bibinfo{person}{Yi Zhou},
  \bibinfo{person}{Jon Howell}, \bibinfo{person}{Bryan Parno}, {and}
  \bibinfo{person}{Chris Hawblitzel}.} \bibinfo{year}{2023}\natexlab{}.
\newblock \bibinfo{booktitle}{\emph{{Verus: Verifying Rust Programs using
  Linear Ghost Types -- Supplementary Material}}}.
\newblock
\urldef\tempurl%
\url{https://doi.org/10.5281/zenodo.7718486}
\showDOI{\tempurl}
\newblock
\shownote{The copy of record of the supplementary material is available in the
  ACM DL.}.


\bibitem[Leino(2010)]%
        {DBLP:conf/lpar/Leino10}
\bibfield{author}{\bibinfo{person}{K.~Rustan~M. Leino}.}
  \bibinfo{year}{2010}\natexlab{}.
\newblock \showarticletitle{Dafny: An Automatic Program Verifier for Functional
  Correctness}. In \bibinfo{booktitle}{\emph{Logic for Programming, Artificial
  Intelligence, and Reasoning - 16th International Conference, LPAR-16, Dakar,
  Senegal, April 25-May 1, 2010, Revised Selected Papers}}
  \emph{(\bibinfo{series}{LNCS}, Vol.~\bibinfo{volume}{6355})}.
  \bibinfo{publisher}{Springer}, \bibinfo{pages}{348--370}.
\newblock
\urldef\tempurl%
\url{https://doi.org/10.1007/978-3-642-17511-4\_20}
\showDOI{\tempurl}


\bibitem[Leino and Rümmer(2010)]%
        {poly-boogie}
\bibfield{author}{\bibinfo{person}{Rustan Leino} {and} \bibinfo{person}{Philipp
  Rümmer}.} \bibinfo{year}{2010}\natexlab{}.
\newblock \showarticletitle{A Polymorphic Intermediate Verification Language:
  Design and Logical Encoding}. In \bibinfo{booktitle}{\emph{Conference: Tools
  and Algorithms for the Construction and Analysis of Systems, 16th
  International Conference, TACAS 2010, Held as Part of the Joint European
  Conferences on Theory and Practice of Software, ETAPS 2010, Paphos, Cyprus,
  March 20-28, 2010.}}
\newblock
\urldef\tempurl%
\url{https://doi.org/978-3-642-12002-2_26}
\showDOI{\tempurl}


\bibitem[Li et~al\mbox{.}(2022)]%
        {DBLP:journals/pacmpl/LiLZCHPH22}
\bibfield{author}{\bibinfo{person}{Jialin Li}, \bibinfo{person}{Andrea
  Lattuada}, \bibinfo{person}{Yi Zhou}, \bibinfo{person}{Jonathan Cameron},
  \bibinfo{person}{Jon Howell}, \bibinfo{person}{Bryan Parno}, {and}
  \bibinfo{person}{Chris Hawblitzel}.} \bibinfo{year}{2022}\natexlab{}.
\newblock \showarticletitle{Linear types for large-scale systems verification}.
\newblock \bibinfo{journal}{\emph{Proc. {ACM} Program. Lang.}}
  \bibinfo{volume}{6}, \bibinfo{number}{{OOPSLA}} (\bibinfo{year}{2022}),
  \bibinfo{pages}{1--28}.
\newblock
\urldef\tempurl%
\url{https://doi.org/10.1145/3527313}
\showDOI{\tempurl}


\bibitem[Matsakis and Klock(2014)]%
        {the-rust-language}
\bibfield{author}{\bibinfo{person}{Nicholas~D. Matsakis} {and}
  \bibinfo{person}{Felix~S. Klock}.} \bibinfo{year}{2014}\natexlab{}.
\newblock \showarticletitle{The {Rust} Language}.
\newblock \bibinfo{journal}{\emph{Ada Lett.}} \bibinfo{volume}{34},
  \bibinfo{number}{3} (\bibinfo{date}{Oct.} \bibinfo{year}{2014}),
  \bibinfo{pages}{103--104}.
\newblock
\showISSN{1094-3641}
\urldef\tempurl%
\url{https://doi.org/10.1145/2692956.2663188}
\showDOI{\tempurl}


\bibitem[Matsushita et~al\mbox{.}(2022)]%
        {DBLP:conf/pldi/MatsushitaDJD22}
\bibfield{author}{\bibinfo{person}{Yusuke Matsushita}, \bibinfo{person}{Xavier
  Denis}, \bibinfo{person}{Jacques{-}Henri Jourdan}, {and}
  \bibinfo{person}{Derek Dreyer}.} \bibinfo{year}{2022}\natexlab{}.
\newblock \showarticletitle{{RustHornBelt}: A Semantic Foundation for
  Functional Verification of {Rust} Programs With Unsafe Code}. In
  \bibinfo{booktitle}{\emph{{PLDI} '22: 43rd {ACM} {SIGPLAN} International
  Conference on Programming Language Design and Implementation, San Diego, CA,
  USA, June 13 - 17, 2022}}. \bibinfo{publisher}{{ACM}},
  \bibinfo{pages}{841--856}.
\newblock
\urldef\tempurl%
\url{https://doi.org/10.1145/3519939.3523704}
\showDOI{\tempurl}


\bibitem[Matsushita et~al\mbox{.}(2020)]%
        {DBLP:conf/esop/0002T020}
\bibfield{author}{\bibinfo{person}{Yusuke Matsushita}, \bibinfo{person}{Takeshi
  Tsukada}, {and} \bibinfo{person}{Naoki Kobayashi}.}
  \bibinfo{year}{2020}\natexlab{}.
\newblock \showarticletitle{{RustHorn}: {CHC}-Based Verification for Rust
  Programs}. In \bibinfo{booktitle}{\emph{Programming Languages and Systems -
  29th European Symposium on Programming, {ESOP} 2020, Held as Part of the
  European Joint Conferences on Theory and Practice of Software, {ETAPS} 2020,
  Dublin, Ireland, April 25-30, 2020, Proceedings}}
  \emph{(\bibinfo{series}{LNCS}, Vol.~\bibinfo{volume}{12075})}.
  \bibinfo{publisher}{Springer}, \bibinfo{pages}{484--514}.
\newblock
\urldef\tempurl%
\url{https://doi.org/10.1007/978-3-030-44914-8\_18}
\showDOI{\tempurl}


\bibitem[Morrisett et~al\mbox{.}(2005)]%
        {morrisett2005l3}
\bibfield{author}{\bibinfo{person}{Greg Morrisett}, \bibinfo{person}{Amal
  Ahmed}, {and} \bibinfo{person}{Matthew Fluet}.}
  \bibinfo{year}{2005}\natexlab{}.
\newblock \showarticletitle{L3: A Linear Language with Locations}. In
  \bibinfo{booktitle}{\emph{Typed Lambda Calculi and Applications}}.
\newblock
\urldef\tempurl%
\url{https://doi.org/10.1007/11417170_22}
\showDOI{\tempurl}


\bibitem[M{\"{u}}ller et~al\mbox{.}(2016)]%
        {DBLP:conf/vmcai/0001SS16}
\bibfield{author}{\bibinfo{person}{Peter M{\"{u}}ller}, \bibinfo{person}{Malte
  Schwerhoff}, {and} \bibinfo{person}{Alexander~J. Summers}.}
  \bibinfo{year}{2016}\natexlab{}.
\newblock \showarticletitle{Viper: {A} Verification Infrastructure for
  Permission-Based Reasoning}. In \bibinfo{booktitle}{\emph{Verification, Model
  Checking, and Abstract Interpretation - 17th International Conference,
  {VMCAI} 2016, St. Petersburg, FL, USA, January 17-19, 2016. Proceedings}}
  \emph{(\bibinfo{series}{LNCS}, Vol.~\bibinfo{volume}{9583})}.
  \bibinfo{publisher}{Springer}, \bibinfo{pages}{41--62}.
\newblock
\urldef\tempurl%
\url{https://doi.org/10.1007/978-3-662-49122-5\_2}
\showDOI{\tempurl}


\bibitem[O'Hearn(2007)]%
        {DBLP:journals/tcs/OHearn07}
\bibfield{author}{\bibinfo{person}{Peter~W. O'Hearn}.}
  \bibinfo{year}{2007}\natexlab{}.
\newblock \showarticletitle{Resources, concurrency, and local reasoning}.
\newblock \bibinfo{journal}{\emph{Theor. Comput. Sci.}} \bibinfo{volume}{375},
  \bibinfo{number}{1-3} (\bibinfo{year}{2007}), \bibinfo{pages}{271--307}.
\newblock
\urldef\tempurl%
\url{https://doi.org/10.1016/j.tcs.2006.12.035}
\showDOI{\tempurl}


\bibitem[Pearce(2021)]%
        {pearce-rust-toplas2021}
\bibfield{author}{\bibinfo{person}{David~J. Pearce}.}
  \bibinfo{year}{2021}\natexlab{}.
\newblock \showarticletitle{A Lightweight Formalism for Reference Lifetimes and
  Borrowing in Rust}.
\newblock \bibinfo{journal}{\emph{ACM Transactions on Programming Languages and
  Systems (TOPLAS)}} \bibinfo{volume}{43}, \bibinfo{number}{1}, Article
  \bibinfo{articleno}{3} (\bibinfo{date}{apr} \bibinfo{year}{2021}),
  \bibinfo{numpages}{73}~pages.
\newblock
\urldef\tempurl%
\url{https://doi.org/10.1145/3443420}
\showDOI{\tempurl}


\bibitem[Reynolds(2002)]%
        {DBLP:conf/lics/Reynolds02}
\bibfield{author}{\bibinfo{person}{John~C. Reynolds}.}
  \bibinfo{year}{2002}\natexlab{}.
\newblock \showarticletitle{Separation Logic: {A} Logic for Shared Mutable Data
  Structures}. In \bibinfo{booktitle}{\emph{17th {IEEE} Symposium on Logic in
  Computer Science {(LICS} 2002), 22-25 July 2002, Copenhagen, Denmark,
  Proceedings}}. \bibinfo{publisher}{{IEEE} Computer Society},
  \bibinfo{pages}{55--74}.
\newblock
\urldef\tempurl%
\url{https://doi.org/10.1109/LICS.2002.1029817}
\showDOI{\tempurl}


\bibitem[Smith et~al\mbox{.}(2000)]%
        {smith2000aliastypes}
\bibfield{author}{\bibinfo{person}{Frederick Smith}, \bibinfo{person}{David
  Walker}, {and} \bibinfo{person}{J.~Gregory Morrisett}.}
  \bibinfo{year}{2000}\natexlab{}.
\newblock \showarticletitle{Alias Types}. In
  \bibinfo{booktitle}{\emph{Proceedings of the 9th European Symposium on
  Programming Languages and Systems}} \emph{(\bibinfo{series}{ESOP '00})}.
\newblock


\bibitem[Swamy et~al\mbox{.}(2016)]%
        {mumon}
\bibfield{author}{\bibinfo{person}{Nikhil Swamy},
  \bibinfo{person}{C\u{a}t\u{a}lin Hri\c{t}cu}, \bibinfo{person}{Chantal
  Keller}, \bibinfo{person}{Aseem Rastogi}, \bibinfo{person}{Antoine
  Delignat-Lavaud}, \bibinfo{person}{Simon Forest},
  \bibinfo{person}{Karthikeyan Bhargavan}, \bibinfo{person}{C\'{e}dric
  Fournet}, \bibinfo{person}{Pierre-Yves Strub}, \bibinfo{person}{Markulf
  Kohlweiss}, \bibinfo{person}{Jean-Karim Zinzindohou\'e}, {and}
  \bibinfo{person}{Santiago {Zanella-B\'eguelin}}.}
  \bibinfo{year}{2016}\natexlab{}.
\newblock \showarticletitle{Dependent Types and Multi-Monadic Effects in {F*}}.
  In \bibinfo{booktitle}{\emph{Proceedings of the ACM Symposium on Principles
  of Programming Languages (POPL)}}.
\newblock
\showISBNx{978-1-4503-3549-2}
\urldef\tempurl%
\url{https://doi.org/10.1145/2837614.2837655}
\showDOI{\tempurl}


\bibitem[Vaughan-Nichols(2022)]%
        {rustlinux}
\bibfield{author}{\bibinfo{person}{Steven Vaughan-Nichols}.}
  \bibinfo{year}{2022}\natexlab{}.
\newblock \bibinfo{title}{Linus Torvalds: Rust will go into Linux 6.1}.
\newblock
\newblock
\urldef\tempurl%
\url{https://www.zdnet.com/article/linus-torvalds-rust-will-go-into-linux-6-1/}
\showURL{%
\tempurl}


\bibitem[Wadler(1990)]%
        {DBLP:conf/ifip2/Wadler90}
\bibfield{author}{\bibinfo{person}{Philip Wadler}.}
  \bibinfo{year}{1990}\natexlab{}.
\newblock \showarticletitle{Linear Types can Change the World!}. In
  \bibinfo{booktitle}{\emph{Programming concepts and methods: Proceedings of
  the {IFIP} Working Group 2.2, 2.3 Working Conference on Programming Concepts
  and Methods, Sea of Galilee, Israel, 2-5 April, 1990}}.
  \bibinfo{publisher}{North-Holland}, \bibinfo{pages}{561}.
\newblock


\bibitem[Weiss et~al\mbox{.}(2019)]%
        {DBLP:journals/corr/abs-1903-00982}
\bibfield{author}{\bibinfo{person}{Aaron Weiss}, \bibinfo{person}{Daniel
  Patterson}, \bibinfo{person}{Nicholas~D. Matsakis}, {and}
  \bibinfo{person}{Amal Ahmed}.} \bibinfo{year}{2019}\natexlab{}.
\newblock \showarticletitle{Oxide: The Essence of Rust}.
\newblock \bibinfo{journal}{\emph{CoRR}}  \bibinfo{volume}{abs/1903.00982}
  (\bibinfo{year}{2019}).
\newblock
\showeprint[arXiv]{1903.00982}
\urldef\tempurl%
\url{http://arxiv.org/abs/1903.00982}
\showURL{%
\tempurl}


\bibitem[Yanovski et~al\mbox{.}(2021)]%
        {DBLP:journals/pacmpl/YanovskiDJD21}
\bibfield{author}{\bibinfo{person}{Joshua Yanovski},
  \bibinfo{person}{Hoang{-}Hai Dang}, \bibinfo{person}{Ralf Jung}, {and}
  \bibinfo{person}{Derek Dreyer}.} \bibinfo{year}{2021}\natexlab{}.
\newblock \showarticletitle{{GhostCell}: Separating Permissions from Data in
  {Rust}}.
\newblock \bibinfo{journal}{\emph{Proc. {ACM} Program. Lang.}}
  \bibinfo{volume}{5}, \bibinfo{number}{{ICFP}} (\bibinfo{year}{2021}),
  \bibinfo{pages}{1--30}.
\newblock
\urldef\tempurl%
\url{https://doi.org/10.1145/3473597}
\showDOI{\tempurl}


\bibitem[Zhu and Xi(2005)]%
        {zhu2005atsviews}
\bibfield{author}{\bibinfo{person}{Dengping Zhu} {and} \bibinfo{person}{Hongwei
  Xi}.} \bibinfo{year}{2005}\natexlab{}.
\newblock \showarticletitle{Safe Programming with Pointers through Stateful
  Views}. In \bibinfo{booktitle}{\emph{Proceedings of the 7th International
  Conference on Practical Aspects of Declarative Languages}}
  \emph{(\bibinfo{series}{PADL'05})}.
\newblock
\urldef\tempurl%
\url{https://doi.org/10.1007/978-3-540-30557-6_8}
\showDOI{\tempurl}


\end{thebibliography}

\end{document}